\documentstyle[aps,prd,eqsecnum,floats]{revtex}
\input{psfig}

\def\no{\nonumber}
\def\a{\alpha}
\def\b{\beta}
\def\e{\epsilon}
\def\g{\gamma}

\def\p{\phi}
\def\t{\theta}

\def\be{\begin{equation}}
\def\bea{\begin{eqnarray}}
\def\eea{\end{eqnarray}}
\def\ee{\end{equation}}
\def\bi{\begin{itemize}}
\def\ei{\end{itemize}}
\def\cross{\times}
\def\d{\delta}
\def\D{\Delta}

\def\vt{\vartheta}
\def\vp{\varphi}

\begin{document}
\draft

\preprint{\vbox{\baselineskip=12pt
\rightline{IUCAA-33/2000}
\rightline{AEI-2000-055}
\rightline{}
\rightline{gr-qc/0009078}
}}
\title{A data-analysis strategy for detecting gravitational-wave signals from 
inspiraling compact binaries with a network of laser-interferometric detectors}
\author{
Archana Pai,${}^1$ \footnote{Electronic
address: {\em apai@iucaa.ernet.in}} Sanjeev Dhurandhar,${}^{1,2}$\footnote{
Electronic address: {\em sdh@iucaa.ernet.in}}
and Sukanta Bose${}^{2,3}$\footnote{Electronic address:
{\em bose@aei-potsdam.mpg.de}}}
\address{${}^1$Inter-University Centre for Astronomy and Astrophysics, Post 
Bag 4, Ganeshkhind,\\ Pune 411007, India}
\address{${}^2$Max Planck Institut f\"{u}r Gravitationsphysik,
Albert Einstein Institut, Am M\"{u}hlenberg 1,\\ Golm, D-14476, Germany}
\address{${}^3$Department of Physics and Astronomy, P. O. 
Box 913, Cardiff University, CF24 3YB,\\ United Kingdom}

\maketitle
\begin{abstract}

A data-analysis strategy based on the maximum-likelihood method (MLM)
is presented for the detection
of gravitational waves from inspiraling compact binaries with a network
of laser-interferometric detectors having arbitrary orientations and 
arbitrary 
locations around the globe. For simplicity, we restrict 
ourselves to the Newtonian inspiral waveform. 
However, the formalism we develop 
here is also applicable to a waveform with post-Newtonian (PN) corrections. 
The Newtonian waveform depends on eight parameters:
the distance $r$ to the binary, the phase  $\delta_c$ of the waveform at the 
time of final coalescence, the polarization-ellipse 
angle $\psi$, the angle of inclination $\epsilon$ of the binary orbit to the
line of sight, the source-direction angles $\{\theta, \phi\}$, the time of
final coalescence $t_c$ at the fiducial detector, and the chirp time $\xi$. 
All these parameters are relevant for a chirp search with multiple detectors,
unlike the case of a single detector. 
The primary construct on which the MLM is based is the network 
likelihood ratio (LR). We obtain this ratio here. For the Newtonian inspiral 
waveform, the LR is a function of the eight signal-parameters.
In the MLM-based detection strategy, the LR must be maximized over all 
of these parameters. Here, we show that it is possible 
to maximize it {\em analytically} with respect to four of 
the eight parameters, namely, $\{r, \delta_c, \psi, \epsilon\}$. 
Maximization over the time of arrival is handled most efficiently by using the 
Fast-Fourier-Transform  algorithm, as in the case of a single detector. This 
not only allows us to scan the parameter 
space continuously over these five parameters but also cuts down {\it
substantially} on 
the computational costs. The analytical maximization over the four parameters 
yields the optimal statistic on which the decision must be based.
The value of the statistic also depends on the nature of the noises in the 
detectors. Here, we model these noises to be mainly Gaussian, stationary, 
and uncorrelated for every pair of detectors. Instances of non-Gaussianity,
as are present in 
detector outputs, can be accommodated in our formalism by implementing vetoing
techniques similar to those applied for single detectors. Our formalism not 
only allows us to express the likelihood ratio for the network in a very 
simple and compact form, but also is at the basis of giving an elegant 
geometric interpretation to the detection problem. Maximization of the LR over 
the remaining three parameters is handled as follows. Owing to the arbitrary 
locations of the detectors in a network, the time of arrival of a signal 
at any detector will, in general, be different from those at the others and,
consequently, will result in signal time-delays. {}For a given network, these
time delays are determined by the source-direction angles $\{\theta,\phi\}$.
Therefore, to maximize the LR over the parameters $\{\theta,\phi\}$ 
one needs to scan over the possible time-delays allowed by a network. 
We opt for obtaining a bank of templates for the chirp time and the 
time-delays. This means that
we construct a bank of templates over $\xi$, $\theta$, and
$\phi$. We first discuss ``idealized'' networks with all the detectors 
having a common noise curve for simplicity. Such an exercise nevertheless 
yields useful estimates about computational costs, and also tests the 
formalism developed here. We then consider realistic cases of networks 
comprising of the LIGO and VIRGO detectors: These include two-detector 
networks, which pair up the two LIGOs or VIRGO with one of the LIGOs, and
the three-detector network that includes VIRGO and both the LIGOs.
For these networks we present the computational speed 
requirements, network sensitivities, and source-direction resolutions.

\end{abstract}


\vfil
\pagebreak

\section{Introduction}
\label{sec:intro}

The existence of gravitational waves, which is predicted in the theory of 
general relativity, has long been verified `indirectly' through the 
observations of Hulse and Taylor \cite{HT1}. The inspiral of the members of the
binary pulsar system named after them has been successfully accounted for in 
terms of the back-reaction due to the radiated gravitational waves 
\cite{HT1,HT2}. However, detecting such waves with man-made `antennas' has not
been possible so far. Nevertheless, this problem has received a lot of 
attention this decade, especially, due to the arrival of laser-interferometric 
detectors, which are expected to have sensitivities close to that required for detecting 
such waves.

A gravitational-wave (GW) source that is one of the most promising candidates 
for detection by Earth-based interferometric GW detectors is the
inspiraling compact binary \cite{Thorne}. Present estimates show a significant 
number of coalescence events every year of such binaries that produce waves 
strong enough to be detectable by current detectors during their inspiral 
phase, a few seconds before the onset of coalescence. Moreover, the 
time-evolution of these waveforms (chirps) is well 
understood in the frequency band where the present interferometric detectors
are most sensitive. 

In the past, a sizable amount of research has been done on the problem of 
detecting gravitational waves using a single bar or interferometric detector. 
However, very little work has been devoted to develop techniques to
analyze the data from a network of such detectors. Searching for chirps using 
a network of detectors is gaining importance due to (a) its superior 
sensitivity {\it vis \`{a} vis} that of a constituent detector  and (b) 
improving feasibility for a real-time computational search. As has been argued 
before (see, e.g., Ref \cite{BFS}), for a given false-alarm probability, 
the threshold for detection is lowered as the number of detectors is increased.
This increases the probability of detection by `coherently' analyzing the 
signals from a network rather than a single detector. One can think of 
simpler approaches to the network problem where one matches event
lists from different detectors in the network and sets up thresholds on the
estimated parameter differences. Indeed, in the past, a formalism 
for interpreting coincidences of burst events in a pair of detectors has been 
suggested by Schutz and Tinto \cite{ST}. However, such approaches, even if they
were extended to the case of chirps, would be non-optimal, because they do not 
use the phase information of a signal at different detectors. The coherent 
search strategy described here crucially uses phase information.
 
One of the early papers that came close to discussing the problem 
of detecting a Newtonian chirp using a network optimally
was that of Finn and Chernoff \cite{FC}. This paper observed that
since the orientations of the two LIGO detectors were very similar, their 
joint sensitivity was larger than any one of them. 
Bhawal and Dhurandhar also addressed the issue of detection using multiple 
detectors \cite{BD}. Their main aim was to 
find the optimal recycling mode of operation of the planned laser 
interferometric detectors for which a meaningful coincidence detection of 
broadband signals could be performed. 
However, the issue of how a network of 
detectors with arbitrary orientations and arbitrary locations on the globe
can be optimally used as a ``single" detector of sensitivity higher than that 
of any of its subsets of individual detectors was not addressed in these 
earlier papers. 

Use of a detector network has nevertheless received considerable attention in 
the context of the parameter estimation problem. A formalism for using
the responses of multiple detectors, in the absence of noise, to infer the 
parameters of a chirp (also known as the ``inverse problem'') was developed
by Dhurandhar and Tinto \cite{DT1,DT2}. Some of the other notable works that 
address this issue in the presence of noise are Refs. \cite{GT,JK1,JK2}.
The prime motivation behind using a network in this regard is that the larger 
the number of detectors, the smaller are the errors in estimated values of 
the binary parameters. However, the starting point in these approaches is 
the assumption that the problem  of detection has already been addressed and 
the detector-specific chirp-templates that result in ``super-threshold" 
cross-correlations with the individual detector outputs, have been picked.

A formalism was developed in Ref. \cite{BDP} that sought an optimal detection
strategy for chirps in the simplifying case of a network with closely located 
laser-interferometric 
detectors and with idealized detector noise. This work was based on a coherent
search. Its main result was that 
the optimal statistic for a network of up to three such detectors was 
proven to be the sum of the signal-to-noise (SNR) ratios of the individual
detectors. It was also shown that the sensitivity of such a network improved
as roughly the square-root of the number of detectors in the network. This 
formalism was extended to the case of arbitrarily located detectors in 
Ref. \cite{BPD}, which showed a way to reduce the network statistic such that
the number of chirp parameters over which a numerical search is required 
for detecting chirps drops from eight to three. This paves the way for
a vast reduction in computational speed requirements and makes a multiple
detector search for chirps much more feasible. One of the main aims of this 
paper is to formulate a data-analysis strategy that implements these
formal findings in the case of existing and upcoming networks and to provide
estimates on the required computational speeds, etc.

As the members of a binary orbit around their center of mass, they lose 
energy in the form of gravitational waves. This results in their inspiral. 
Consequently, they emit gravitational waves with monotonically increasing 
amplitude and frequency \cite{PM}. Although
the gravitational waveform originating from an inspiraling binary is known 
accurately up to the 2.5 post-Newtonian (PN) order \cite{BDIWW}, 
nevertheless
as a first calculation we limit our study to the detection of the Newtonian 
chirp. This is because our primary aim here is to develop the new formalism, 
namely, that of optimally using the data from a {\em network} of detectors to 
detect a chirp. We find evidence of the applicability of our formalism to 
higher post-Newtonian orders. We also find for the Newtonian signal that
there is essentially no correlation between the parameters describing the
masses and the direction angles to the source when the noise curves are 
assumed to
be identical for all the detectors in the network. This has the following
important implications: The total number of templates is then just a 
product of number of templates for a single detector and the number of
templates needed to scan source directions. 
If this property holds also for the PN case, then, in effect, we need to 
obtain the number of templates for the directions
only and club together with this the information we have on the number of
templates for a PN signal in a single detector. We hope to address this issue 
in detail in future work.

In our analysis, we assume that the noise in each detector is predominantly 
stationary and Gaussian, with occasional contamination from non-Gaussian 
events. Indeed, the real data stream from the detectors is not expected 
to be purely stationary and Gaussian, unlike what is assumed in most of the GW 
data-analysis literature thus far. In fact, the data from the Caltech 40 meter 
prototype interferometer have the expected broadband noise spectrum, but 
superposed on this are several other noisy features \cite{Allen}, such as 
long-term sinusoidal
disturbances arising from suspensions and electric-main harmonics, which
have been studied in other works \cite{SiSc,MF}. There are also
transients occurring every few minutes, typically due to
servo-controls instabilities or mechanical relaxation in suspension
systems, etc. Adaptive methods are being explored to combat high amplitude 
ring-downs and sinusoids occurring in the data \cite{CD}, by effectively 
removing them from the data, so that the data are `cleaned' from these 
non-Gaussian 
features. In the improved detectors of the future, it is expected that the 
noise will tend to Gaussianity, and may only be occasionally contaminated by 
non-Gaussian events. It is in this spirit that our above assumption about the 
noise must be taken. The strategy we adopt in dealing with such a detector 
noise is to assume it to be stationary and Gaussian for the purposes of our 
statistical analyses. Such an assumption is justified in practice provided 
one vetoes out detections due to the occasional non-Gaussian events occurring 
in the data obtained from the detectors. The vetoing 
criterion we propose is an extension of the $\chi^2$ criterion used in 
Ref. \cite{Allen}. This model of the noise is simple enough so that
analytical methods can be used for the approach we take.

We also assume that noises to be uncorrelated among different pairs of 
detectors.
When the detectors are widely separated around the globe, correlations among
the noises of different detectors are expected to be negligible, and our 
assumption remains valid for such a case. For most networks of proposed  
detectors this is true, unless it consists of the two coincident detectors at 
Hanford.\footnote{The Hanford site has two detectors with arm-lengths of 4 km 
and 2 km, respectively.} In that case, a more general analysis is
necessary that takes into account possible correlations between the noises
in different detectors. Such an approach is being pursued by Finn \cite{Finn}.
The only other assumption we make on the detector noise is that it is 
additive.

We use the maximum-likelihood method for 
optimizing the detection problem. The problem is formulated by obtaining a 
single likelihood-ratio (LR) for the entire network. 
We define the ``network output'', ${\sf x}$,
as a single network-vector, the components of which are the individual 
detector outputs. Similarly, the ``network signal'', ${\sf s}$, is defined 
to be a single network-vector, the components of which are the individual 
detector signals. Since we assume the noise in different detectors to be 
independent, the probability density function (PDF) for the noise of the 
network is just a product of the PDFs of noise in the individual detectors. 
The LR is then a
simple expression in terms of the norm of ${\sf s}$ and the inner product of 
${\sf x}$ and ${\sf s}$. In this form, the LR is a function of a complete set 
of eight independent parameters that characterize the Newtonian chirp signal
of an inspiraling binary. 

{}For the assumptions made on the detector noises, a network's 
logarithmic likelihood ratio (LLR) turns out to be the sum of 
the individual detector's LLRs. The form of LLR allows us to deduce the 
network matched-filter in a straightforward way; it turns 
out to be an $M$-dimensional vector with components that are just the 
normalized single detector templates of Sathyaprakash and Dhurandhar \cite{SD} 
(henceforth referred to as SD). Inferring this network-template is the second 
result of this paper. The problem of detection then reduces to the 
maximization of LLR over the parameters and comparing its maximized value 
with the 
pre-determined detection threshold. We argue that this step can be implemented
in a way similar to the one suggested in SD. 
\par
To obtain the maximum likelihood ratio (MLR), the LR has to be 
maximized over the eight parameters: the distance to the binary system $r$, 
the phase of the waveform  at the time of coalescence $\delta_c$,   
the polarization-ellipse angle $\psi$, the inclination of the binary orbit 
$\epsilon$, the direction angles $\{\theta, \phi\}$, the time of final 
coalescence at the fiducial detector $t_c$, and the chirp time $\xi$.
In principle, this can always be done numerically using 
a grid in the eight dimensional parameter space. In practice, such a strategy 
is not only computationally infeasible but, as we show in this paper, is also 
wasteful. The first important result in this paper is a new representation for
the signal, which is expressed here in terms of the complex expansion 
coefficients of the wave and the detector tensor in a basis of symmetric, 
trace-free (STF) tensors of 
rank $2$. Such a representation of the signal not only allows us to 
express the LR for the network in a very simple and compact form,
but also brings out the symmetries in the response functions of the detectors 
and is at the basis of giving a novel geometric interpretation to the 
detection problem. Maximization of the LR over four of the eight parameters can
be performed analytically using the symmetries in the responses, which are 
clearly brought out when the responses are expressed in terms of the 
Gel'fand functions. {}Further, the Fast Fourier transform (FFT) can be used to 
maximize the LLR over the parameter $t_c$, as in the case of a single detector.
The network template is constructed by taking into account appropriate
time-delays at the individual detector sites.

The analytic maximization and the FFT have the following advantages:

\begin{enumerate}

\item They allow us to scan {\em continuously} the parameter space for the 
five parameters $r, \delta_c, \epsilon, \psi$ and $t_c$.

\item They save substantially on the computational cost. 

\end{enumerate}

We are then left with the three parameters, namely, $\xi$, $\theta$, and 
$\phi$. A full sky search over $(\theta,\phi)$ maps to a time-delay window,
consisting of all possible time-delays, for a given network. The search 
over the time-delay window may be performed by using the samples of the 
cross correlations between the signal and the detector outputs or 
by constructing a template bank. The latter approach has the 
advantage of incorporating the desired mismatch related to the fractional 
loss in
signal-to-noise ratio. Turning the argument around, the template bank 
can also be used as guideline to re-sample the data at a rate that is
consistent with the desired mismatch and then scan over all samples in
the time-delay windows.
We, therefore, opt to 
construct a template bank 
in $\xi$, $\theta$, and $\phi$. This is efficiently obtained by first
computing the metric as given by Owen \cite{Owen}. We then obtain the volume
of the parameter space, given the metric, and divide this volume by the volume
spanned by a template to obtain the number of templates. {}From this 
information, we can easily evaluate the computational costs for the search.
Secondly, the metric is essentially the Fisher information matrix and its 
inverse yields the covariance matrix from which errors in the parameters can 
be obtained. 
 
We apply our formalism to analyze several networks of detectors. First, we
examine idealized networks, with all detectors having the LIGO-I noise. Such 
an exercise nevertheless yields useful estimates of  
computational costs while at the same time simplifying the calculations and  
providing invaluable insights.  We then consider the LIGO-VIRGO network
with their respective noise curves. The computations for this
case are done numerically. {}For these networks we estimate the computational
speed requirements, sensitivities, and the 
resolution in the direction to the source. We find that the computational
costs are high even for the two-detector network. The online data processing
speeds required are in terms of Gflops and for a 3 detector network the
online speeds needed escalate to few tens of Tflops. The costs would go up 
further when PN corrections are incorporated into the waveform.
For example, for LIGO-I noise and allowing a maximum mismatch of $3 \%$ between
the signal and the template, the number of templates required increases by a 
factor of about 4 or 5 \cite{MD1}. Hence, even for a network searching for
PN-corrected waveforms, one may 
expect the computational costs to increase by similar factor. Clearly, our 
results show that use of hierarchical search methods are called for. 
Assuming the individual masses to be greater
than $0.5 M_{\odot}$, and with LIGO-I noises in the detectors at Hanford and
Louisiana, the online computing speed requirement is $12$ Gflops, for a 
$3 \%$ mismatch between the signal and the template. The corresponding figure 
for one
of the LIGO detectors and VIRGO is $170$ Gflops. For the three detector 
LIGO-VIRGO network, the cost rises to few tens of Tflops.
The sensitivity roughly scales as $\sqrt M$ or a little less, where
$M$ is the number of detectors. The resolution in direction is about a
fraction of a degree for the networks where we have assumed LIGO-I noise and 
a signal-to-noise ratio of 12.

The paper is organized as follows. We begin by setting up in Sec.
\ref{sec:math} the basic mathematical framework required for our formalism.
This includes a discussion of the various relevant coordinate frames and
their relationships with one other. We also introduce the wave and detector 
tensors, using which we define the signal at a detector. The signal at a 
detector is then used to define the network signal and infer the network 
statistic. In Sec. \ref{chirp}, we present the Newtonian 
chirp in its familiar form. This allows us to define the role of each 
parameter influencing the waveform. It also 
introduces important notations that we follow in the rest of the paper. We 
then derive a less familiar expression for the signal induced by a chirp in 
a detector. This new representation for the chirp-signal simplifies the 
analysis associated with a network-based detection strategy.  In Sec. 
\ref{sec:detect}, we show how the detection problem can be optimally addressed 
using the maximum-likelihood method. 
We present a single likelihood ratio for 
the entire network. It has a very simple form owing to our use of the new 
representation for the signal. The LR is analytically maximized 
first with respect to $r$ and $\delta_c$ in the well established way 
\cite{SD,BDP}, and then with respect to $\psi$ and $\epsilon$ using the 
symmetry properties of the detector responses \cite{BPD}. 
We then show how the FFT can be used to efficiently maximize over the time of 
final coalescence or, alternatively, the time of arrival at a fiducial 
detector. This is followed by a construction of the
network template and the network correlation-vector. 
The window consisting of all possible time-delays is discussed.
In Sec. \ref{filbank},
we construct the template bank on the rest of the parameter space,
i.e., on $\{\xi,\t,\p\}$ by extending
Owen's \cite{Owen} method and present a way of arriving at the number of 
templates required. We then give expressions for the computational
costs and the resolutions achievable in parameter values. 
Sec. \ref{difnet} is devoted to the discussion of several networks
including the realistic network of LIGOs and VIRGO. 
In Sec. \ref{FDV}, we discuss the statistical properties of the optimal 
network statistic.
We calculate the false-alarm and the detection probabilities associated with 
the network statistic, and obtain 
a relation between the network sensitivity
and the number of detectors in a network. We also discuss the case 
where the detector noise is contaminated by non-Gaussian noise events and 
suggest a 
vetoing criterion based on the $\chi^2$ - test. 

We use the following convention for symbols in this paper, unless otherwise 
specified. When it is useful to keep track of the complex nature of a 
network-based or individual detector-based {\em variable} we denote it by an 
uppercase Roman letter, whereas the lower case letters are reserved 
for real variables.\footnote{Note that quantities such as the gravitational 
constant, $G$, though written in upper case, are not complex since they do not 
represent any inherent characteristic of the network or an individual 
detector. On the other hand, we shall not use an uppercase letter to denote a
complex quantity when its complex nature is apparent from other means, such as 
by the use of a tilde, e.g., in $\tilde{n}$, which denotes the, in general, 
complex Fourier transform of the real quantity $n$.} Network-based vectors are 
displayed in the Sans Serif font. The 
{\em label} $I$ in the superscript or subscript of a variable denotes a (real) 
natural number that associates it with a particular detector. It ranges from 
$1$ to $M$, where $M$ is the total number of detectors in a network. It can 
be considered as a vector index over detectors. We use the index I for several 
of the network variables. However, certain quantities that do not obviously 
display a vector character, but still pertain to a detector, $I$, are denoted 
by enclosing the index in parentheses. Einstein summation convention over 
repeated indices is used for brevity, unless explicitly stated. 

\section{Mathematical framework}
\label{sec:math}
\subsection{Reference Frames}
\label{subsec:Ref}

Our first aim is to obtain a quantity that would define the {\em response}
of an arbitrary network of broadband detectors to an incoming gravitational
wave. In this quest, it is important to understand how the responses of 
arbitrarily oriented and arbitrarily located individual detectors to such a 
wave relate to one another. This is aided by introducing the three different
frames of reference that arise naturally in such a problem, namely,
{\it $(i)$} the wave frame, {\it $(ii)$} the network frame, 
and {\it $(iii)$} the frame of a representative detector
in the network. We define these reference frames in terms of the following 
right-handed, orthogonal, three-dimensional Cartesian coordinates:

\noindent {\it $(i)$}
{\it Wave frame}: We associate with this frame the coordinates $(X,Y,Z)$.
The gravitational wave, which is assumed to be weak and planar, is taken to 
travel along the positive $Z$-direction; then $X$ and $Y$ denote the axes of 
the polarization ellipse of the wave. 

\noindent {\it $(ii)$} 
{\it Network frame}: There is no unique definition of this frame. 
For Earth-based detectors being discussed here, if the network has a large
number of detectors (say, $M > 3$), a convenient choice is a frame attached
to the center of the Earth. Let the coordinate system that defines this 
frame be $(x_E ,y_E ,z_E )$. The 
$x_E$ axis lies along the line joining Earth's center and the equatorial point 
that lies on the meridian passing through Greenwich, England. It points 
radially outwards. The $z_E$ axis is chosen to lie in the direction of the line
passing through the center of Earth and the north pole. The $y_E$ axis is
chosen to form a right-handed coordinate system with the $x_E$ and $z_E$
axes. 

For a network consisting of $M \le 3$ detectors, certain calculations can be
simplified by using the fact that the corner stations (or hubs) of all the 
detectors will lie on a single plane. Specifically, for $M=3$ we define the  
network frame such that one of the detectors is at its origin, a
second detector is on one of the coordinate axes, say, $z$, and the
third lies on one of the coordinate planes containing the $z$ axis,
say, the $x-z$ plane. Later in the text when we consider various examples of
three-detector networks, we choose this as the network frame.

\noindent {\it $(iii)$} 
{\it Detector frame}: Let $(x_{(I)} ,y_{(I)} ,z_{(I)} )$ 
(with $I$ = 1,2,...,M) denote the orthogonal coordinate frame attached to the 
detector labeled $I$. The 
$(x_{(I)} ,y_{(I)} )$ plane contains the detector arms and is assumed to be
tangent to the surface of the Earth. The $x_{(I)}$ axis bisects the angle 
between the detector's arms. The direction of the $y_{(I)}$ axis is chosen in 
such a way that $(x_{(I)} ,y_{(I)} ,z_{(I)} )$ form a right-handed coordinate
system with the $z_{(I)}$ axis pointing radially out of Earth's surface.

Apart from the above choices for frames, we define a fourth frame, namely, 
the frame of a ``fiducial'' detector (henceforth referred to as the ``fide'').
This frame serves as a reference frame with respect to which the locations
or orientations of each of the detectors in a network shall be specified.
Indeed, we will develop our whole formalism for a general network using the
fide frame as a reference. It is only towards the end, when we consider 
specific cases of networks, shall we identify the fide frame with one of the
three frames defined above, depending upon suitability. 

Physical quantities in these frames are related by orthogonal 
transformations that
rotate one frame into another. These orthogonal transformations are defined in
terms of three sets of Euler angles that specify the orientation of one frame
with respect to another. To understand these relations, let ${\cal O}$ be an
orthogonal transformation matrix. 
Let $(\p,\t,\psi)$ be the Euler angles through which one must rotate 
the fide frame to the align with the wave frame. Then
\be
{\sf x}_{\rm wave} = {\cal O}(\p, \t, \psi) {\sf x}_{\rm fide} \ \ ,
\ee
where ${\sf x}_{\rm fide}$ denote the axes of the fide frame and 
${\sf x}_{\rm wave}$, those of the wave frame. The transformation is an 
orthogonal matrix given by \cite{HG}
\be
{\small
{\cal O}(\p, \t, \psi) = \left(\begin{array}{ccc}
\cos\psi\cos\phi -\cos\theta\sin\phi\sin\psi & 
\cos\psi\sin\phi + \cos\theta\cos\phi\sin\psi & \sin\psi\sin\theta \\
-\sin\psi\cos\phi - \cos\theta\sin\phi\cos\psi & 
-\sin\psi\sin\phi + \cos\theta\cos\phi\cos\psi & \cos\psi\sin\theta \\
\sin\theta\sin\phi & -\sin\theta\cos\phi & \cos\theta
\end{array}\right) \,.}
\ee
Similarly, if $(\a_{(I)},\b_{(I)},\g_{(I)})$ are the Euler angles that rotate 
the fide frame to the frame of the $I$-th detector, then
\be 
{\sf x}_{\rm detector_{(I)}} = {\cal O}(\a_{(I)}, \b_{(I)}, \g_{(I)} ) {\sf x}_{\rm fide} \ \ ,
\ee
where ${\sf x}_{\rm detector_{(I)}}$ denote the 
axes of the $I$-th detector frame.

One can imagine yet another frame attached to the source whose $z$ axis
is along the orbital angular-momentum vector of the binary. The angle, $\e$, 
between this vector and our line of sight to the binary is termed as the 
inclination angle and has the range $[0,\pi]$. The associated $x-y$ plane 
specifies the plane of the binary. 
It is then possible to orient the $x$ and $y$ axes on this plane in such a 
way that a rotation of the wave frame through the Euler angles, $(0,\e,0)$, 
aligns it with the source frame.\footnote{However, since we will be expressing 
the gravitational-wave metric fluctuations, $h_{\mu\nu}$, in the 
transverse-traceless gauge (see below), in addition to this rotation we must 
also project $h_{\mu\nu}$ orthogonal to the direction of the wave in order to
obtain its components in the wave frame. In that case the 
polarization-ellipse angle $\psi$ can be included as an Euler angle in the 
transformation ${\cal O} (\psi,\epsilon, 0)$ of the wave frame to the source 
frame, instead of including it in the rotation from the fide frame to the 
wave frame, as is traditionally done. This observation will be used in Sec.
\ref{sec:detect} to obtain a reduced statistic for the detection of chirps.}

\subsection{Wave tensor, detector tensor, and beam-pattern functions}
\label{subsec:wave}

A gravitational wave can be represented by metric tensor fluctuation, $h_{\mu
\nu}$, about the background space-time which we take to be flat. 
The subscripts $\mu$ and $\nu$ denote space-time
indices. In the transverse trace-free (TT) gauge, the non-vanishing components
of  $h_{\mu \nu}$ in the wave frame are $h_{xx} = - h_{yy} \>{\equiv}\> h_+ $, 
$h_{xy} = h_{yx}\>{\equiv}\>h_{\cross}$. Here, $h_+$ and $h_\cross $ are the 
two linear-polarization components of the wave. When a metric fluctuation 
specifically represents a gravitational wave, its spatial part is identified 
as twice the wave tensor, $w_{ij}$, where $i$ and $j$ refer to space indices 
and take values $1,2$, and $3$ (see Ref. \cite{DT1}). In the TT gauge, 
the wave tensor is a symmetric trace-free (STF) tensor of rank 2.\footnote{See 
appendix \ref{app:stf} for properties of such tensors.} In any arbitrary 
frame, the wave tensor can be expressed in terms of 
its circular-polarization components as,
\be \label{wcp}
w^{ij}(t) = {1\over 2} \left[ (h_+(t)+i h_\times (t)) e_R^{ij} + (h_+(t)-i
h_\times (t)) e_L^{ij} \right] \ \ ,
\ee
where $e_{R,L}^{ij}$ are the right and left-circular polarization tensors, 
respectively. The $e_{R,L}^{ij}$ are both second rank STF tensors and
obey the orthonormality conditions,
\be\label{polprop}
e_{L,\> R}^{ij} \>e_{L,\> R\> ij}^* =1 \>,\quad e_{L,\> R}^{ij} 
\>e_{R,\>L\> ij}^* = 0 \ \ ,
\ee
where a star denotes complex conjugation. The reality of the wave tensor 
ensures that 
\be\label{LRdual}
e_L^{ij\>*} = e_R^{ij} \,.
\ee 
Using (\ref{LRdual}), the wave-tensor expression (\ref{wcp}) simplifies to
\be \label{wS}
w^{ij} (t)= \Re \left[ \left(h_+(t) + ih_\times (t)
\right) e_R^{ij}\right] \ \ ,
\ee
where $\Re [A]$ denotes the real part of a complex quantity $A$.

In an arbitrary reference frame, the polarization tensor can be expressed as
\be
e_L^{ij} = m^i  m^j \ \ ,
\ee
where the $m^k$ is the $k$-th component of a complex null vector $\sf m$ in 
that reference frame. It is defined as 
\be
m^k = \frac{1}{\sqrt 2} (e_X^k + i e_Y^k) \ \ , 
\ee
where ${\bf e}_X$ and ${\bf e}_Y$ are unit vectors in the $X$ and $Y$ 
direction of the wave axes, respectively \cite{DT1}. The above expression shows
that the $e_{R,L}^{ij}$ are STF tensors of rank 2. Hence, they can be expanded 
in a basis of STF-2 tensors, ${\cal Y}^{ij}_{2n}$, 
(see appendix \ref{app:stf}):
\be\label{polt}
e_R^{ij} = \sqrt{8\pi\over 15} T_{-2}{}^n  {\cal Y}^{ij}_{2n}
\quad {\rm and} \quad 
e_L^{ij} = \sqrt{8\pi\over 15} T_{2}{}^n  {\cal Y}^{ij}_{2n}
\ \ , \ee
where the expansion coefficients, $T_{\pm 2}{}^n$, with $n=0,\pm 1,\pm 2$, are 
Gel'fand functions \cite{GMS}. (For a more elaborate discussion on these 
functions see Refs. \cite{DT1,DT2}.) These functions depend on the Euler 
angles through which one must rotate the reference frame to the wave frame. 
If the reference frame is chosen to be the fide, then these angles are just 
$(\phi,\theta,\psi)$. While implementing more than a single 
frame-transformation in relating these two frames, the addition theorem 
(\ref{add}) for Gel'fand functions is used to obtain the 
required wave tensor. In such a case, the wave tensor depends on more than
one set of Euler angles. 

The $I$-th detector tensor, $d_{ij}^I$, is given by
\be \label{det}
d_{ij}^I = n_{(I)1i} n_{(I)1j} -  n_{(I)2i} n_{(I)2j} \ \ ,
\ee
where ${\sf n}_{(I)1}$ and ${\sf n}_{(I)2} $ are the unit vectors along 
the arms of
the $I$-th interferometer, which is taken to have orthogonal 
arms.\footnote{See appendix \ref{app:poldeten} for a more general expression.}
Like the polarization tensors, even $d_{ij}^I$ is an STF tensor of rank 2. 
Hence, in any frame it can be expanded in a basis of STF-2 tensors. In such a 
basis, the components of $d_{ij}^I$ can be expressed in terms of Gel'fand 
functions, $T_{\pm 2}{}^{n}$ (see Eq. (\ref{dtcomp})). In the fide frame, these
functions depend on the Euler angles $\{\a_{(I)},\b_{(I)},\g_{(I)} \}$, which 
specify the relative orientation of the $I$-th detector with respect to the 
fide.

When detectors are distributed around the globe there are, in general, 
relative delays in the arrival times of a particular phase of a given wave at 
different locations. Let $\tau_{(I)} (\t,\p)$ be the relative delay between 
the arrival times at the $I$-th detector and the fide, where the source 
direction is given by $(\t,\p)$. If ${\hat {\sf n}}(\t,\p)$ is the unit vector 
along the direction of the wave, i.e., ${\hat {\sf n}}(\t,\p) 
= {\hat {\sf Z}}$, then
\be \label{time}
\tau_{(I)}(\t,\p) = {{({\sf r}_{(I)} - {\sf r}_{(f)}) \cdot {\hat {\sf n} (\t,\p)}}
\over {c}} \ \ ,
\ee
where ${\sf r}_{(I)}$ and ${\sf r}_{(f)}$ are the position vectors of the $I$-th
detector and fide, respectively, with respect to any given reference frame.
Note that $\tau_{(I)} (\t,\p)$ can take positive as well as negative values.

The signal in the $I$-th detector is the scalar
\be\label{sig1}
s^I(t) = w^{ij}(t-\tau_{(I)}) d_{ij}^I \ \ ,
\ee
which, by definition, is invariant under coordinate transformations. Above, 
$w^{ij}(t)$ is the wave tensor at the location of the 
fide at time $t$. It is a function of 
$h_{+}(t)$ and $h_{\times}(t)$, which define the amplitudes of the two 
polarization components at time $t$ and at the location of the fide. 
The above definition shows that the signal depends 
on the projections of the polarization tensors, $e_{L,\> R}^{ij}$, onto the 
$I$-th detector tensor, $d_{ij}^I$. These projections are
\be \label{fI}
F^I = e_{L}^{ij} d_{ij}^I,   \hspace{1.in} F^{I*} = e_{R}^{ij} d_{ij}^I
\ \ ,\ee
which are the beam-pattern functions for the left- and right-circular 
polarizations, respectively. They depend on the direction of the wave 
and the orientation
of the detector. Owing to any motion of the source with respect to the detector
this orientation can change with time. Hence, in general, $F^I$ are functions
of time. Since we will be concerned here with only short-duration signals,
we will assume these functions to be independent of time (which is valid
to a very good approximation). Using the above definition of the beam-pattern 
functions and the wave-tensor expression (\ref{wS}) in Eq. (\ref{sig1}),
we find the signal to be
\be
s^I(t) = \Re \left[\left(h_+^I(t) + ih_\times^I (t)\right) F^{I*}
\right] \ \ ,
\ee
where $h_+^I(t) \equiv h_+ (t-\tau_{(I)})$ and 
$h_\times^I (t) \equiv h_{\times}(t-\tau_{(I)})$ 
are the time-delayed amplitudes of the two polarizations of the
wave at detector $I$. 

\subsection{Network signal and network statistic}
\label{match}

The signal from an inspiraling binary will typically not stand above the 
broadband noise of the interferometric detectors; the concept of an 
absolutely certain detection does not exist in such a case. Only 
probabilities can be assigned to the presence of an expected signal. In the 
absence of prior probabilities, such a situation demands a decision strategy 
that maximizes the 
detection probability for a given false alarm probability. This is termed as 
the Neyman-Pearson criterion \cite{Hels}. Such a criterion 
implies that the decision must be based on a statistic called 
the likelihood ratio (LR). It is defined as the ratio of the probability that 
a signal is present in an observation to the probability that it is
not. This is the criterion we employ in formulating our detection 
strategy.

In order to define
a strategy to search for signals in a noisy environment, it is important to 
recognize the characteristics of the noise. Here, we assume that the noise, 
$n^I(t)$, in the $I$-th detector (a) has a zero mean and (b) is mostly 
stationary\footnote{In reality, detector noise contains non-Gaussian and 
non-stationary 
components. To accommodate such features in our treatment, we use vetoing 
techniques, which are discussed in Sec. \ref{subsec:veto}.} and
statistically as well as algebraically independent of the noise in any other
detector. These requirements are mathematically summarized, respectively, as:
\begin{mathletters}%
\label{noise}
\begin{eqnarray} 
{\overline{n^I(t)}} &=& 0, \label{noisea}\\
{\overline{{\tilde n}_I^*(f){\tilde n}^J(f')}} &=& s_{h(I)}(f) \d(f-f') 
\d_I^J \ \ , \label{noiseb}
\eea
\end{mathletters}%
where the over-bar on a symbol denotes the ensemble average of that quantity
and the tilde denotes the Fourier transform, 
\be \label{FTdef}
{\tilde n}^I(f) = \int_{-\infty}^{\infty} n^I(t) e^{-2\pi i f t} dt \,.
\ee
Also, $s_{h(I)}(f)$ is the one sided power-spectral-density (PSD) of the $I$-th
detector. Note that $s_{h(I)}(f)$ is the Fourier transform of the 
auto-covariance of the noise in detector $I$. We also assume the noise to be 
additive. This implies that 
when a signal is present in the data, then $x^I(t)$ is given by
\be
x^I(t) = s^I(t) + n^I(t) \ \ ,
\ee
otherwise $x^I(t) = n^I(t)$.

As we shall see below, an important tool in the theory of detection of 
known signals in noisy environments is the cross correlation between a signal 
template and a detector's output. In order to define it, consider two real, 
sufficiently smooth, and absolutely integrable functions of time, namely, 
$a(t)$ and $b(t)$. {}For the purposes of this paper, we can assume that the 
signal template, $s^I(t)$, and the detector outputs, $x^I(t)$, belong to this 
category of functions to a good approximation.  A cross correlation can be 
represented in terms of an inner product, which is defined as
\be \label{ip}
\langle a ,\>b \rangle_{(I)} = 2 \Re \int_{0}^{\infty} \! df\>
{\tilde{a}^* (f) \tilde{b}(f) \over s_{h(I)}(f)} \ \ , 
\ee
where $\tilde{a}(f)$ and $\tilde{b}(f)$ are the Fourier
transforms of $a(t)$ and $b(t)$, respectively.
In order to obtain the correlation between 
a complex function $A(t) = a_1(t) + i a_2(t)$ and a real function $b(t)$, 
we adopt the following convention to define the inner product
\be
\langle A, b \rangle \equiv \langle a_1,b \rangle - i \langle a_2,b
\rangle \,.
\ee
This definition is consistent with the convention of (\ref{ip}) where
the complex conjugation is performed on the first entry in the inner product.

{}For a network of $M$ detectors, the data consist of $M$ data trains, 
$\{x^I(t)|\>I=1,2,...,M$ and $t \in [0,T]\}$. The network matched-template can 
be obtained naturally by the maximum-likelihood method, where the 
decision whether
the signal is present or not is made by evaluating the likelihood ratio (LR)
for the network \cite{Hels}. Under the assumptions made on the noise, the 
network LR, denoted by $\lambda$, is just a product of the individual 
detector LRs. In addition, for Gaussian noise, the logarithmic likelihood 
ratio (LLR) for the network is just the sum of the LLRs of the individual 
detectors \cite{BPD}, 
\be
\ln\lambda = \sum_{I=1}^M \ln\lambda_{(I)} \ \ ,
\ee
where \cite{BDP}
\be
\ln\lambda_{(I)} = \langle s^I, x^I \rangle_{(I)} -
{1 \over 2} \langle s^I, s^I \rangle_{(I)} \,.
\ee
The network LLR takes a compact form in terms of the network inner-product,
\be
\langle {\sf s},{\sf x}\rangle_{NW} = \sum_{I=1}^M \langle s^{I}(t), x^{I}
(t)\rangle_{(I)} \ \ ,
\ee
where 
\be
\label{sNW}
{\sf s} (t) =\left(s^{1}(t), s^{2}(t),\cdots , s^{M} (t) \right)
\ee
is the network template-vector, which comprises of individual 
detector-templates as its components, and
\be
\label{xNW}
{\sf x} (t) =\left(x^{1} (t), x^{2}(t),\cdots , x^{M} (t) \right)
\ee
is the network data-vector. It can be shown by using the Schwarz inequality 
that the network template, ${\sf s}$, defined 
above yields the maximum signal-to-noise (SNR) amongst all linear templates 
and, hence, is the matched template. As shown in Ref. \cite{BDP}, in terms of 
the above definitions, the network LLR takes the following simple form:
\be\label{MLRNW}
\ln\lambda = \langle {\sf s},{\sf x}\rangle_{NW} -
{1 \over 2} \langle {\sf s}, {\sf s}\rangle_{NW} \ \ ,
\ee
which is a function of the source parameters that determine ${\sf s}$. Given 
${\sf s}$, different selections of source-parameter values and, therefore,
different values of ${\sf s}$ result in varying magnitudes of the LLR. The
selection that gives the maximum value stands the best chance for beating the
pre-set threshold on the LLR. Since scanning the complete source-parameter 
manifold for the maximum of LLR is computationally very expensive, we propose
to perform its maximization analytically over as many parameters as possible.
This requires the knowledge of the analytic dependence of the network
matched-template on source parameters. This is what we seek below.

\section{The signal}
\label{chirp}

Assume that the binary is at a luminosity distance of $r$ from the 
Earth\footnote{Here, $r$ is not to be confused with the magnitude of a detector
position-vector, which always carries as an index the label of the detector, 
i.e., $(I)$ or $(f)$.}. 
{}Further, let
$m_1$ and $m_2$ be the masses of the individual stars. Then, in the Newtonian 
approximation the two corresponding GW linear-polarization 
components in the wave frame, at the location of the fide, are
\begin{mathletters}%
\label{h}
\begin{eqnarray} 
h_+(t;r,\d_c,t_{c},\xi)&=&{2{\cal N} \over r} a^{-1/4}(t;t_{c},\xi) 
{1+\cos^2 \e \over 2} \>\cos[\chi(t;t_{c},\xi) + \d_c]\ \ ,\label{hplus} \\
h_\times(t;r,\d_c,t_{c},\xi)&=&{2{\cal N} \over r} a^{-1/4}(t;t_{c},
\xi) \cos\e \>\sin [\chi(t;t_{c},\xi) + \d_c] \ \ , \label{hcross}
\end{eqnarray}
\end{mathletters}%
where 
\be \label{N}
{\cal N}\equiv \left[ { 2G^{5/3} {\cal M}^{5/3} ({\pi}f_s)^{2/3}\over c^4} 
\right] \ee
is a constant appearing in the chirp amplitude having the dimensions
of length. It depends on the binary's
`chirp' mass, ${\cal M} \equiv (m_1 m_2)^{3/5}/ (m_1 + m_2)^{1/5}$, 
and a fiducial
chirp frequency, $f_s$. Usually, $f_s$ is taken to be the lowest frequency
in the bandwidth of a detector - the seismic cut-off - hence the
reason for the subscript $s$. This choice of the fiducial frequency
maximizes the duration of tracking the chirp because the chirp frequency
increases monotonically with time.
Here we set $f_s = 40$ Hz, which is the seismic cut-off for LIGO-I, because 
every network we consider below has at least one detector with 
LIGO-I noise. Note that a general network might include several detectors with 
different seismic cut-offs, $f_{s(I)}$. Even in such a case, it is convenient 
to use the fiducial frequency as a reference. This is apparent in Eq. 
(\ref{newtmomfunc}) where the noise moments for different detectors are merely 
scaled by appropriate powers of $(f_{s(I)}/f_s)$.
 
A quantity closely related to the chirp mass is the so-called chirp time,
\be \label{chirpp}
\xi = 34.5 \left({{\cal M} \over M_\odot }\right)^{-5/3} \left({f_s \over 40
\>{\rm Hz}} \right)^{-8/3} \> \> {\rm sec.} \ \ ,
\ee
which equals the time duration for which the chirp exists in a detector's 
sensitivity window from the time of arrival until the time of final 
coalescence. The time of arrival, $t_a$, is defined as the time when the 
instantaneous frequency of the chirp equals the fiducial frequency, i.e., 
$f(t_a) = f_s$. {}Formally, the coalescence time, $t_c$, is the time at which 
the chirp frequency diverges (see Eq. (\ref{ft})). The corresponding phase of
the wave-form at $t_c$ is $\d_c$. We define the quantity
\be\label{at}
a(t;t_{c},\xi) = {t_{c}-t\over \xi} \ \ ,
\ee
and the instantaneous frequency,
\be\label{ft}
f(t;f_s,t_{c}, {\xi}) = f_s\> a^{-3/8}(t;t_{c},\xi) 
= f_s \left( {t_{c}-t\over \xi} \right)^{-3/8} \ \ ,
\ee
which diverges at final coalescence. The above expression also shows 
that 
\be
t_c = t_a +\xi \,.
\ee
{}Finally, the instantaneous phase of the waveform is $\chi(t) +\d_c$, where
\be\label{chi}
\chi(t;f_s,t_c,\xi) \equiv - 2\pi \int_{t}^{t_c} f(t';f_s,{t_c},\xi) dt' 
= -{16\over 5}\pi f_s \xi a^{5/8}(t;t_{c},\xi)
\,.\ee
The two GW polarization amplitudes at the $I$-th detector site are obtained 
by substituting $t$ with $(t-\tau_{(I)})$ in Eqs. (\ref{h}), (\ref{at}),
(\ref{ft}), and (\ref{chi}).

A chirp signal registers itself in a detector's output
only after its instantaneous frequency crosses the seismic cut-off of that
detector. Thus, a signal arrives in the $I$-th detector's bandwidth when its 
instantaneous frequency reaches $f=f_{s(I)}$ and it lasts there for a
time duration equaling $\xi_{(I)} = \xi \left(f_{s(I)}/f_s\right)^{-8/3}$. 
Alternatively put, the chirp waveform at the $I$-th detector starts at 
$t=t_c + \tau_{(I)} - \xi_{(I)}$ and ends at $t=t_c + \tau_{(I)}$.

In order to formulate a strategy for detecting a chirp, it helps to isolate 
the factors in the two polarizations, $h_{+,\cross}$, that 
are time dependent from those that are not. To this end, we define
two mutually orthogonal normalized waveforms $s_0^I$ and $s_{\pi/2}^I$, 
with $s_{0,\pi/2}^I(t) = s_{0,\pi/2}(t-\tau_{(I)})$, and
their complex combination $S^I = s_0^I + i s_{\pi/2}^I$ - the normalized
complex signal - as
\be \label{SI}
S^{I}(t;{t_c},\xi) \equiv  {{a^{-1/4}(t-\tau_{(I)};
\xi)} 
\over {g_{(I)} \sqrt{\xi}}} e^{i \chi(t-\tau_{(I)};\xi)} \,.
\ee
Here, $g_{(I)}$ is a normalization factor such that 
\be \label{sisi}
\langle S^I,S^I \rangle_{(I)} =1 \,.
\ee 
We now obtain an expression for the normalization factor, $g_{(I)}$. In the
stationary-phase approximation (SPA), the Fourier transform of $S^I(t)$
for positive frequencies is,
\begin{eqnarray}
{\tilde S}^I(f;t_c,\xi) &=& \int_{-\infty}^{\infty} S^I(t;t_c,\xi)
e^{-2\pi i f t}dt\ \  \nonumber \\
&=&{2 \over {g_{(I)}}} \sqrt{2 \over {3 {f_s}}} \left({f \over {f_s}}\right)
^{-7/6} \exp \left[i \Psi_{(I)}(f;f_s,t_c,\xi)\right] \ \ ,
\end{eqnarray}
where 
\begin{eqnarray} \label{psiI} 
\Psi_{(I)}(f;f_s,t_c,\xi) &=& -2\pi f_s \left[{f \over f_s} t_c + {f \over f_s}
\tau_{(I)} + {3 \over 5} \xi \left({f \over f_s}\right)^{-5/3} \right] + 
{\pi \over 4}  \nonumber \\
&\equiv& \Psi(f;f_s,t_c,\xi) - 2 \pi f \tau_{(I)} \ \ , 
\end{eqnarray}
for the Newtonian chirp. Note that $\Psi_{(I)} = \Psi$ for vanishing 
time-delay ($\tau_{(I)} = 0$). Thus, $\Psi$ defines the phase in the FT of the 
normalized complex signal at the fide, in the SPA. The normalization 
condition (\ref{sisi}) implies that,
\be \label{g}
g_{(I)}^2 = {8 \over 3} {f_s}^{4/3} \int_{f_{s(I)}}^{\infty} 
{df \over {f^{7/3} {s_{h(I)}(f)}}} \ \ ,
\ee
where $f_{s(I)}$ is the seismic cut-off for the $I$-th detector.

\subsection{The signal at a detector}
\label{detesig}

The signal due to a Newtonian chirp at the $I$-th detector can be expressed 
in terms of $s^I(t)$. It is obtained by the chirp-specific components 
$h_{+,\times}^I$ from (\ref{h}). We now express the GW circular-polarization
components, $(h_+^I + i h_{\times}^I)$, in terms of the normalized complex 
signal $S^I$ and the overall amplitude $\kappa$. In the special
case of the face-on binary (i.e., $\e = 0$), the signal at the detector is 
given by
\be \label{sIt}
s^I(t) = 2 \kappa \Re \left[g_{(I)} F_I^* S^I (t) e^{i \d_c}\right] \ \ ,
\ee
where $\kappa \equiv {{\cal N} {\sqrt \xi}/ r}$.
Note that $\delta_c$ is detector independent and separates out as a phase 
factor in the expression for the complexified $s^I(t)$.

The generalization of (\ref{sIt}) for arbitrary value of $\e$ is 
straightforward. In this case, $s^I$ can be expressed as follows:
\be \label{primsig}
s^I(t) = 2 \kappa \>\Re \left[ ( E_{I}^{*} S^{I})e^{i\delta_c} \right]\ \ ,
\ee
where we have defined the extended beam-pattern functions
\be \label{ext}
E^I = g_{(I)} \left[{{1+{\cos^2 \e}} \over 2} {\Re} (F^I) + 
i \cos\e {\Im} (F^I)
\right] \,.
\ee
Here, $\Re (F^I)$ and $\Im (F^I)$ are the real and imaginary parts of the 
detector beam pattern functions, respectively. In the limit $\epsilon \to 0$,
the signal in Eq. (\ref{primsig}) reduces to that in (\ref{sIt}).
In terms of the Gel'fand functions, we have
\be \label{ext1}
E^I = g_{(I)} T_2{}^p (\psi,\e,0) D^I_p, \hspace{0.3in} p = \pm 2 \ \ ,
\ee
where, for a detector with orthogonal arms (see Eq. (\ref{DIp})),
\be \label{ext2}
D^I_p = -i {T_p{}^s(\p,\t,0)} 
\left( T^{2*}{}_s (\a_{(I)},\b_{(I)},\g_{(I)}) 
- T^{-2*}{}_s (\a_{(I)},\b_{(I)},\g_{(I)})\right) \ \ ,
\ee
which obeys $D_p^{I*} = D_{-p}^I$.
Thus, $E^I$ depends on the source-direction angles, $\{\t,\p\}$, the 
angles, $\{\e,\psi\}$, as well as on the orientation of 
the $I$-th detector relative to the fide, given by the Euler angles 
$(\a_{(I)},\b_{(I)},\g_{(I)})$. Also, $E^I$ depends on the signal-normalization
factor $g_{(I)}$, which expresses the sensitivity
of the detector to the incoming signal. As we find in the next section,
the fact that the dependence of $E^I$ on $\{\e,\psi\}$ factors out in 
each summand in (\ref{ext1}) will turn out to be a useful property 
in obtaining the optimal statistic for the detection problem. Thus, the
signal at the detector depends on a total of eight 
independent parameters, viz., $\{r, \d_c, \psi,\e, \t_c,\xi,\t, \p\}$. The 
ranges of the four angles are as follows: $\psi\in [0,2\pi]$, 
$\e\in [0,\pi]$, $\p\in [0,2\pi]$, and $\t\in [0,\pi]$.

{}From Eqs. (\ref{ext1}) and (\ref{ext2}), it is clear that $E^I$ can be 
resolved into various 
factors (using the addition theorem (\ref{add}) for Gel'fand functions). One 
may interpret $E^I$ as the {\it combined amplitude gain} of the source and 
the $I$-th detector. As was shown in Ref. \cite{BDP}, up to an $r$-dependent 
factor, $|E^I|^2$ can be interpreted as the total power transferred to the 
$I$-th detector.\footnote{In Ref. \cite{BDP}, the symbol $W_{(I)}$ denotes a 
quantity analogous to $E^I$.} More appropriately, it is the gain factor 
associated with the $I$-th detector. It can be decomposed into a sum of the 
fractions of power transferred from a signal to the detector by each of its 
two polarizations. 

Expression (\ref{primsig}) shows that the contribution of the extended 
beam-pattern function, $E^I$, to the signal, $s^I$, comes from its magnitude 
as well as its phase. In the case of a single detector, these contributions 
cannot be separated from the overall amplitude and the ``effective'' initial 
phase of the signal. Thus, one cannot obtain precise information, even in the 
absence of any kind of noise, about the parameters $\{\e,\psi,\t,\phi\}$, 
which affect the signal only through $E^I$. As a result, for data analysis 
involving a single detector, it is more meaningful to
resolve $E^I$ and express the signal in the form
$s (t) =  \varrho (t) \cos(\chi (t) + \varpi)$
where $\varrho$ and $\varpi$ are overall amplitude and the effective initial 
phase of the signal, which get contributions from $E$. This is precisely what 
was done by SD. While using a network with multiple detectors such a 
degeneracy
in parameters can be broken. Indeed, information about source direction, i.e.,
$(\t,\phi)$, can be obtained from the time-delays $\tau_{(I)}(\t,\p)$, 
by using the triangulation method. More pertinently, even when all the
detectors are coincident in a network, one can use a set of independently 
oriented detectors to recover information about the different $E^I$'s and,
consequently, about the 
parameters $\{\e,\psi,\t,\phi\}$ \cite{DT1,DT2}. Hence, in 
data analysis with a network it is crucial to track the effect of $E^I$ on 
the signal explicitly. This is where we shall find the form of the signal 
given in (\ref{primsig}) useful in rest of the paper.

\subsection{Network signal normalization}
\label{netnorm}

The total energy in a signal that is accessible to a network is just the
network scalar $\langle {\sf s}, {\sf s} \rangle_{NW}$, and is given by
\begin{eqnarray}
\langle {\sf s}, {\sf s}\rangle_{NW} &=& \sum_{I=1}^M \langle s^{I}(t), s^{I}
(t)\rangle_{(I)}\ \, \nonumber \\
 &=& 4\kappa^2 \sum_{I=1}^{M} {E^*}_{I} {E^I} \equiv {\sf b}^2\label{bsq}.
\end{eqnarray} 
The quantity $\sum_{I=1}^{M} {E^*}_{I}{E^I} \equiv {{\sf E} \cdot {\sf E}} =
\parallel {\sf E} \parallel^2$ is the ${\cal L}^2$ norm of 
$E^I$ in ${\cal C}^{M}$. 
To understand the significance of $\|E\|^2$ consider a network comprising of 
detectors with identical noise PSDs and, therefore, identical $g_{(I)}$.
For simplicity, let $g_{(I)} = 1$ for all $I$. Then, for a given set of 
values for $\{\psi,\epsilon,\phi,\theta\}$, $\|E\|^2$ is a pure number. It 
defines the factor by which the energy accessible to such a network is larger 
or smaller than the maximum energy that is accessible from an identical 
source, but with $\psi=0=\epsilon$, to a (favorably oriented) single 
detector. This maximum energy is $4\kappa^2$. Therefore, ${\sf b}^2$
represents the total energy in the signal that is accessible to a network. It 
is just the sum of the signal energies accessible to each individual 
detector in the network. The quantity ${\sf b}$ can be regarded as the signal 
strength accessible to a network and has the following properties:

$(\it{i})$ If the detectors have identical noise PSDs and are oriented 
identically, then we have $\parallel {\sf E} \parallel ^2 \propto M$ and,
therefore, the strength obeys ${\sf b} \propto {\sqrt M}$. This clearly shows that 
for a given
$\kappa$, a network of detectors can probe deeper than a single detector,
by a distance that is $\sqrt M$ times larger.

$(\it{ii})$ If the detectors are oriented identically but have different 
noise PSDs, then the amount of energy accessible to each detector is 
proportional to the optimal SNR of that detector, namely, $g_{(I)}$. The 
detector having maximum SNR will contribute the most in terms the energy 
accessible to the entire network.

\begin{figure}[!hbt] 
\caption{Plots of the quantity $\parallel {\sf E} \parallel^2$ for a 
two-detector 
network as a function of the direction to the source, with $\e = 0$ and 
$\psi = 0$. In (a), the two detectors are identically
oriented with arms lying in the X-Y plane, but have different noise PSDs, say
$g_{(1)} = 1$ and $g_{(2)} = 0.25$. Whereas in (b), the two detectors have 
identical noise PSDs, i.e., $g_{(1)} = g_{(2)} = 1$, but have different 
orientations, say, $\a_{(1)} = \b_{(1)} = \g_{(1)} = 0$ and 
$\alpha_{(2)} = \gamma_{(2)} = 0$, $\beta_{(2)} = 90^{\circ}$,
with respect to a fide frame.}
\centerline{\psfig{file=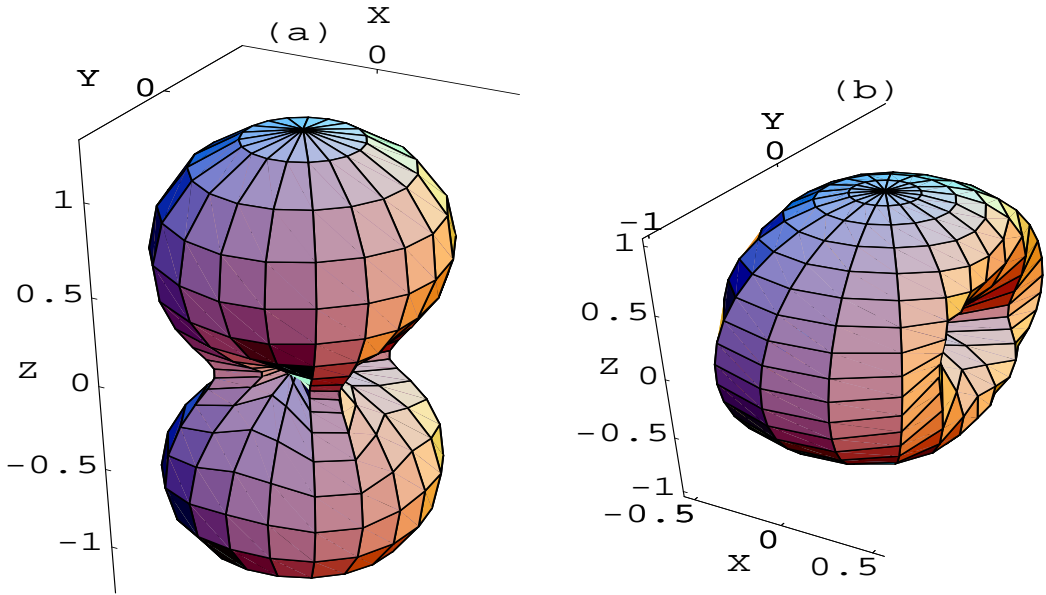,height=4.in,width=6.in}}
\label{fig1}
\end{figure}

$(\it{iii})$ If the detectors have different orientations but identical
noise PSDs,
then the amount of energy accessible to each detector is proportional to the
modulus square of the extended antenna-pattern function of the individual
detector. In Fig. \ref{fig1}, for a network of two detectors, ${\sf b}^2$ is 
plotted as a function of $\t$ and $\phi$ (for $\e = 0$ and $\psi = 0$).

Another important quantity of physical interest is the signal power averaged 
over all the directions and the orientations of the source, i.e., $\t$, $\p$, 
$\e$, and $\psi$.
\begin {eqnarray} \label{Pav}
P_{av} &\equiv& \langle {\sf b}^2 \rangle_{\t,\p,\e,\psi} \nonumber \\
&=& {1 \over {16 \pi^2}} \int_0^{\pi} \sin \e d \e \int_0^{2 \pi} d \psi
\int_0^{\pi} \sin \t d \t \int_0^{2 \pi} d \p \hspace{0.1in} {\sf b}^2.
\end {eqnarray}
This power is clearly independent of the orientations of the detectors. A
detailed calculation shows that 
\be \label{Pav2}
P_{av} = \left({{2 \kappa}\over 5} \right)^2 \sum_{I=1}^M g_{(I)}^2 \,.
\ee
The factors $g_{(I)}$ are then important in deciding the average signal 
strength.

The above analysis
also suggests the normalization for the network
signal. The signal vector with unit norm is defined by $\hat{\sf s} \equiv
{{\sf s}/{\sf b}}$. Its components are 
\be \label{zI}
\hat{s}^{I} = {\Re} \left[({Q_{I}}^* S^{I}) e^{i\d_c}\right]\ \ ,
\ee
where 
\be
\label{QI}
Q^{I} \equiv {E^{I} \over {\parallel {\sf E} \parallel}}\,.
\ee
Note that the network vector ${\sf Q} = ( Q^1, Q^2, ...., Q^M)$ 
lies in ${\cal C}^{M}$ and 
has a unit norm, i.e., ${\parallel {\sf Q} \parallel^2} = 1$.

\section{Maximizing the LLR}
\label{sec:detect}

In the case of a single detector, the LLR is a functional of the data as 
measured by that detector. {}For a network of $M$ detectors, one needs to 
compute the statistic in terms of the network data-vector $\sf x$. When our
assumptions about the statistical properties of detector noise are valid, 
the appropriate network LLR is given by Eq. (\ref{MLRNW}). The optimal network 
statistic is obtained by maximizing this LLR over the eight physical 
parameters that define the signal. It is this maximized LLR that must be
compared with a predetermined threshold, corresponding to a given false alarm 
probability. In the following two subsections, we show how such a maximization 
over four of the parameters can be performed analytically. Subsequently, we 
describe an efficient way of maximizing over the time of final coalescence or, 
analogously, over the time of arrival (at the fide) of the signal.

\subsection{Maximizing the LLR over ${\sf b}$ and $\delta_c$}

We begin by analytically maximizing the network LLR with respect to two
parameters that are simplest to handle, namely, $r$ and $\delta_c$. Note that
the network LLR obtained in Eq. 
(\ref{MLRNW}) can be expressed as an explicit function of ${\sf b}$:
\be\label{LRb}
\ln \lambda = {\sf b} \sum_{I=1}^M \langle \hat{s}^{I}, x^{I}\rangle_{(I)} - 
{1\over 2} {\sf b}^2 \,.
\ee
Above, the luminosity distance, $r$, appears only 
through ${\sf b}$. Maximizing $\ln \lambda$ with respect to ${\sf b}$ gives 
\be \label{hatb}
\hat{\sf b} = \sum_{I=1}^M \langle \hat{s}^{I}, x^{I}\rangle_{(I)} \ \ ,
\ee  
where a hat denotes the value of a variable at which the LLR is a maximum as a
function of that variable, keeping all other variables fixed.
Here, the value of LLR at ${\sf b} = \hat{\sf b}$ is
\begin{eqnarray}
\ln \lambda |_{\hat{\sf b}} &=& {1\over 2} \left(\sum_{I=1}^M \langle 
\hat{s}^{I}, x^{I} \rangle_{(I)} \right)^2  \nonumber \\
&=&{1\over 2}\left({\Re}\left[ e^{-i\d_c} \left( {\sf C} \cdot {\sf Q} \right)
\right]\right)^2
\ \ , \label{LLRhatb}
\end{eqnarray}
where we have defined,
\be \label{Cstar}
C_I^* = c_0^I - i c_{\pi/2}^I \equiv \langle S^{I}, x^{I}\rangle_{(I)},
\ee
with $c_0^I = \langle s_0^I, x^{I}
\rangle_{(I)}$ and $c_{\pi/2}^I = \langle s_{\pi/2}^I, x^{I} \rangle_{(I)}$.
$C^I$ is a complex quantity that combines the correlations of the two 
quadratures of the normalized template with the data.
We proceed further and maximize the LLR in (\ref{LLRhatb}) with respect 
to $\delta_c$. This yields, 
\be
\label{hatd}
\hat{\d_c} = \arg \left( {\sf C} \cdot {\sf Q} \right) \ \ ,
\ee
and the LLR maximized over ${\sf b}$ and $\d_c$ as,
\be
\label{LLRhatbd}
\ln \lambda |_{\hat{\sf b},\hat{\d}_c} = {1\over 2}
\left| {\sf C} \cdot {\sf Q} \right|^2 
\,. \ee
Now the maximized LLR is a function of six parameters, namely,
$\{\epsilon,\psi,t_c,\xi,\t,\p\}$. 

When all the detectors are ``closely" located or coincident, it is only the 
$Q^{I}$'s that depend on four angles $\{\t,\p,\epsilon,\psi\}$; the $C^{I}$
then depend only on $\{t_c,\tau_{(I)},\xi\}$, with all the times of arrival 
being equal. We refer to this situation as the ``same-site" approximation. 
Such a case was dealt with in Ref. \cite{BDP}. When the detectors are 
spatially well separated or non-coincident, the $C^{I}$ depend on 
$\{\t,\p\}$ as well. In such a case, maximization over the remaining parameters
is not as simple as in the same-site approximation. However, in Ref. 
\cite{BPD} it was found that analytic maximization over two of the angular 
variables, $\{\psi,\epsilon\}$, is possible even in the case of non-coincident
detectors. This useful observation allowed further reduction of the LLR to 
obtain a network statistic. In the following, we briefly mention this 
analytic maximization before discussing how the reduced statistic can be used 
for searching chirps.

\subsection{Further maximization over $\epsilon$ and $\psi$}
\label{sec:MSS2}

{}For spatially separated detectors the cross-correlation, $C^{I}$, is strongly
dependent on the time of final coalescence, ${t_c}$, and the time delay, 
$\tau_{(I)}$. Since $\tau_{(I)}$ depends on the source-direction, $(\t,\p)$, 
so does $C^I$. This prompts us to recast the statistic, (\ref{LLRhatbd}), in 
such a way that its dependence on the angles $\{\psi,\e\}$ is isolated 
as shown below. This aids in the analytic maximization of 
$\ln \lambda |_{\hat{\sf b},\hat{\d_c}}$ over the angles $\{\e,\psi\}$. We 
note that using Eqs. (\ref{ext1}) and (\ref{QI}), the network vector $\sf Q$ 
can be re-expressed as 
\begin{eqnarray}\label{Q}
{\sf Q} &=& {1 \over {\parallel {\sf E} \parallel}}
            \left[ T_{2}{}^{-2} (\psi,\e, 0) {\sf D}_{-2}
              +  T_{2}{}^{2} (\psi,\e, 0) {\sf D}_{+2} \right] \no\\
        &\equiv& Q^{-2} {\hat {\sf D}}_{-2} + Q^{+2} {\hat {\sf D}}_{+2}\ \ ,
\end{eqnarray}
where ${\sf D}_p$ ($p=\pm 2$) define two network vectors with the 
components $g_{(I)} {D}^I_p$, respectively. 
The vectors ${\hat {\sf D}}_p$ are their normalized counterparts. 
Since ${\sf Q}$ has a unit norm, the above expression implies that 
\be \label{QPM}
Q^{-2} = {{\parallel {\sf D} \parallel} \over {\parallel {\sf E} \parallel}}
T_2{}^{-2}(\psi,\e,0), \hspace{0.3in} Q^{+2} = {{\parallel {\sf D} \parallel} 
\over {\parallel {\sf E} \parallel}}
T_2{}^{2}(\psi,\e,0) \ \ ,
\ee
where ${\parallel {\sf D} \parallel} \equiv {\parallel {{\sf D}_{+2}} 
\parallel} = {\parallel {{\sf D}_{-2}} \parallel}$.

The following relations hold among $Q^{\pm 2}$, $Q_{\pm 2}$, and 
${\hat {\sf D}}_{\pm2}$:
\begin{eqnarray}\label{QPM1}
Q_{+2}  &\equiv& {\hat {\sf D}}_{+2} \cdot {\sf Q} = Q^{+2} + Q^{-2}{\hat 
{\sf D}}_{+2} \cdot {\hat {\sf D}}_{-2} \ \ , \no\\
Q_{-2} &\equiv& {\hat {\sf D}}_{-2} \cdot {\sf Q} = Q^{+2} {\hat
{\sf D}}_{-2} \cdot {\hat {\sf D}}_{+2}+ Q^{-2} \,.
\end{eqnarray}
The pair $\{{\hat {\sf D}}_{+2},{\hat {\sf D}}_{-2}\}$
defines a two-dimensional complex subspace in ${\cal C}^M$, on which a metric 
$G_{pq}$ can be defined.
${\hat {\sf D}}_{+2}$ and ${\hat {\sf D}}_{-2}$ depend only on the direction
of the source and the orientation of the detectors, that is, on 
$\{\t,\p,\a_{(I)},\b_{(I)},\g_{(I)}\}$ and not on $\{\e,\psi\}$.
The metric components in the ${\hat {\sf D}}_{\pm 2}$ basis are given by,
\be
{\small
{G_{pq}} = \left(\begin{array}{ccc}
1 & {\hat {\sf D}}_{+2} \cdot {\hat {\sf D}}_{-2}\\
{\hat {\sf D}}_{-2} \cdot {\hat {\sf D}}_{+2}& 1
\end{array}\right) \,.}
\ee
This metric can be used to `raise' and `lower' indices of vectors 
lying in this complex subspace, e.g., one has $Q_p = G_{pq} Q^q$, where $p$, 
$q=\pm 2$. We observe that, in general, ${\hat {\sf D}}_{\pm 2}$ are not 
orthogonal. {}For a face-on binary (i.e., $\e = 0$) revolving anti-clockwise 
(or clockwise), ${\sf D}_p$ itself is proportional to the network vector
${\sf E}$ for $p=-2$ (or for $p=2$). Hence, we may call the two-dimensional 
subspace as the ``helicity'' plane, ${\cal H}$.

The $M$-dimensional complex correlation vector ${\sf C}$, in general, lies 
outside ${\cal H}$. However, ${\sf Q}$ lies totally in ${\cal H}$. Thus, the 
statistic reduces to
\be \label{lamb}
2\ln \lambda |_{\hat{\sf b},\hat{\d}_c} = 
\left| {\sf C} \cdot {\sf Q} \right|^2 = \left| {\sf C}_{\cal H} \cdot 
{\sf Q} \right|^2 \ \ , \ee
where ${\sf C}_{\cal H}$ is the projection of ${\sf C}$ on ${\cal H}$.
Maximization of the above statistic over $\{\e, \psi\}$ is achieved by 
aligning  
${\sf Q}$ along ${\sf C}_{\cal H}$ by a proper choice of $\psi$ and $\e$. 
To show that this is always possible, we expand ${\sf C}_{\cal H}$ in the 
two-dimensional basis $({\hat {\sf D}}_{+2},{\hat {\sf D}}_{-2})$:
\be \label{C}
{\sf C}_{\cal H} = C^{-2}_{\cal H} {\hat {\sf D}}_{-2} + C^{+2}_{\cal H} 
{\hat {\sf D}}_{+2} \ \ ,
\ee
where $C^p_{\cal H} \equiv G^{pq} ({\hat {\sf D}}_q \cdot {\sf C})$.
Then the maximization condition is
\be \label{MAX}
{{Q^{+2}} \over  {Q^{-2}}} = {{C^{+2}_{\cal H}} \over {C^{-2}_{\cal H}}} \,.
\ee
In principle, ${\sf C}$ can take any value in ${\cal C}^M$, so the
RHS of (\ref{MAX}) can take any value in the complex plane.
The task is to prove that the ratio ${{Q^{+2}}/{Q^{-2}}}$ obeys
the same property. In other words, we must show that
the ratio ${{Q^{+2}}/{Q^{-2}}}$ too can span the entire complex plane.
This follows immediately from the observation that 
\be \label{MAX1}
{Q^{+2} \over {Q^{-2}}} = 
{{T_2{}^{-2} (\psi,\e, 0)} \over {T_2{}^{2} (\psi,\e, 0)}} = 
\left ({{1 - \cos \e} \over {1 + \cos \e}}\right)^2 \exp({4i \psi}) \,.
\ee
{}For $\e \in [0,\pi]$ and $\psi \in [0,2\pi]$, the RHS of (\ref{MAX1})
can indeed attain any value in the Argand plane. One set\footnote{There exist 
three other values of $\hat {\psi}$.} of values of $\hat{\psi}$ and
$\hat{\e}$ that maximizes the statistic are
\be
{\hat \psi} = {\arg (\zeta) \over 4}, \hspace{0.3in}
{\hat \e} = \cos^{-1} \left[{1-\sqrt{|\zeta|}} \over {1+\sqrt{|\zeta|}}
\right] \ \ ,
\ee
where $\zeta \equiv C^{+2}_{\cal H} / C^{{-2}}_{\cal H}$. 
Thus, the LLR maximized over the four parameters is 
\be \label{st}
2\ln \lambda |_{{\hat{\sf b}},{\hat{\d}_c},{\hat \psi},{\hat \e}} = 
{\parallel {\sf C}_{\cal H} \parallel}^{2} \equiv \Lambda \,.
\ee

Geometrically, we summarize the above maximization over the angles $\{\epsilon,
\psi\}$ as follows: Choosing a given source direction fixes the orientation of 
the helicity plane in the network space. After making this choice, one 
projects the correlation vector ${\sf C}$ onto this plane. The vector ${\sf Q}$
inevitably lies in this plane. Thus, the values of $\{\epsilon,\psi\}$ that 
maximize the statistic are those that align the vector
${\sf Q}$ along the projected vector ${\sf C}_{\cal H}$.

It is always possible to choose in ${\cal H}$, a two-dimensional basis 
comprising of a pair of orthonormal real vectors. In such a basis the 
components of any vector in ${\cal H}$ will, in general, be complex.
{}For the sake of concreteness, we define one such basis, 
$({\hat {\sf v}}^+, {\hat {\sf v}}^-)$, in the following way.
We split ${\sf D}_{+2}$ into its real and imaginary parts
\be \label{F1}
{\sf D}_{+2} \equiv {\sf d}_1 + i {\sf d}_2 \ \ ,
\ee
where ${\sf d}_1$ and ${\sf d}_2$ are real vectors. We then define
\be
{\hat {\sf v}}^\pm = ({\hat {\sf d}}_1 \pm {\hat {\sf d}}_2)/\parallel 
{\hat {\sf d}}_1 \pm {\hat {\sf d}}_2 \parallel \,.
\ee
Taking the projection of ${\sf C}$ on ${\hat {\sf v}}_{\pm}$, i.e., $C^\pm
\equiv {\hat {\sf v}}^{\pm} \cdot {\sf C} = c_0^\pm + ic_{\pi/2}^\pm$, 
we re-express Eq. (\ref{st}) as
\bea \label{SSnew}
{\parallel {\sf C}_{\cal H} \parallel}^{2} &=& {|{C}^{+}|}^{2} 
+{|{C}^{-}|}^{2} 
 = (c_0^+)^2 + (c_{\pi/2}^+)^2 + (c_0^-)^2 + (c_{\pi/2}^-)^2 \nonumber \\
&\equiv& \L^2 \,. 
\eea
It can be verified that the statistic is, therefore, a sum of the 
squares of four Gaussian random variables with constant variance. With an
appropriate choice of normalization, these variances can be made unity. As we
show in Sec. \ref{FDV}, this simplifies the computation of detection 
thresholds and probabilities associated with the above statistic. Instead of 
using the squared norm of $\sf{C}_{\cal H}$, we will 
find it convenient to use  as our statistic $\L \equiv
\parallel {\sf C}_{\cal H} \parallel$, in what follows. We note that $\L$ then
scales linearly with the amplitude of the signal vector, rather than its 
square. We will call $\L$ as our network statistic. This statistic was first 
obtained in Ref. \cite{BPD}.

\subsection{Maximizing $\L$ over the time of arrival}

Given a network data-vector, ${\sf x}(t)$, which may or may not contain
a chirp, it is necessary to first compute the correlation vector, 
${\sf C}$, before one can obtain ${\sf C}_{\cal H}$ and, therefore, the 
network statistic. In order to do so, we need $C^I$, for all $I$. We compute a 
$C^I$ (or, rather, $C_I^*$) by first calculating the Fourier 
transform of the cross-correlation (\ref{Cstar}) 
for an individual detector by using FFTs.
Taking its inverse FFT then gives us the $C_I^*(\tau)$ at all the 
time lags, $\tau$, in a cost effective way. Thus we get,
\be \label{CI}
C_I^*(\tau;t'_c,\xi',\t',\p') = \langle S^I(t;t'_c+\tau,\xi'), x^I(t;t_c,\xi)
\rangle_{(I)} \ \ ,
\ee
where the primed parameters define the detector template. If the values 
chosen for $\{\xi',\t',\p'\}$ match those of a chirp in the data, then 
$|C^I|$ is likely to peak when $\tau$ exactly compensates for the 
difference $(t'_c - t_c)$. Indeed, the correct ${\sf C}$ for a network of
coincident detectors is just 
\be \label{corcoin}
{\sf C}(\tau; \mbox{\boldmath $\vt$}') 
\equiv \{C^I (\tau; \mbox{\boldmath $\vt$}') \} \ \ ,
\ee  
where $\mbox{\boldmath $\vt$}'$ is the four-dimensional template 
parameter-vector. Also, $I$ takes values from $1$ to $M$.

Construction of ${\sf C}$ for a network of non-coincident detectors is 
somewhat more involved owing to non-vanishing time-delays, 
$\tau_{(I)}(\t,\p)$, 
that may arise for a given source direction. Recall that the time of arrival 
at the $I$-th detector is $t_a + \tau_{(I)}(\t,\p)$. If the detectors are 
spread 
around the globe, the times of arrival at any pair of detectors can at most 
differ by $2R_{\oplus}/c \sim 40$ ms, where 
$R_{\oplus}$ is the radius of the Earth. {}For the two LIGO detectors, the 
maximum time difference is $\sim 10$ ms; for the network of LIGO-VIRGO,
it is $\sim 27$ ms. We note that $\tau_{(I)}(\theta,\p)$ can take positive as 
well as negative values. Its range depends on the location of the fide. 
If the fide is chosen at the center of the Earth, then $|\tau_{(I)}| \leq 
R_{\oplus}/c$; but if it is chosen to be one of the detectors on the surface 
of the Earth, then $|\tau_{(I)}| \leq 2R_{\oplus}/c$. This contingency is 
dealt with by using the appropriate set of $\tau_{(I)}$'s in Eq. (\ref{SI}) 
to obtain the $S^I(t;t'_c,\xi')$. With this in place, the network 
correlation-vector is given by the same expression as in Eq. (\ref{corcoin}).

One can obtain the same value for ${\sf C}$ by an alternative construction,
which may be simpler to implement in practice. In this method, one first  
obtains the $S^I$ by setting the arrival time at every detector to equal that 
at the fide. This is the same as computing the $S^I(t+\tau_{(I)};t'_c,\xi')$, 
for all $I$. (Note from Eq. (\ref{SI}) that, despite appearances, 
a knowledge of $\tau_{(I)}$'s is actually not needed for this 
computation.) 
With these templates one constructs the following inner products: 
\be\label{CIalt}
\bar{C}_I^*(\tau;t'_c,\xi') 
\equiv \langle S^I(t+\tau_{(I)};t'_c+\tau,\xi'), x^I(t;t_c,\xi) \rangle_{(I)} 
\ \ , \ee
which are independent of the time delays. Indeed, a choice of the time delays 
is not made thus far in this alternative construction of ${\sf C}$. To 
construct the network correlation-vector from the above quantities, one begins
by choosing a source direction, $\{\theta',\phi'\}$, for the template. This
direction is used to compute a consistent set of time delays, 
$\tau_{(I)}(\theta',\phi')$, for a given network. The network 
correlation-vector can then be defined as
\be \label{cor}
{\sf C}(\tau; \mbox{\boldmath $\vt$}') = \{\bar{C}^I (\tau 
+\tau_{(I)}(\theta',\phi'); t'_c,\xi' ) \} \ \ ,
\ee  
where the appropriately shifted time lags for each value of $I$ 
compensates for the time-delay at each detector. In the rest of this section, 
we discuss the construction of network templates in greater detail. There,
for a network of non-coincident detectors, we choose this latter prescription 
for such a construction.

\subsubsection*{\textsf{Network template construction}}
\label{NF}

Based on the above discussion, we construct a network template as follows.
Given a chirp, consider the detector with the least seismic cut-off 
frequency. Label that detector as $I=1$. The other detectors in the
network are labeled such that $f_{s(1)} \leq f_{s(2)} \leq \cdots f_{s(M)}$. 
Then
from Sec. \ref{chirp} we have ${\xi}_{(1)} \geq {\xi}_{(2)} \geq \cdots {\xi}
_{(M)}$. Now, 
consider the signal in the first detector, $I=1$. It lasts $\xi_{(1)}$ 
seconds. As in the case of a single-detector ``network'', an individual 
detector 
template, which is an array of numbers, is constructed to be much longer 
than the signal: It comprises of a sub-array that stores the signal being 
searched for, followed by a padding of the requisite number of 
zeros \cite{BFS1}. In the single-detector case, it has been shown that 
a padding factor of $75 \%$, that is, $75 \%$ zeros and $25 \%$ signal, is a 
good choice in the sense that it optimizes the computational cost arising from 
the computation of the FFT's. Accordingly, here too we pad the template for 
the first detector with zeros for a time duration of $\sim 3 \xi_{(1)}$. 

In the sensitivity window of any of the other detectors, that is, for 
$(I\neq 1)$, the signal effectively lasts for a time-duration equal to or 
shorter than $\xi_{(1)}$. Nevertheless, the simplest way to construct their 
templates is to let them contain the same signal as in the first detector,
and for the same duration, namely, $\xi_{(1)}$. Such a construction does not 
restrict the network statistic in any way. Its only pertinent implication is 
that any part of the signal that has an instantaneous frequency below 
$f_{s(I)}$
will be ineffective in contributing to the SNR in the $I$-th detector, which
conforms with what we expect. In the case of a network with coincident 
detectors, the individual detector-templates so constructed define the 
components of the network template-vector. Using it in Eq. (\ref{corcoin})  
yields the relevant network correlation-vector.

Obtaining the template-vector of a network with non-coincident detectors is
trickier. This is essentially due to the possibility of the occurrence in a 
given detector of negative time-delays. We deal with this possibility by 
splitting the padding into two parts of durations $\tau_d$ and
$(3\xi_{(1)} -\tau_d)$, respectively. Since the maximum magnitude that a 
time-delay can have is less than $50$ ms, a choice of $\tau_d = 50$ ms 
satisfies all requirements at negligible cost. This is the value we assume for 
$\tau_d$ in our simulations. Thus, the template of any one of the detectors in 
such a network is an array of numbers that begins with a ``pre-padding'' (with 
zeros) of duration $\tau_d$ preceding the signal of interval $\xi_{(1)}$, 
which in turn is followed by a ``post-padding'' of duration 
$(3\xi_{(1)} - \tau_d)$. The network template is then the Cartesian 
product of all these individual templates. Note that this can be taken to 
define the network template for the most general network, regardless of whether
the detectors in it are coincident or not. For non-coincident detectors, the
relative time-delays are accounted for in the construction of the network 
statistic via Eq. (\ref{cor}).

In order to construct the network correlation-vector, ${\sf C}$, one utilizes 
the above templates as follows. Using these individual-detector templates 
one first obtains the correlations $\bar{C}^I(\tau ;t'_c,\xi')$ 
for all values of $\tau$ (after setting $t'_a = 0$ or, equivalently, $t'_c =
\xi'$, without any loss of generality) by using the FFT 
algorithm, such as in the single-detector case. Note that the range 
of $\tau$, for a data train of length $T$, is $\tau_d \leq \tau
\leq T - \xi_{(1)}-\tau_d$. Next one selects a source direction, $(\theta',
\phi')$. One evaluates the time-delays, $\tau_{(I)}(\theta',\phi')$, 
corresponding to this direction.\footnote{Alternatively, one may first select a
consistent set of $\tau_{(I)}$'s and then deduce $\{\theta', \phi'\}$ from 
them.} These time delays are then used in Eq. 
(\ref{cor}) to compute ${\sf C}$, which in turn is projected on ${\cal H}$
(defined by the same selection of $\{\theta', \phi'\}$) in order to obtain
${\sf C}_{\cal H}$. The network statistic can be easily recovered from this
using Eq. (\ref{st}).

\begin{figure}{
\centerline{\psfig{file=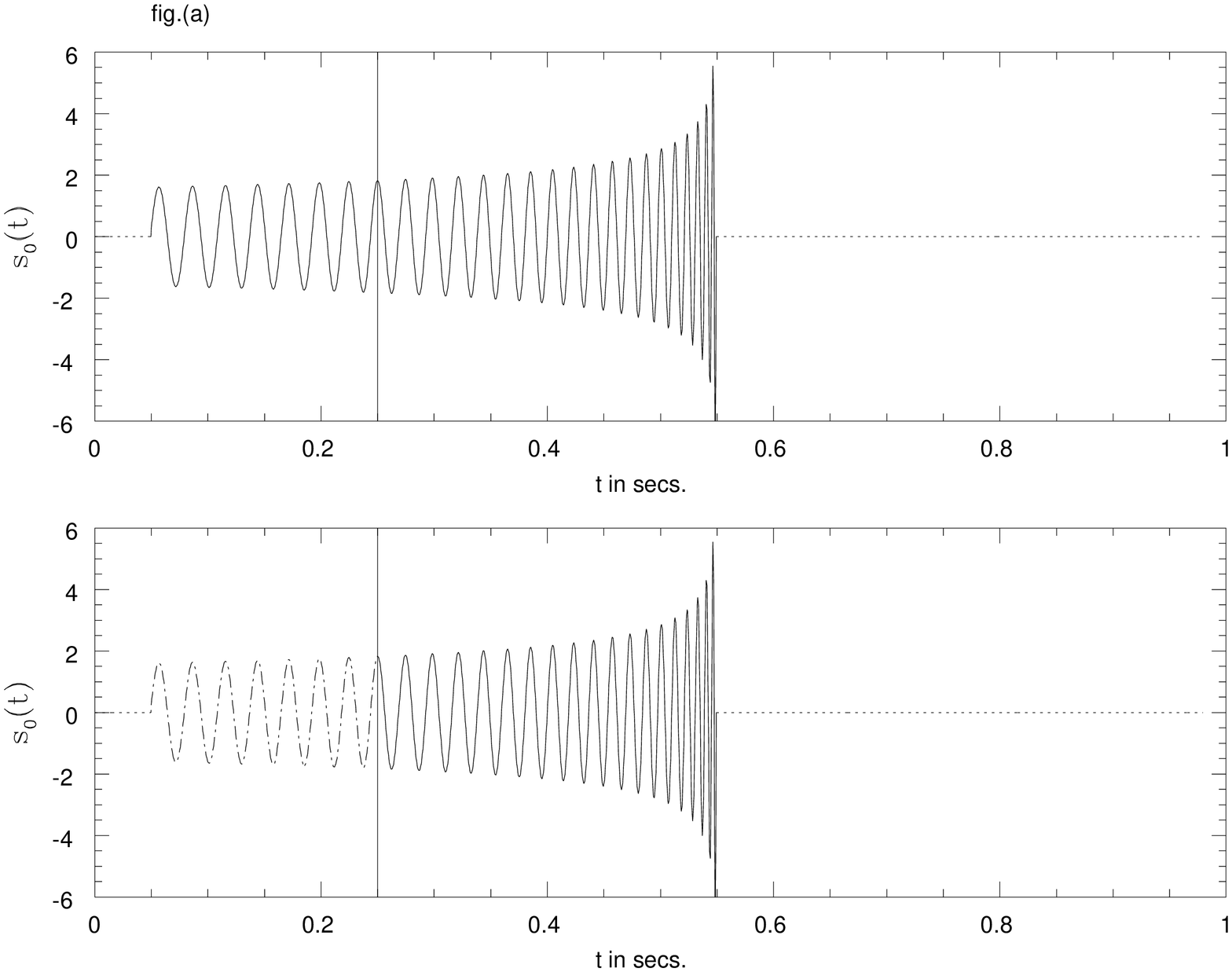,height=3.2in,width=6.0in,angle=270}} 
\centerline{\psfig{file=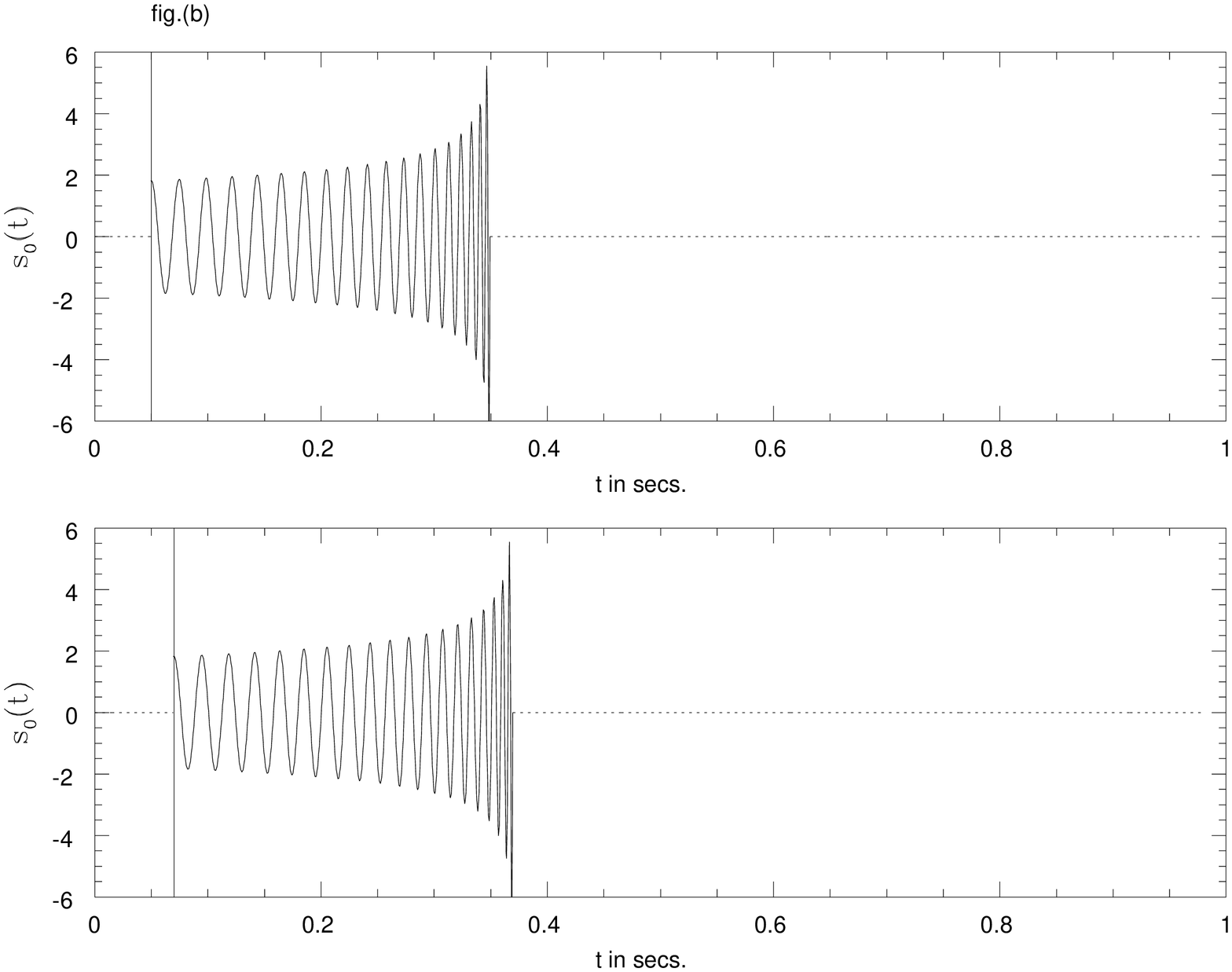,height=3.2in,width=6.0in,angle=270}}
\caption{Network template for a two-detector network. We choose the fiducial 
frequency as $f_s = 40$ Hz. The instant at which the signal reaches $f_s$ is 
shown by a vertical line. We take the chirp time corresponding to the fide to 
be $\xi = 0.3$ sec. The dots represent the padding with zeros, which is
introduced before and after the signal. The two panels in (a) depict 
the two individual detector-templates constituting the network template. These
two panels correspond to detectors with different seismic cut-offs, viz., 
$f_{s(1)} = 33$ Hz and $f_{s(2)} = 40$ Hz, respectively. The respective 
chirp times are $\xi_{(1)} = 0.5$ sec and $\xi_{(2)} = 0.3$ sec. The padding 
before (i.e., on 
the left-hand side of) the signal is of a duration $\tau_d=50$ ms. The part of 
the curve for detector 2 that is shown in dots and dashes is ineffective in 
contributing to the SNR. Panels (b) show the relative positions of the signal 
in the individual detector-templates for which $\|{\sf C}\|$ has a maximum 
when the second detector has a relative time-delay of $(\tau_{(2)} -\tau_{(1)})= 20$ ms.
Each of the detectors has its seismic 
cut-off equal to $40$ Hz, which gives $\xi_{(1)} = \xi_{(2)} = \xi = 0.3$ sec.}
}
\label{fig2} 
\end{figure}

In Fig. \ref{fig2} (a), we show a network template for a network of two 
detectors. In all the panels, the dots represent the padding (with zeros), 
which is introduced before and after the signal. The two panels in (a) depict 
the two individual detector-templates constituting the network template. These
two panels correspond to detectors with different seismic cut-offs, viz., 
$f_{s(1)} = 33$ Hz and $f_{s(2)} = 40$ Hz, respectively. The padding before 
(i.e., on 
the left-hand side of) the signal is of a duration $\tau_d=50$ ms. The part of 
the curve for detector 2 that is shown in dots and dashes is ineffective in 
contributing to the SNR. Panels (b) show the relative positions of the signal 
in the individual detector-templates for which $\|{\sf C}\|$ has a maximum 
when the second detector has a relative time-delay of $(\tau_{(2)} -\tau_{(1)})
= 20$ ms. Here, we have included the time delay in the network template.

Above, the chirp time in the detector with the least $f_{s(I)}$ decided the 
durations of the padding and the chirp-signal in all the individual
detector-templates. 
Indeed, these durations are the same in all of them. It may be 
possible to optimize on computational costs by varying these 
durations in different templates. However, in this work we do not pursue this 
point any further.

\subsection{Maximization of $\L$ over $\xi$, $\theta$, and $\phi$.}
\label{sec:com}

Consider the network correlation-vector, ${\sf C}(\tau)$, for a fixed value of 
$\xi'$, but for a range of values for $\tau$. As remarked before, such a vector
is constructed for specific values of $\tau$ and source direction 
$(\theta',\phi')$ by taking into account the time delays, 
$\tau_{(I)}(\theta',\phi')$, appropriate for the network under consideration.
The network statistic for these chosen values of $(\theta', \phi')$ and $\tau$ 
can be obtained by first projecting ${\sf C}$ on the helicity plane and then
computing the norm of the projection. A chirp search over $\{\theta',\phi'\}$ 
for a given configuration of network leads to a ``window'' of time delays. In 
any network of non-coincident detectors a window, $W$, is a bounded region in 
the space of time-delays that arises from the restrictions on each of the time 
delays to lie within certain limits. As we illustrate below, these limits, in 
turn, originate owing to the maximum light-travel time between pairs of 
detectors. 
Since the data are sampled discretely, a window is a bounded region of a 
lattice in the space of possible time-delays.
Consequently, the number of (lattice) points in a window of finite ``volume'' 
is also finite. 
We now discuss the shape and size of a window for networks with two, three, 
four, and more than four detectors. 

\begin {itemize}
\item{{\it Two-detector network}:\\
{}For a network of two detectors, there exists effectively a single time-delay
function of significance. It is the difference in the times of arrival of the 
wave at the two detectors. For such a network we choose one of the detectors,
say, $I=1$, as the fide. Then $\tau_{(1)} = 0$ and the time delay between
the two detectors is reflected solely in $\tau_{(2)}$, which is restricted to
lie within the range $[-d_{12}/c, d_{12}/c]$, where $d_{12}$ is the distance 
between the two detectors. Let $\Delta$ be the sampling
interval, which is typically $0.5$ ms. Then the ``width'' of the window, 
expressed in terms of the number of time-sampled points, is 
$n_S^{\Omega} = 2 \tau_{(2)}/\Delta$, where the subscript $S$ stands for
sampling and $\Omega$ for the direction in sky, $(\theta,\phi)$.
If we denote the LIGO detector at Hanford by H, the LIGO detector at Louisiana 
by L, and the VIRGO detector by V, then for the two-detector networks formed 
from pairs among these, we have
$n_S^{\Omega}$($LH$) $\sim 40$, $n_S^{\Omega}$($HV$) $\sim 108$, and 
$n_S^{\Omega}$($LV$) $\sim 105$.}

\item{{\it Three-detector network}:\\
In the case of a network with three detectors, we once again take the first 
one (i.e., $I=1$) to be fiducial. Also, we can always imagine all of them to 
lie on a single plane. In such a case, there arise two nontrivial time-delays, 
namely, $\tau_{(2)}$ and $\tau_{(3)}$.
The allowed values of these two time-delays are easily shown to be
restricted within a bounded region of a plane; this region is circumscribed by
an ellipse. Any point in this region represents a pair of 
time-delay values, $(\tau_{(2)},\tau_{(3)})$, corresponding to a given pair of 
values for the source-direction angles, $\{\theta', \phi'\}$. The equation of 
this ellipse is given by \cite{BD}
\be \label{ellipse}
(\tau_{(2)})^2 + (\tau_{(3)})^2(p^2 + q^2) - 2 p \tau_{(2)} \tau_{(3)} - q^2 (d_{12}/c)^2 = 0 \ \ ,
\ee
where $p = (d_{13}/d_{12}) \cos {\alpha_{23}}$ and 
 $q = (d_{13}/d_{12}) \sin {\alpha_{23}}$, with $d_{12}$ (or $d_{13}$)
being the distance between the first and the second (or third) detectors. Also,
${\alpha_{23}}$ is the angle subtended by the hubs of detectors 2 and 3
at that of the first detector. 
{}From (\ref{ellipse}), the ``area'' of the elliptical window,
given in terms of the number of time-sampled points, is $n_S^{\Omega} = 2 \pi 
\times {\cal A} /(c^2 \Delta^2)$ where $\cal A$ is the area of the triangle 
formed by the hubs of the three detectors in the network. We find that the 
number of possible time-delays for the three-detector 
network of LIGOs-VIRGO ($LHV$) is $n_S^{\Omega} \sim 3 \times 10^3$.}

\item{{\it Four and more detectors}:\\
In the case of a three-detector network, the two time-delays 
produce two circles on the
celestial sphere that intersect at two points, which give the possible
directions to the source. This two-fold degeneracy is broken when we 
introduce a fourth detector lying outside the plane of the three detectors.
In such a case, the possible time-delays can be represented in a 
three-dimensional space of the three time-delays, which now lie on the 
surface of an ellipsoid. The number of possible time-delays is exactly 
doubled compared to that of the three-detector network. Thus, its maximum 
possible value is $n_S^{\Omega} = 4 \pi {\cal A}/(c^2 \Delta^2)$, where
${\cal A}$ is the area of the smallest of all possible triangles formed 
subsets of three of the four detectors. When there are more than 4 detectors 
there is redundant information on the direction to the source, but
the $n_S^{\Omega}$ is the same as for four detectors. In the presence of 
noise this redundant information may be used to reduce the errors in the
direction to the source. Here, we do not pursue this point any further.}

\end{itemize}

The sampling interval naturally provides the most simplistic 
discretization in carrying out the search in time-delays.
In searching over $\theta$ and $\phi$ one does not need to compute additional 
Fourier transforms; rather, one needs to combine the individual 
detector-correlations with the correct time delays to construct the optimal 
statistic, $\L$. This gives rise to two components to the computational cost: 
the cost involved in computing Fourier transforms and the cost arising due to 
the arithmetic operations involved in computing $\L$ over all possible 
time-delays. As shown later, the latter cost can be considerable and may 
dominate over the cost in computing Fourier transforms while searching over 
$\xi$. 

It is important to note that sampling can introduce an arbitrary 
mismatch between the actual source direction, $(\theta$, $\phi)$, and the 
direction in the template, $(\theta'$, $\phi')$. 
The mismatch, $\mu$, is the fractional loss in SNR when the signal
and the template parameter differ slightly. In agreement with most 
investigations carried out so far, we decide to tolerate a mismatch
to a maximum of $3 \%$. The sampling can lead to
a mismatch either greater or smaller than $\mu$.
If the sampling gives a mismatch less than
the desired one, then this simplistic procedure of scanning $W$
leads to unnecessary extra computational costs. On the other hand,
if the mismatch is more, then one is likely to miss out more events than 
desired. The question
whether the sampling is adequate one way or the other can be resolved by
constructing a template-bank for the desired mismatch $\mu$.
In the next
section, we proceed to construct a bank of templates in the parameter
$\xi$, $\theta$, and $\phi$. The template bank, in general, will
produce time-delays that do not fall exactly at the sampled values of the
correlation vector. However, we can easily interpolate to obtain the 
intermediate
values by applying Shannon's theorem \cite{Brace}, which essentially
states that a band-limited function can be constructed in the time domain 
from its discretely sampled values at the Nyquist rate. Or in other words,
the template bank provides the rate at which the output can be re-sampled
so as to obtain the desired maximum mismatch. 

\section{Template bank in $\xi$, $\theta$, and $\phi$} 
\label{filbank}

Recall that the LLR is a function of eight parameters,
namely $\{ r,\delta_c,\epsilon,\psi,t_c,\xi,\theta,\phi\}$ for the
Newtonian chirp. As mentioned earlier, we adopt the 
maximum-likelihood method for the detection problem. It implies 
that the LLR must be maximized over all the parameters to 
obtain the MLR. We have shown that the
maximization of LLR can be carried out analytically with respect to
the four of the eight parameters, $\{r,\delta_c,\epsilon,\psi\}$. Also,
we can deal with the time of final coalescence at the fide $t_c$ efficiently 
by using the FFT algorithm. Therefore, we now need to formulate a strategy to 
search over the rest of the 3-D parameter space formed by $\xi$, $\theta$, 
and $\phi$. Ideally, one should scan the whole range of 
the 3-D parameter space over the physically allowed parameter values.
This, however, is impractical due to computational limitations.
Therefore, a prerequisite for such a maximization is an estimate for the 
magnitude of grid discretization. The grid spacing in the parameter space 
depends upon the fractional loss in the SNR that one is prepared to tolerate. 

To estimate the number of templates in the 3-D parameter space, we take the
differential-geometric approach \cite{BSD} and use Owen's method of 
introducing a metric on that space \cite{Owen} and extend his formula for the 
one detector case to that of the network. Also the inverse of the metric 
is just the covariance matrix scaled by the square of the SNR. Thus the metric 
also provides information on the errors in estimating the parameters.  
In this geometric method, the signal vector is characterized by $P+1$ 
parameters, $\vartheta^\alpha$, where $\alpha = 0,1,...,P$. The signal vector 
lies in a $(P+1)$-dimensional manifold denoted by ${\cal P}$. We define the
metric on $\cal{P}$ by $g_{\alpha \beta}$, which is related to the fractional 
loss in SNR, denoted by $\mu$, when there is a mismatch between the signal and 
the template parameters. Since here we consider the Newtonian chirp as our 
signal, we have $P=3$, $\vartheta^0 \equiv t_c$, and $\vartheta^i \equiv \{\xi,
\theta,\phi \}$. As $\L$ can be maximized over $t_c$ numerically via the 
FFT, we only need to lay the templates in the rest of the $P$-dimensional 
parameter space, comprising of $\{\xi,\theta,\phi \}$. 
In other words, we need to compute the metric $\gamma_{ij}$ in the 
$P$-dimensional subspace. It is determined by projecting 
the metric $g_{\alpha \beta}$
onto the subspace orthogonal to $t_c$. We then obtain
\be \label{metric}
 \gamma_{ij} = g_{ij} - {{g_{0i}g_{0j}} \over {g_{00}}} \,.
\ee 
The number of templates is obtained as follows. We compute the proper volume
 of the parameter space with the metric $\gamma_{ij}$ and multiply the 
volume by the number density of the templates. {}Fixing the value of
$\mu$ determines the grid spacing of the network templates in the
parameter space. The number density, $\rho_{P}(\mu)$, which is the number of 
templates per unit proper volume, is given by
\be \label{vol}
\rho_{P}(\mu) = \left({1 \over 2} \sqrt{P \over \mu} \right)^{P} \,.
\ee
It is defined to be uniform over the whole parameter space and, therefore, its
use is applicable as long as the curvature of the astrophysically interesting 
region of the manifold described by $\gamma_{ij}$ is sufficiently small, and 
the effects arising from the boundary of the region are negligible. 

The total volume of the parameter space is
\be \label{vol2}
{\cal V} = \int_{\cal P} \sqrt{\det \parallel \gamma_{ij} \parallel} 
\hspace{0.1in}  d^{P} \vt \,.
\ee
Thus, the total number of templates required is , $n_f = {\cal V} \times 
\rho_{P}(\mu)$ . In general, the total number of templates depends on the 
source parameters $\{ \e, \psi \}$. In order to scan the parameter space (for
a given mismatch $\mu$) for each pair of values $\{ \e,\psi \}$,
we must maximize the volume ${\cal V}$ over $\{ \e,\psi \}$. This is
tantamount to choosing the finest bank of templates. {}For simple
cases, this is straightforward and has been implemented in some examples
in Sec. \ref{difnet}. In general, however, such a maximization is 
non-trivial to perform. 

\subsection{The network metric}
\label{filter}

We apply the method described above to obtain the metric in the
four-dimensional parameter space $\{t_c,\xi,\theta,\phi\}$. 
When the parameters of the network
template and that of signal mismatch, the network statistic given by
(\ref {SSnew}) drops below the maximum value. The metric $g_{\alpha \beta}$
defined on this four-dimensional space is related to the amount of drop in the 
statistic, ${\L}$, and is obtained by expanding the statistic about the 
maximum. Using Eq. (\ref{SSnew}) the squared statistic can be rewritten as,
\bea
\L^2 &=& |{\hat {\sf v}}^{+'} \cdot {\sf C} |^2 + |{\hat {\sf v}}^{-'} 
         \cdot {\sf C}| ^2 \no \\ 
  &\equiv& p'_I{}^J C^I C_J^* \,.
\eea
The quantity $p'_I{}^J$ is a projection tensor given by, 
\be
p'_I{}^J \equiv {v'}^{+J} {v'}^{+}_I + {v'}^{-J} v'^{-}_I \ \ , 
\ee
which projects a vector in ${\cal C}^M$ on the helicity plane 
spanned by
${\hat {\sf v}}'^\pm$. It obeys the identities 
\be
p'_I{}^J p'_J{}^K = p'_I{}^K \>, \hspace{0.5in} p'_I{}^J p'_J{}^I = 2 \ \ ,
\ee
which are consistent with its being a projection tensor on a two-dimensional
plane. The primed coordinates refer to the template.

Let $\vartheta^\alpha$ and $\vartheta'^\alpha =
\vartheta^\alpha + \Delta \vartheta^\alpha$ be 
the parameters corresponding to the signal and
the network template, respectively. {}For computing the metric one takes 
normalized templates for the signal as well as the template, so that the
maximum value of $\L$ is unity when the parameters of the signal
and template match. In the absence of noise, Eq. (\ref{Cstar}) yields
\be
C_I^* = \langle S'^I, s^I \rangle_{(I)} 
\simeq e^{i\delta_c} Q_I^* \langle S'^I,S^I \rangle_{(I)} \ \ ,
\ee
where, the $S'^I$ denotes the template. The above expression is exact within
the SPA. So the statistic can be written as,
\be \label{lsq}
\L^2 = p'_I{}^J Q^I Q_J^* \Theta_{(I)(J)} \ \ ,
\ee
where 
\begin{eqnarray} \label{th}
\Theta_{(I)(J)} &\equiv& \langle S^I(t;\vt'^\mu), S^I(t;\vt
^\mu)\rangle^*_{(I)} \langle S^J(t;\vt'^\mu), S^J(t;\vt^\mu)\rangle_{(J)} 
\ \, \nonumber \\
&=& \int_{f_{s(I)}}^{\infty} \! df {\cal A}_{(I)}(f) \exp \left(-i \Phi_{(I)} (f;\vt^\mu,
\Delta \vt^\mu) \right) \int_{f_{s(J)}}^{\infty} \! df {\cal A}_{(J)}(f) \exp 
\left(i \Phi_{(J)} (f;\vt^\mu,\Delta \vt^\mu) \right) 
\end{eqnarray}
is the product of the individual ambiguity functions of the $I$-th and $J$-th
detectors. It is a measure of how distinguishable the two wave-forms, i.e., 
the signal and the template, are. Here, 
\be
{\cal A}_{(I)} (f) \equiv 
{8 \over {3 f_s g_{(I)}^2}} {1 \over {s_{h(I)}(f)}} \left( {f \over f_s} 
\right)^{-7/3} \ \ ,
\ee
which satisfies the normalization condition $\int_{f_{s(I)}}^\infty 
{\cal A}_{(I)} (f) df = 1$. Note that in the limit of 
$\D \t ^ \a \rightarrow 0 $ the projection tensor of the filter is same as 
that of the projection tensor of the signal, {\it{i.e.}} 
$p'_I{}^J \rightarrow p_I{}^J$. In this limit, the projection tensor of the 
signal obeys the relation $p_I{}^J Q^I Q_J^* = 1$ and noting that $S^I$'s 
are normalized to unity, we can see from Eq.~({\ref{lsq}}) that  
$\L^2 \rightarrow 1$ as desired. 

The correlation phase, $\Phi_{(I)} (f;\vt^\mu,
\Delta \vt^\mu)$, is given by (see Eq. (\ref{psiI})):
\be \label{phiI}
\Phi_{(I)} (f;\vt^\mu, \Delta \vt^\mu) = \Psi_{(I)}(f;f_s,t_{c},\xi) - 
\Psi_{(I)}(f;f_s,t'_{c},\xi') \,.
\ee 
The correlation phase includes the contribution from the differential
time-delay between the signal and the template. Instead of using $(\t,\p)$
to specify the direction to the source we use the components $n_1$ and $n_3$
of the unit vector ${\hat{\sf n}} \equiv (n_1,n_2,n_3)$ to do so. The time 
delays in units of $f_s^{-1}$ are,
\be
f_s \tau_{(I)}(n_3,n_1) = \left[ r_{(I)1} n_1 + r_{(I)2} (1-n_3^2-n_1^2)^{1/2} 
+ r_{(I)3} n_3 \right] \ \ ,
\ee 
where ${\sf r}_{(I)}$ is the position vector of the $I$-th detector's hub and 
is henceforth measured in units of the ``fiducial wavelength'',
$\lambda_s \equiv c/f_s$. Since we choose to measure the time delays with 
respect to the fide, we must have ${\sf r}_{(f)} = 0$. Thus, we may write the 
correlation phase as
\be\label{Phicompact}
\Phi_{(I)} (f;\vt^\mu, \Delta \vt^\mu) \equiv 2 \pi \vp_{(I)\a}(f;\vt^\mu) 
\D\vt^\a \ \ ,
\ee
where  $\vt^\a \equiv \{f_st_c,\>f_s\xi,\>n_3,\>n_1\}$ is 
a quartet of dimensionless parameters and
\be \label{quart}
{\small \vp_{(I)\a} (f;\vt^\a) =
\left\{ \left({f\over f_s}\right),
\>{3\over 5}\left({f\over f_s}\right)^{-5/3}\!\!\!\!,
\> \left[r_{(I)3} - r_{(I)2} {{n_{3}} \over 
{\left(1-n_3^2-n_1^2\right)^{1/2}}}\right]{f\over f_s},
\> \left[r_{(I)1} -r_{(I)2}{{n_{1}} \over 
{\left(1-n_3^2-n_1^2\right)^{1/2}}}\right]{f\over f_s}\right\} \,.}
\ee

To obtain $g_{\alpha \beta}$ we Taylor expand the network
statistic about the peak at $\Delta \vt^\mu = 0$ to obtain
\begin{displaymath}
\L(\mbox{\boldmath $\vt , \Delta \vt$}) 
\approx 1 + {1 \over 2} {\left({{\partial^2 {\L}}\over {\partial \Delta \vt^\alpha \partial \Delta \vt^\beta}}\right)} \Bigg|_{\D\mbox{\boldmath $\vt$} =0}
 \Delta \vt^\alpha \Delta \vt^\beta \ \ ,
\end{displaymath}
where $\mbox{\boldmath $\vt$}$ is the four-dimensional signal parameter-vector.
Note that the first-order term in the Taylor expansion gives vanishing 
contribution. This is because $\L$ has a maximum there at
${\D\mbox{\boldmath $\vt$} =0}$, 
\be
{\partial \L\over \partial \vt^\a}\Bigg|_{\D\mbox{\boldmath $\vt$} =0} =0 \ \ ,
\ee
for any $\a$.
Then, the metric 
$g_{\alpha \beta} (\sf {\vt}, \Delta{\sf \vt})$ is defined as
\be \label{metg}
g_{\alpha \beta} (\vt) = -{1 \over 2} {\left({{\partial^2 {\L}}\over 
{\partial \Delta \vt^\alpha \partial \Delta \vt^\beta}}\right)} 
\Bigg|_{\D\mbox{\boldmath $\vt$} =0} \,.
\ee
The above differentiations can be performed. But first we study the effect of
a mismatch of signal and template parameters on the network statistic. 

{}For a perfect match between the signal and template parameters, the 
correlation vector ${\sf C} \propto {\sf Q}$, lies in the signal helicity
plane ${\cal H} (\theta,\phi)$. When mismatched, however, each component of 
${\sf C}$ gets multiplied by the weight factor $\langle S'^I,S^I 
\rangle_{(I)}^*$, i.e., $C^I = Q^I \langle S'^I,S^I \rangle_{(I)}^*$. Since 
$\langle S'^I,S^I \rangle_{(I)}^*$ depends
on the noise PSD of the detector and the time delay, which are different
for each $I$, the components of the correlation vector get scaled 
differently for each $I$ which makes the vector ${\sf C}$ move out of 
${\cal H}$. Owing to this mismatch ${\sf C}$ may lie outside ${\cal H}
(\theta, \phi)$ as well as 
${\cal H}(\theta',\phi')$. However, the maximization over 
$\epsilon$ and $\psi$ requires projecting ${\sf C}$ onto the template helicity 
plane ${\cal H}(\theta',\phi')$ in order to obtain the network statistic, 
$\parallel {\sf C}_{\cal H}\parallel$. Thus, the value of the computed
$\parallel {\sf C}_{\cal H} \parallel$ can decrease due to two effects:\\
(a) reduction in the norm of ${\sf C}$, \\
(b) ${\sf C}$ moving out of the signal helicity plane.

We assume that the orientation of the helicity plane changes slowly as 
compared to the effect of the time delays. This means that we treat 
$p'_I{}^J$ as effectively constants in (\ref{lsq}), and equal to the
corresponding tensor for the source parameters, namely, $p_I{}^J$. The 
validity of this assumption, to a good approximation, is supported by 
the extensive numerical computations that we have performed for the
networks and parameters that we have considered. Since we consider the 
mismatch to be quite small ($3\%$), the templates are closely spaced in 
the direction angles and hence the approximation is valid to 
about few parts in $10^3$ or even better.
Thus, from Eqs. (\ref{lsq}) and (\ref{metg}) we obtain the metric to be 
\be
g_{\alpha \beta} (\vt) \approx {1 \over 4} p_I{}^J \Re [Q^I Q_J^*] g_
{(I)(J)\alpha \beta} \ \ ,
\ee
where we used the fact that both $p^J_I$ and $g_{(I)(J)\alpha \beta}$ are
symmetric under the interchange of $I$ and $J$. Also, $g_{(I)(J)\alpha \beta} 
= -{\left({{\partial^2 {\Theta_{(I)(J)}}}/ {\partial \Delta \vt^\alpha 
\partial \Delta \vt^\beta}}\right)} |_{\Delta \vt^\gamma = 0}$.
The reality of $g_{\a \b}$
is now manifest. Owing to the linearity of $\Phi$ in $\Delta \vt^\alpha$,
the metric $g_{\alpha\beta}$ depends only on its first derivatives.
Therefore, it can be easily shown that
\be \label{gI}
g_{(I)(J)\alpha \beta} = \langle \Phi_\alpha \Phi_\beta \rangle_{(I)} + 
\langle \Phi_\alpha \Phi_\beta \rangle_{(J)}
        - \langle \Phi_\alpha \rangle_{(I)} \langle \Phi_\beta \rangle_{(J)}
        - \langle \Phi_\beta \rangle_{(I)} \langle \Phi_\alpha \rangle_{(J)}
\ \ , \ee
where the suffix $\alpha$ denotes the derivative with respect to $\Delta 
\vt^\alpha$. The angular bracket denotes the average over a given frequency 
range. {}For the frequency range of
$[f_s,\infty]$, the average value of the function $X_{(I)}(f)$ is denoted as,
\be\label{XIf}
\langle X \rangle_{(I)} = \int_{f_{s(I)}}^{\infty} {\cal A}_{(I)}(f) 
X_{(I)}(f) df \ \ ,
\ee
where within the angular brackets we have dropped the subscript on $X$ simply
because the same subscript appears outside those brackets. In other words, we
have reduced redundancy by 
introducing the notation: $\langle X \rangle_{(I)} = 
\langle X_{(I)} \rangle_{(I)}$.
We observe that (\ref{gI}) is a generalization of Owen's formula in Ref.
\cite{Owen}, wherein the metric for the single-detector case was derived.
It is not difficult to understand the origin of the different factors in the
expression for $g_{\a \b}$. This metric gets contribution from every pair of
detectors in a network, including the diagonal terms (i.e., terms with $I=J$), 
through the ``coupling'' metric $g_{(I)(J)\alpha \beta}$. The magnitude of 
each of these contributions is determined by their respective coupling 
strengths in the form of coefficients,
$p_I{}^J \Re [\left(Q^I Q_J^* \right)]$, which depend on the four angles
$\{\epsilon,\psi,\t,\p\}$. This is because these coefficients essentially
arise from the extended beam-pattern functions of the detectors, which, apart
from depending on the signal amplitude through $\epsilon$, determine how 
sensitive a given detector is to a source direction and $\psi$.

The above expressions allow one to calculate the parameter space metric for
any Earth-based network. However, since the metric is non-flat (as opposed to 
a flat metric for a single-detector ``network''), the template spacings 
$\D\mbox{\boldmath $\vt$}$ will depend on the location, $\mbox{\boldmath 
$\vt$}$, of the template. 
The general expressions for the moment functionals are:
\be
\langle \Phi_\a \rangle_{(I)} = 2\pi \vp_{(I)s \a} j_{(I)} (7-3m_\a) \ \ ,
\ee
where $\vp_{(I)s \a} \equiv \vp_{(I) \a}(f_{(I)s};\vt^\a)$ and $m_\a$ 
is the power of $f$ on which $\vp_{(I)\a}$ depends and $j_{(I)} (q)$ 
is $q$-th noise moment of the noise-curve corresponding to the $I$-th 
detector and is defined in appendix \ref{app:noise}.
Similarly,
\be
\langle \Phi_\a \Phi_\b \rangle_{(I)} = 4\pi^2 \vp_{(I)s \a}
\vp_{(I)s \b} j_{(I)} (7-3(m_\a+m_\b)) \,.
\ee
In terms of the above expressions, the coupling metric is 
\bea \label{gIJfs}
g_{(I)(J)\alpha \beta} = 4\pi^2 \big\{ &\big[& \vp_{(I)s \a}
\vp_{(I)s \b} j_{(I)} (7-3(m_\a+m_\b)) \no\\
-&& \vp_{(I)s \a} \vp_{(J)s \b} j_{(I)}(7-3m_\a)j_{(J)} (7-3m_\b)
\big] +\left[ I\leftrightarrow J\right]\big\} \,.
\eea

Another form of the parameter-space metric that proves useful in later
computations is obtained by taking projections on the vectors ${\sf D}_p$ 
($p = \pm 2$), which were first used in Eq. (\ref{Q}). Such projections allow
us to isolate the dependence of the metric on the parameters 
$\{\epsilon,\psi\}$. This simplification arises because ${\sf Q}$ lies in the 
plane of ${\sf D}_{+2}$ and ${\sf D}_{-2}$. To see this, let us define the 
following four quantities,
\be \label{gpm}
g_{\pm 2,\pm 2 \a \b} = {1 \over 4} p_{IJ} D^{I}_{\pm 2} D^{J}_{\pm 2}
g_{(I)(J)\a \b} \,.
\ee 
It follows from aforementioned properties that
\be
g_{-2,2 \a \b} = g_{2,-2 \a \b}, \hspace{0.2in} g_{-2,-2 \a \b} =
(g_{2,2 \a \b})^* \ \ ,
\ee
and that $g_{-2,2 \a \b}$ is real. The metric is then given by
\be \label{gpmmet}
\parallel {\sf E} \parallel^{2} g_{\a \b} = {{1 + \cos^2 \epsilon}
\over 2} g_{-2,2 \a \b} + {{1 - \cos^2 \epsilon} \over 2} 
\Re [ e^{-4i\psi} g_{2,2 \a \b}] \,.
\ee
We make use of this form in computing the cost for the LIGOs-VIRGO network.

Before we proceed to the discussion of various cases of networks, we mention a
scaling property of the phase $\Phi$. If $\Phi$ is scaled by a constant $a$, 
i.e., if $\Phi = a \tilde \Phi$, then the components of the metric get
scaled by $a^2$, viz., $\gamma_{\alpha \beta} = a^2 \tilde \gamma_{\alpha 
\beta}$. Assuming the dimension of the search parameter-space to be $P$, one
finds that $\det \parallel {\gamma}_{\alpha \beta} \parallel = a^{2P} \det 
\parallel {\tilde \gamma}_{\alpha \beta} \parallel$. Thus, the proper volume 
scales as ${\cal V} = a^P \tilde {\cal V}$. 

We now obtain the errors in determining the direction to the source. The
errors are obtained via the covariance matrix. Note that $\gamma$ is the
metric on the three-dimensional parameter subspace with $\{\xi,\t,\p\}$ as
the coordinates. {}From a
statistical point of view, $\gamma_{ij}$ is the Fisher information
matrix for a signal of unit amplitude \cite{Hels}. The covariance
matrix for a signal of unit amplitude is just $\gamma^{ij}$, the
inverse of $\gamma_{ij}$. The covariance matrix for a signal of 
arbitrary amplitude is obtained from $\gamma^{ij}$ by dividing it by 
the square of the signal amplitude. 
The diagonal elements of the covariance matrix are the variances in the errors 
of the estimated parameters. The errors in the estimates of the parameters
are given by the corresponding standard deviations,
\be \label{resol}
\sigma_{(i)} \equiv {\sqrt{\gamma^{ii}} \over {\sf b}} \ \ ,
\ee 
where ${\sf b}$ is the amplitude of the signal. 
In the next section, we list
$\sigma_{(i)}$ in Table \ref{tab:3} for various networks.

\subsection{Computational costs}
\label{subsec:costs}

We now estimate the computational cost involved in 
searching over the parameters $\theta$, $\phi$ and $\xi$.
We consider data trains of duration $T$ seconds sampled at the rate
$\Delta^{-1}$. The number of sampled points in each of the $M$ data trains is 
denoted by $N = T /\Delta$. As remarked before, the computational 
cost involved in obtaining the statistic has two important contributions:
\begin{itemize}
{\item The cost involved in computing Fourier transforms, denoted by $C^{FT}$.}
{\item The cost in computing the optimal statistic while searching over 
the time delays or, analogously, over the source-direction angles, 
$\{\theta$, $\phi\}$. We denote this cost by $C^{\Omega}$, where $\Omega 
\equiv (\theta,\phi)$.}
\end{itemize}
Also, we consider two different ways of searching over the time delays:
\begin{enumerate}
{\item Scanning all the time-delays in a window, $W$.}
{\item Using selected values of time-delays from the template bank.}
\end{enumerate}
{}For the quantities relevant to $(1)$, we attach a subscript $S$ (for 
sampling) and to $(2)$ a subscript $B$ (for bank-of-templates).
Let the number of templates required in searching over $\xi$ in these two
methods be denoted by 
$n^{\xi}_S$ and $n^{\xi}_{B}$, respectively. Then $n_B^{\xi}$ is obtained by
placing the templates in the $P$-dimensional space, while $n_S^{\xi}$ is
obtained by placing the templates just in the $\xi$ coordinate. 
Thus, $n_S^{\xi}$ is the proper length associated with the parameter range
of the $\xi$-coordinate, multiplied by $\rho_1(\mu) = 1/(2 {\sqrt{\mu}})$.
This yields,
\be
n_S^{\xi} = {{\pi f_s} \over {\sqrt{\mu}}} {\sqrt {\tilde \gamma_{11}}}
(\xi_{\max} - \xi_{\min}) \,.
\ee
{}For detectors with identical noise curves, $n^{\xi}_S$ is
just the number of templates required for searching over $\xi$ in the
one-detector case for a given coalescence phase, $\delta_c$. The
quantity $n^{\xi}_B$ for a detector network, with all detectors having 
identical
noise, is just $\sqrt P$ times the number of templates in $\xi$ for a single
detector. The factor of $\sqrt P$ comes from the fact that the template must
reach out to signals that mismatch with templates in all the parameters.
{}For a two-detector network, since the bank is in $\xi$ and $\theta$, we have
$P=2$. {}For networks with three or more detectors, the search is over three 
parameters and, therefore, $P=3$.

Let us assume that the FT of the templates are stored in memory and the FT
of the data has been taken. Then the computing cost in FT for the two
cases is \cite{BFS1},
\be
C_{S,B}^{FT} = 6 n_{S,B}^{\xi} M N \log_2 N \ \ ,
\ee  
where we have included a factor of 2 for the two sets of templates 
corresponding to $\delta_c = 0,
\pi/2$. We assume that the vectors ${\hat {\sf v}}_{\pm}(\theta,\phi)$ are 
stored in the memory for each pair of $(\theta,\phi)$ in the window/template 
bank, typically few thousand for the $LHV$ network. The number of real 
floating 
point operations (``fl-pt ops'' for short) for constructing the statistic 
$\Lambda$ is $8M+3$ 
(which essentially behaves as $8M$ when $M$ is large) for each point in the
sky, $(\theta,\phi)$. Hence the costs in two cases are,
\bea \label{costw1}
C_S^{\Omega} &\simeq& 16 M N n_S^{\Omega} n_S^{\xi} \ \ , \nonumber \\
C_B^{\Omega} &\simeq& 8 M N n_{tot} \equiv 16 M N n_B^{\Omega} n_B^{\xi}\ \ , 
\eea
where $n_{tot} = 2 n_B^{\Omega} n_B^{\xi}$.
This equation defines $n_B^{\Omega}$. The total costs 
are given by,
\be \label{cost}
C_{S,B} = 2 n_{S,B}^{\xi} M N (8 n_{S,B}^{\Omega} + 3 \log_2 N) \,.
\ee
In (\ref{cost}), we have ignored the overhead costs. Also, in the template
bank case, we have ignored the cost involved in computing ${\sf C}$
at non-sampled values, which can be obtained from Shannon's theorem.
The length of the data that is effectively processed is equal to the
length of the zero padding. This is because if the time-of-arrival of the
signal exceeds the padding duration, the longest chirp can extend out of
the data train. Thus we need to consider times of arrival only up to  the
padding length. The next data train must begin at this instant leading to
an overlap between successive data trains. To process the data online requires 
that the processing rate be at least equal to the rate of data acquisition. 
Since the length of the processed data is just the length of the
padding with zeros in the template, the online computing speed $S_S$ 
and $S_B$ are given by dividing the computing costs by 
the padding duration, $3\xi_1$. In the next section, we obtain 
the template bank and the computational costs for various networks.

\section {Examples of detector networks: Results}
\label{difnet}

Before considering the case of the actual network of laser-interferometric
detectors being built around the globe, we shall first consider some idealized
cases that are simple to analyze. Such an exercise is meant to
provide us with some useful estimates on the number of templates required, 
computational costs, etc. We begin by verifying that in the case of a single 
detector the formalism developed in the previous sections yields the results
expected from earlier studies \cite{SD,Owen}. Then we apply our formalism
to cases of networks with two and three identical detectors, respectively, 
with identical noise PSDs. Assuming a common noise simplifies computation of  
the metric on the signal parameter space. Purely for the purposes of obtaining 
estimates we choose the noise PSDs in these cases to be that of LIGO-I, with 
$f_s = 40$ Hz. Here, cases of non-coincident as well as arbitrarily oriented 
detectors are also studied. Subsequently, we generalize these analyses to 
obtain estimates for realistic cases of networks comprising of the LIGO and 
VIRGO detectors: These include two-detector networks, which pair up the two 
LIGOs or VIRGO with one of the LIGOs, and the three-detector network that 
includes VIRGO and both the LIGOs. In Tables \ref {table2} and \ref {tab:3}
we summarize our
results, which include the computational speed requirements, network 
sensitivities, and source-direction resolutions for all of these networks. 

\subsection{The one-detector 'network'}

It is instructive to start with the one detector case since it lays the
foundation for the $M$-detector case, the analysis of which would be the 
final goal. More pertinently, the number of templates in $\xi$ for the 
single-detector case is required for computing $n_\xi$ in the case of networks 
with more than one detector. We first verify
that our solution gives the expected estimates for $M=1$. 
{}For a single detector, we have $\vt^0 = f_s t_c$ and $\vt^1
= f_s \xi$. The network statistic is
\begin{eqnarray}
\L &=& |C_1^* Q^1| = |C_1| \nonumber \\
&=& \left|\int_{f_s}^{\infty} {\cal A}(f) \exp \left(-i \Phi (f;\vt^\mu,
\Delta \vt^\mu) \right) df \right| \ \ ,
\end{eqnarray}
where the phase $\Phi$ can be derived from Eqs. (\ref{phiI}) and (\ref{psiI}) 
by setting $I=1$ and the time-delay term to zero in those equations,
respectively. 

For the above statistic, the {\em exact} metric $g_{\alpha \beta}$ can be 
obtained from Eq. (\ref{metg}). Using the scaling $\Phi = 
2 \pi \tilde \Phi$, the scaled metric is 
\be
{\tilde g}_{\alpha \beta} = {1 \over 2}[\langle \tilde \Phi_\alpha \tilde 
\Phi_\beta \rangle 
        - \langle \tilde \Phi_\alpha \rangle \langle \tilde \Phi_\beta 
\rangle] \ \ ,
\ee
where the moment functionals can be expressed in terms of the noise moments 
listed in Appendix \ref{app:noise}. The metric then reduces to
\be
{\small
{{\tilde g}_{\alpha \beta}} = {1 \over 2} \left(\begin{array}{cc}
k_1& k_3\\ k_3& k_2 \end{array}\right) \ \ ,}
\ee
where $k_1$, $k_2$, and $k_3$ are certain useful combinations of noise 
moments and are defined in Eqs. (\ref{k1k2k3}) (here we have dropped the 
detector index $I$ from those combinations for obvious reasons). Projecting 
orthogonal to $\vt^0$, we find
\be
{\tilde \gamma}_{11} = {1 \over 2} \big(k_2 - {k_3^2 \over k_1}\big)\equiv 
{1 \over 2} k^2 \,.
\ee
The parameter-space in this case is just one-dimensional. Its volume is the 
proper length 
\begin{eqnarray} \label{vol3}
{\cal V} = 2 \pi \tilde {\cal V} &=& 2 \pi f_s \int_{\xi_{min}}^{\xi_{max}}
\sqrt{{\tilde \gamma}_{11}}  \> d\xi \nonumber \\
&=& \sqrt{2} \pi f_s k (\xi_{max} - \xi_{min}) \,.
\end{eqnarray}
Since the number density of templates here is $\rho_1 (\mu) = 1/(2\sqrt{\mu})$,
the number of templates (including 2 sets of templates for searching over
$\d_c$) is just
\be 
n_{tot}= \sqrt {2 \over \mu} \pi f_s k (\xi_{max} - \xi_{min}) \,.
\ee 
For LIGO-I noise we have $k = 0.062$, and the corresponding $n_{tot}$ 
is given in Table \ref {table2} for the parameter ranges listed there.

\subsection{Two-detector networks}

\subsubsection*{\textsf{(a) Two non-coincident, identical detectors with 
identical noise PSDs and identical orientations}}

We consider a network of two identical detectors with identical
noise PSDs. We make the following choice of coordinates. We 
choose one of the two detectors to be the fide. The $z$ axis of the fide is 
chosen along the line joining the two detectors. Then the second detector is
taken to be located at $(0,0,z_2)$, with an orientation identical to that of 
the fide, i.e., $\alpha_{(2)} = \beta_{(2)} = \gamma_{(2)} = 0$. Owing to the 
same orientations, the beam-pattern functions of the two detectors are 
identical. If the detectors were located at the same place, then the
resulting network would have mimicked a single detector, but
with a higher sensitivity. Here, however, we consider spatially separated 
detectors, where the relative time delay, $\tau_{(2)}$, provides partial 
information about the source-direction, namely, $\theta$. The time delay in 
units of $f_s^{-1}$ is given by 
\be
f_s \tau_{(2)} = z_2 (\hat {\sf z} \cdot \hat {\sf n}) \ \ ,
\ee
where we measure $z_2$ in units of the fiducial wavelength, $\lambda_s$
(which is $\approx$ 7500 km for $f_s = 40$ Hz). Thus, 
the network statistic in this case is
\begin{eqnarray} \label{L2}
\L &=& | {\sf C} \cdot {\sf Q'}| \label{stat2}\\
&=& {1 \over {\sqrt 2}} |C_1^* + C_2^*| \,.
\end{eqnarray}
Note that $Q'^1 = Q'^2$ and $|Q'^1| = |Q'^2| = {1 /{\sqrt 2}}$. This means 
that we have no information about $\epsilon$ and $\psi$. {}For a two-detector 
network, any given value of the time delay corresponds to more than one 
source directions, all of which lie on the surface of a cone whose axis 
coincides with the line joining the two detectors. Only when the source lies
on the line passing through the two detectors is the time delay 
single-valued, and is of maximum magnitude for a given pair of detectors  
(note that we have allowed $\tau_{(I)}$ to be negative as well).
The value of the time delay $\tau_{(2)}$ determines the opening angle
of the cone. Thus, the azimuthal direction angle
$\phi$ of the wave remains undetermined in the case of two detectors.
Only $\theta $ can be estimated
from the time delay that appears in the phase difference of the 
detector responses. 

As in case $(a)$, the exact metric $g_{\alpha \beta}$ for the statistic 
(\ref{L2}), can be obtained directly from Eq. (\ref{metg}). The corresponding 
scaled metric is
\be 
{\tilde g}_{\alpha \beta} = {1 \over 4} \sum_{I,J} {1 \over 2}\big[\langle 
{\tilde \Phi}_\alpha {\tilde \Phi}_\beta \rangle_{(I)} 
        - \langle {\tilde \Phi}_\alpha \rangle_{(I)} \langle 
{\tilde \Phi}_\beta \rangle_{(J)}\big] \ \ ,
\ee
where the $\Phi_{(I)}$ are defined in Eq. (\ref{phiI}). The metric on
the three-dimensional parameter space $\{f_s t_c,f_s \xi,n_3\}$ can now be 
given in terms of the noise moments as
\be
{\small
{\tilde g}_{\alpha \beta} = {1 \over 2} \left(\begin{array}{ccc}
k_1&\hspace{0.05in} k_3&\hspace{0.05in} z_2 k_1/2\\ k_3&\hspace{0.05in} k_2& 
\hspace{0.05in} z_2 k_3/2\\
z_2 k_1/2 &\hspace{0.05in} z_2 k_3/2&\hspace{0.05in} {z_2^2} (k_1 + j(1))/4
\end{array}\right) \ \ ,}
\ee
where $j(1)$ is a noise moment defined in Eq. (\ref{jq}).\footnote{We have 
dropped the detector index $I$ from the noise moment combinations $k_1$, 
$k_2$, and $k_3$ (see Appendix \ref{app:noise}) since here the noise moments 
are identical for the two detectors.}
After maximization ${\tilde g}_{\alpha \beta}$ over the time of coalescence, 
we deduce the metric ${\tilde \gamma}_{ij}$ to be
\be
{\small
{{\tilde \gamma}_{ij}} = {1 \over 2} \left(\begin{array}{ccc}
{{(k_1 k_2 - k_3^2)}/k_1}&0\\0& z_2^2 j(1)/4\ \end{array}
\right) \,.}
\ee
Here, a vanishing $\gamma_{12}$ implies that there is no covariance between 
$\xi$ and $\theta$. This is however not true when the noise curves are assumed 
to be different for the two detectors. This is so in case $(c)$, which we 
discuss below.

The volume on the 2-D parameter space $\{f_s \xi,n_3\}$ is obtained by 
integrating $\sqrt{\det \parallel {\tilde \gamma} \parallel}$ over $n_3$ and 
$\xi$. The proper volume of interest is 
\be
{\cal V} = (2 \pi)^2 \tilde {\cal V} = 2 \pi^2 f_s k' {z_2} (\xi_{max} - 
\xi_{min}) \ \ ,
\ee 
where $k' \equiv \sqrt {{j(1) (k_1 k_2 - k_3^2)}/k_1}$.
{}For the two-detector network, 
the number density of templates is $\rho_2(\mu) = 
1/(2 \mu)$. Therefore, the number of templates is
\be
n_{tot} = {2 \pi^2 \over \mu} f_s k' {z_2} (\xi_{max} - \xi_{min})\,.
\ee
For LIGO-I noise, $k' = 0.288$. The value of $n_{tot}$ for this case is given 
in Table \ref {table2}.

\subsubsection*{\textsf{(b) Two non-coincident, identical detectors with 
identical noise PSDs, but with different orientations}}
\label{subsec:2Ddifforient}

We make the choice of coordinates identical to that in case $(a)$ above. Since 
the two detectors have different orientations, the beam-pattern functions for 
the two detectors differ, i.e., $Q^1 \neq Q^2$. This has the implication that 
more information about the signal parameters, namely,  
$\epsilon$ and $\psi$, can be obtained. Since here we have only two dimensions 
on the network space to contend with, the network correlation-vector, 
${\sf C}$, always lies in ${\cal H}$ and no projection is required. Therefore,
the problem of maximization of the LLR over the angles $\{\epsilon,\psi\}$
reduces to aligning ${\sf Q'}$ along ${\sf C}$. Thus, the network statistic 
simplifies to
\be
\L = \parallel {\sf C} \parallel = \big( |C^1|^2 + |C^2|^2 \big)^{1/2} \,.
\ee
The metric ${\tilde g}_{\alpha \beta}$ on the parameter space with coordinates 
$\{f_s t_c,f_s \xi,n_3\}$ is obtained from Eq. (\ref{metg}) to be exactly
\be \label{scalmet}
{\tilde g}_{\alpha \beta} = {1 \over 2} \sum_{I} |Q^I|^2 \big[\langle 
{\tilde \Phi}_\alpha {\tilde \Phi}_\beta \rangle_{(I)} 
	- \langle {\tilde \Phi}_\alpha \rangle_{(I)} \langle 
{\tilde \Phi}_\beta \rangle_{(I)} \big] \ \ ,
\ee
where the $\tilde \Phi_{(I)}$  are the same as in case $(a)$ above.
Setting $|Q^1|^2 \equiv \eta$ and, therefore, $|Q^2|^2 = 1 - \eta$, the metric 
in terms of $\eta$ and the noise moments is
\be
{\small
{{\tilde g}_{\alpha \beta}} = {1 \over 2} \left(\begin{array}{ccc}
k_1&\hspace{0.05in} k_3&\hspace{0.05in} z_2 (1 - \eta)k_1 \\ k_3&\hspace{0.05in} k_2& \hspace{0.05in} z_2 (1 - \eta)k_3 \\
 z_2 (1-\eta)k_1 &\hspace{0.05in} z_2 (1-\eta)k_3&\hspace{0.05in} 
z_2^2 (1-\eta) k_1 \ \end{array}\right) \,.}
\ee
The associated $\tilde \gamma_{ij}$ is then given by
\be
{\small
{{\tilde \gamma}_{ij}} = {1 \over 2} \left(\begin{array}{ccc}
{{(k_1 k_2 - k_3^2)}/k_1}&0\\0& z_2^2 \eta(1-\eta)k_1\ \end{array}
\right) \,.}
\ee
We note that the $\sqrt{\rm{det} \parallel {\tilde \gamma}_{ij} \parallel}$
depends on $\eta$, which is a function of source-direction  
$(\theta, \phi)$ as well as the angles $\{\epsilon,\psi\}$. Thus, for every 
single source-direction we have a two-parameter family of metrics 
${\tilde \gamma}_{ij}$ (dependent on $\epsilon$ and $\psi$). Clearly, the
template spacings in this case will vary with their locations. For the purpose 
of obtaining estimates, we perform a simplification by opting to choose a bank 
of templates that is the finest over these two parameters. To do so, we first 
maximize ${\rm{det} \parallel {\tilde \gamma}_{ij}\parallel}$ over $\epsilon$ 
and $\psi$ and then compute the volume. The parameters $\{\epsilon,\psi\}$
appear in the determinant only through the factor $\sqrt{\eta (1-\eta)}$. 
The value of $\eta$ for which the determinant is maximized is
$\eta = 1/2$. We prove in Appendix \ref{app:2D} that one can always find a 
physically allowed pair of $\{ \epsilon,\psi\}$ such that the value 
$\eta = 1/2$ is attainable for any given pair of $\{\theta,\phi\}$ and for any 
orientations of the detectors. We use this value of $\eta$ to compute the
parameter-space volume.

The proper volume after multiplying by the appropriate scaling factor is
\be
{\cal V} = 2 \pi^2 k'' f_s(\xi_{max} - \xi_{min}) 
{z_2}\,.
\ee
where $k'' \equiv \sqrt{k_1 k_2 - k_3^2}$, which is equal to $\sim 0.13$ for 
LIGO-I noise. As in case $(a)$, the number of templates is arrived at by 
multiplying the proper volume by the number density of templates. 

\subsubsection*{\textsf{(c) General case of two detectors}}

Here, we typically consider the case of a network comprising of the VIRGO 
detector and one of the LIGO detectors, say, the one at Louisiana, for 
concreteness. (The results do not differ much if we replace in our 
calculations the numbers corresponding to the detector at Louisiana with those
describing the 4 km long Hanford detector.) We assume 
the noise curves of the respective detectors to be those given in Table 
\ref {table:noise} \cite{OS}. Here, we
have a case in which the seismic cut-offs are different.
Labeling the VIRGO detector as 1 and LIGO as 2, we have $f_{s(1)} = 16$ Hz
 and $f_{s(2)} = 40$ Hz. The important implication of 
this is that the signal in the VIRGO detector will last longer by a factor
$(f_{s(1)}/f_{s(2)})^{-8/3} \sim 11.5$. Thus, till
the chirp reaches the frequency of $40$ Hz, essentially it is 
only one detector,
namely, VIRGO that contributes to the SNR. The longest chirp (for $m_1 = m_2 
\approx 0.5 M_{\odot}$)
in the detector output lasts for about $1588$ sec. Accordingly, we choose the
data segments to be of duration 5000 sec. each, and assuming the same sampling 
rate of $2$ kHz for both detectors, we have $N\sim10^7$. We now compute the 
metric and the number of templates for this network.

The expression for the scaled metric in this case is the same as the one in
Eq. (\ref{scalmet}). Let $z_2$ denote the distance between the detectors as in 
case $(a)$ and let $|Q^1|^2 = \eta$ as in $(b)$. Also, we take the fiducial
frequency to be $f_s = f_{s(2)}$, without any loss of generality. Then 
the metric is given by
\be
{\small
{{\tilde g}_{\a \b}} = {1 \over 2}\left(\begin{array}{ccc}
k_{1(2)} + \eta (\varrho^2 k_{1(1)} - k_{1(2)}) & 
k_{3(2)} + \eta(\varrho^{-2/3} k_{3(1)} - k_{3(2)})&
 (1-\eta) z_2 k_{1(2)} \\
 k_{3(2)} + \eta(\varrho^{-2/3} k_{3(1)} - k_{3(2)})\; \; \:& 
 k_{2(2)} + \eta(\varrho^{-10/3} k_{2(1)} - k_{2(2)})\; \; \:
& (1-\eta) z_2 k_{3(2)} \\
 (1-\eta) z_2 k_{1(2)} & 
(1-\eta) z_2 k_{3(2)} & (1 - \eta) z_2^2 k_{1(2)}
\end{array}\right) \ \ ,}
\ee
\noindent where we have used $\varrho_{(1)} = f_{s(1)}/f_{s(2)}\equiv 
\varrho$ (see Eq. (\ref{newtmomfunc})) and $\varrho_{(2)} = 1$.
{}For the network under consideration $\varrho = 0.4$, and $k_{1(I)}$, 
$k_{2(I)}$,
$k_{3(I)}$ for VIRGO and LIGO-I are listed in Table \ref {table:noise}. From 
Eq. (\ref{metric}) we compute ${\tilde \gamma}_{ij}$. We find that
$\sqrt{\det \parallel {\tilde \gamma}_{ij} \parallel} = z_2 B(\eta)$, with
\be
B(\eta) \simeq
{{[\eta (0.23 + 13.33 \eta + 6.51 \eta^2 - 20.07 \eta^3)]^{1/2}}
\over {4.57 + 6.95 \eta}} \ \ ,
\ee
which is a smoothly varying function of $\eta$ that attains its 
maximum value of $\sim 0.22$ at $\eta \sim 0.6$. 

The volume of the parameter space is
\be
{\cal V} \simeq 1.76 \pi^2 z_2 f_s (\xi_{max} - \xi_{min}) \ \ ,
\ee 
and the corresponding number of templates is
\be \label{nLV}
n_{tot} \simeq {{1.76 \pi^2} \over {\mu}} z_2 f_s (\xi_{max} - \xi_{min}) \ \ ,
\ee
where $n_{tot}$ is listed in Table \ref {table2}. 

We do not specify the computing speed for the sampling method in this case 
because it is not clear what noise curve one must choose
to obtain one detector templates in $\xi$.

\subsection{Three-detector networks}

\subsubsection*{\textsf{(a) Three non-coincident identical detectors with 
identical noise PSDs and identical orientations}}

We consider a network of three detectors with identical noise PSDs. The
detectors are spatially separated and have identical orientations.
Such a situation will be difficult, if not impossible to realize on a spherical
Earth. However, in this simple case our goal is to obtain order of magnitudes 
estimates for computational costs, etc. 
We treat one of the three detectors to be
the fide. We choose the coordinate system of the fide as follows:
The $z$-axis of the fide is along the line joining
the fide and one of the remaining detectors. Thus, the second detector
is located at $(0,0,z_2)$. The $x$-axis is chosen such that the plane formed
by the network coincides with the 
$x-z$ plane. The spatial coordinates of the third
detector are $(x_3,0,z_3)$. The detectors have identical orientations, i.e.,
$\alpha_{(I)} = 0 =\beta_{(I)} = \gamma_{(I)}$. Hence, they 
have identical antenna-pattern functions, i.e., $Q'^1 = Q'^2 = Q'^3$. Then, 
the network statistic simplifies to 
\begin{eqnarray}
\L &=& |{\sf C} \cdot {\sf Q'}| \nonumber \\
&=& {1 \over {\sqrt 3}} |C_1^* + C_2^* + C_3^*| \,.
\end{eqnarray}
Note that, the network of three spatially separated detectors provide two
independent relative time delays, $\tau_{(2)}$ and $\tau_{(3)}$, which 
determine the two possible source directions as follows. {}For each pair of
detectors in such a network, the time delays draw a circle in the sky for 
possible source locations. The intersections of 
two such circles determine two possible source directions. Here, the 
time delays in units of $f_s^{-1}$ are
$f_s\tau_{(2)} = z_2 (\hat {\sf z} \cdot {\hat {\sf n}})$ and $f_s \tau_{(3)} =
x_3 (\hat {\sf x} \cdot {\hat {\sf n}}) + z_3 (\hat {\sf z} \cdot {\hat 
{\sf n}})$. We measure $z_2$, $z_3$, and $x_3$ in units of fiducial 
wavelength $\lambda_s$. 

The exact scaled metric is obtained via Eq. (\ref{metg}) to be  
\be 
{\tilde g}_{\alpha \beta} = {1 \over 9} \sum_{I,J} {1 \over 2}\big[\langle 
{\tilde \Phi}_\alpha {\tilde \Phi}_\beta \rangle_{(I)} 
        - \langle {\tilde \Phi}_\alpha \rangle_{(I)} \langle {\tilde \Phi}_
\beta \rangle_{(J)}\big] \ \ ,
\ee
where the $\tilde{\Phi}_{(I)}$ can be obtained from Eq. (\ref{Phicompact}) by
using for the time delays, $f_s \Delta \tau_{(1)} = 0$, $f_s \Delta \tau_{(2)} 
= z_2 \Delta n_3$, and 
$f_s \Delta \tau_{(3)} = x_3 \Delta n_1 + z_3 \Delta n_3$. The resulting
components of the symmetric metric ${\tilde g}_{\alpha \beta}$, on the space of the 
variables $\{f_s t_c,f_s \xi,n_3,n_1\}$, are 
\be
{\small
\tilde{g}_{\a\b} = {1 \over 2} \left(\begin{array}{cccc}
\hspace{0.1in} k_1\hspace{0.1in}& \hspace{0.1in} k_3 \hspace{0.1in}&
\hspace{0.1in} (z_2 + z_3)k_1/3 \hspace{0.1in} &
\hspace{0.1in}  x_3 k_1/3 \hspace{0.1in} \\
\hspace{0.1in}  \cdot \hspace{0.1in} &
\hspace{0.1in} k_2 \hspace{0.1in} & 
\hspace{0.1in}(z_2 + z_3) k_3/3 \hspace{0.1in} &
\hspace{0.1in} x_3 k_3/3 \hspace{0.1in} \\
\hspace{0.1in}  \cdot \hspace{0.1in} &
\hspace{0.1in}  \cdot \hspace{0.1in} &
\hspace{0.1in} [3 (z_2^2 + z_3^2)j(1) - (z_2 + z_3)^2 j(4)^2]/9\hspace{0.1in}&
\hspace{0.1in} x_3 [3 z_3 j(1) - (z_2 + z_3) j(4)^2]/9 \hspace{0.1in}\\
\hspace{0.1in}  \cdot \hspace{0.1in} &
\hspace{0.1in}  \cdot \hspace{0.1in} &
\hspace{0.1in}  \cdot \hspace{0.1in} &
\hspace{0.1in} x_3^2 (3j(1) - j(4))/9 \hspace{0.1in}
\end{array}\right) \,.}
\ee
Maximization over $f_s t_c$ gives
\be
{\small
{{\tilde \gamma}_{ij}} = {1 \over 2} \left(\begin{array}{ccc}
\hspace{0.1in}(k_1 k_2 - k_3^2)/k_1 \hspace{0.1in}&
\hspace{0.5in} 0 \hspace{0.5in}&\hspace{0.3in} 0 \hspace{0.3in}\\
0& 2(z_2^2 + z_3^2-z_2 z_3)j(1)/9&x_3(2 z_3 - z_2)j(1)/9\\
0& x_3(2 z_3 - z_2)j(1)/9& 2 x_3^2 j(1)/9
\end{array}\right)\,.}
\ee
Then the proper volume is 
\be
{\cal V} = (2 \pi)^3 {\tilde {\cal V}} = {{16 \pi^4} \over {3 \sqrt 6}}
j(1) k {\cal A} f_s(\xi_{max} - \xi_{min}) \ \ ,
\ee 
where ${\cal A}$ is the area of the network. {}For the three-detector network
discussed here, the parameter space is three-dimensional and
the number density of templates is 
$\rho_3(\mu) = 3 {\sqrt 3}/(8 \mu^{3/2})$. For typical 
parameters the quantities of interest are listed in 
Table \ref {table2}. 

\subsubsection*{\textsf{(b) The network comprising of LIGO (Livingston), LIGO 
(Hanford), and VIRGO (Pisa) with their respective noise curves.}}
\label{Sec:LLG}

We finally discuss the case of a non-coincident three-detector network 
involving the LIGO and VIRGO detectors with their respective noise curves. We 
denote such a network as $LHV$. The detector noise PSDs are 
represented by analytical fits given in Ref. \cite{OS}. These fits are
reproduced here in Table \ref {table:noise}. We assume LIGO-I noise in both of 
the LIGO detectors. Table \ref {table:loc1} lists the locations and the
orientations of the detectors on the globe \cite{Allen2}. 
In order to compute the metric
we choose the fide frame to be the network frame with $L$ at its origin, 
$H$ lying on the $z$-axis, and $V$ lying in the $x-z$ plane. In units of 
$\lambda_s$, the dimensionless position-vectors of the detectors
are given by
\be
{\bf r}_L = (0,0,0) \ \ , \hspace{0.5in} {\bf r}_H = (0,0,0.40) \ \ ,
\hspace{0.5in}
{\bf r}_V = (1.05,0,0.29) \,.
\ee
Note that since the $y$-component of each of these vectors is
zero in such a frame (i.e., $r_{(I)2} = 0$, for $I=$1, 2, 3), the moment 
functionals given in Eq. (\ref{newtmomfunc}) 
simplify considerably. An inspection of that equation shows that in this case
the noise moments do not depend on the direction to the source
and, hence, the $g_{(I)(J)\a \b}$ are constants. Therefore, this
choice of the fide frame simplifies the computations involved.
The metric $g_{\a \b}$ however, depends on $n_1$
and $n_3$ through ${\sf Q}$. The orientations of the detectors
can easily be obtained from Table \ref {table:loc1}.
In the fide (network) frame, they are given by
\bea
\{\a_L,\b_L,\g_L\} &=& \{38.11^{\circ},256.35^{\circ},107.43^{\circ}\}  
\ \ ,\no\\
\{\a_H,\b_H,\g_H\} &=& \{38.09^{\circ},283.54^{\circ},196.88^{\circ}\}  
\ \ ,\no\\
\{\a_V,\b_V,\g_V\} &=& \{320.34^{\circ},275.92^{\circ},159.02^{\circ}\}  
\,.
\eea

\begin{table}[!htb]
\caption{
Locations and arm orientations of Earth-based interferometric 
gravitational-wave detectors. The length of each arm is given in meters. The
location of the corner station of each detector is given in terms of the 
latitude and longitude there. The orientation of an arm is given by the angle
through which one must rotate it clockwise (while viewing from top) to point 
north.}
\begin{center}
\begin{tabular}{ccccccc} 
Project & Location & Year & Length (m)& Corner
location & Arm 1 & Arm 2 \\ \hline
TAMA-300 & Tokyo, JPN & 1998 & 300 & $35.68^\circ$N $\quad
139.54^\circ$E  & $90.0^\circ$ & $180.0^\circ$ \\
GEO-600 & Hannover, GER & 1999 & 600 & $52.25^\circ$N $\quad
9.81^\circ$E  & $25.94^\circ\>$& $291.61^\circ\>$\\
VIRGO   & Pisa, ITA & 2000  & 3000 & $43.63^\circ$N $\quad
10.5^\circ$E  & $71.5^\circ$ & $341.5^\circ$  \\
LIGO    & Hanford, WA, USA  & 2000  & 4000        & $46.45^\circ$N
$\quad -119.41^\circ$E & $36.8^\circ$ & $126.8^\circ$ \\
LIGO    & Livingston, LA, USA & 2000  & 4000        & $30.56^\circ$N
$\quad -90.77^\circ$E & $108.0^\circ$ & $198.0^\circ$ \\
AIGO    & Gingin (Perth), AUS & TBA   & 80       & -$31.04^\circ$N
$\quad 115.49^\circ$E  & $180^\circ$ & $270^\circ$ \\
\end{tabular}
\end{center}  
\label{table:loc1}
\end{table}

The numerical code that we have developed for this case first computes 
$g_{(I)(J)\a \b}$
from the moment functionals given in Eq. (\ref{newtmomfunc}). Then Eq. 
(\ref{gpm}) is used to compute $g_{\pm 2,\pm 2 \a \b}$ and, subsequently, 
the metric $g_{\a \b}$ is obtained from Eq. (\ref{gpmmet}) as a
function of $n_1$, $n_3$ (analogously, $\{\theta,\phi\}$), $\epsilon$
and $\psi$. This metric is then projected orthogonal to $\vt^0$ to obtain 
$\gamma_{ij}$. Since our goal here is to get estimates (within
a factor of $10$) of the online computational speed requirements, we obtain 
the parameter-space volume of such a search by integrating 
$\sqrt{\det \parallel \gamma \parallel}$ over the parameters $\theta$, $\phi$, 
and $\xi$ for a few chosen values of $\epsilon$ and $\psi$. From this volume 
we derive the number of templates, $n_{tot}$, the computational cost, and the 
online speed needed for this network. 

The parameter-space volume is given by,
\be
{\cal V} \simeq f_s \xi_{max} \int \int \sqrt{\det \parallel \gamma \parallel} 
\> dn_1 dn_3 \,.
\ee
We have evaluated this integral for several values of $\epsilon$ and
$\psi$. Its average value turns out to be $\sim 250$. 
For most of the astrophysically interesting ranges for $\epsilon$ and
$\psi$, the proper volume does not vary by more than a factor of 3 beyond 
this value. As before, taking $f_s = 40$ Hz,
$\xi_{max} \sim 138$ sec., $\rho_3(\mu = 0.03) = 125$, we find the number
of templates to be $n_{tot} \sim {\rm few \hspace{0.2cm}times} \times 10^8$.
The computational cost using a template bank is obtained from Eq.~(\ref{cost}).
This cost is essentially given by the search over the time-delay window and,
hence,
\be
C_B \simeq 24 N n_{tot} \,.
\ee
If we take a data train $5000$ sec. long corresponding to 
$\xi_{(1)max} = 1588$ sec. for VIRGO, we have 
$N \simeq 10^7$, and the computing cost is 
$C_B \sim 10^{17}$ fl-pt ops. For online 
processing, this data must be processed in about $3412$ sec., yielding an 
online speed requirement of about few tens of Tflops.  

\subsection{Numerical results}

In Table \ref {table2}, we summarize the numerical results for various 
networks. We list the total number of templates, $n_{tot}$, required for a 
search over $\xi$, $\theta$, and $\phi$ and also the break up into 
$n_B^{\xi}$ and $n_B^{\Omega}$. In the case of a two-detector network, 
$n_B^\Omega$ corresponds to a 1-dimensional grid in $\theta$. We also give
the corresponding values for $n_S^{\Omega}$. {}Finally in the last two columns 
we provide the online computational speeds where we have taken the data train 
to be of $500$ sec.
duration and the zero-padding is $500 - \xi_{max} = 362$ sec. long, except for 
the $LV$ network. In the $LV$ case, the duration of the data train is $5000$ 
sec., and the zero-padding is $3412$ sec. The
computational speeds $S_B$ and $S_S$ are obtained by dividing the
computational costs by the duration of the padding.

\begin{table}[!hbt]
\begin{center}{
\begin{tabular}{ccccccccc}
Network &  $n_{tot}$ & $n_B^{\xi}$ &  $n_B^{\Omega}$ & 
$n_S^{\Omega}$ &  $C_B$ &  $C_S$ & $S_B$ &  $S_S$ \\
configuration &
{}&{}&{}&{}& (fl-pt ops) & (fl-pt ops) & (Gflops) & (Gflops) \\ 
\hline
$I$&$8.9 \times 10^3$&$4.4 \times 10^3$& {-} & - & $5.3 \times 10^{11}$&
-&${1.5}$&- \\
$II-a$& $1.6 \times 10^6$ &$ 6.2 \times 10^3$ & 129 & 170 &{$2.7\times 
10^{13}$}&{$2.5 \times 10^{13}$}&{75}&{69} \\
$II-b$& $ 7.7 \times 10^5$ &$ 6.2 \times 10^3$ & 62 & 170 &$ 1.4 \times 
10^{13}$& $2.5 \times 10^{13}$&{39} &{69} \\
$LH$& $1.9 \times 10^5$ &$ 6.2 \times 10^3$ &15 & 40 &$4.5 \times 10^{12}$& 
$6.9 \times 10^{12}$ &{12} &{19}\\
$LX_V$& $5.3 \times 10^5$ & $ 6.2 \times 10^3$ &43 &105 &$1.0 \times 
10^{13}$& $1.6 \times 10^{13}$&{28}&{44} \\
$LX_T$& $6.2 \times 10^5$ &$ 6.2 \times 10^3$ &50&128&$ 1.1 \times 10^{13}$
&$ 1.9 \times 10^{13}$&{33}&{52} \\
$LX_A$& $8.0 \times 10^5$ &$ 6.2 \times 10^3$ & 65 & 166 &$ 1.4 \times 
10^{13}$& $ 2.4 \times 10^{13}$&{39}&{69} \\
$LV$&$3.5 \times 10^6$&$1.9 \times 10^4$&92&105&$6.1 \times 10^{14}$& {-}&
170&{-}\\
$III$& $3.6 \times 10^8$ &$ 7.6 \times 10^3$&$2.4 \times 10^4$ &$1.5
\times 10^4$ &$8.6 \times 10^{15}$& $3.2 \times 10^{15}$&{$2.3 \times 10^4$}&{
$8.8\times 10^3$} \\
\end{tabular}
}
\end{center}
\caption{The table lists number of templates, computational costs,
and online computing speeds required for a search using specific networks.
The detector networks are labeled as $I$ for a single detector,
$II-a$ for two identical detectors with identical orientations located 
diametrically opposite on the surface of the earth,
$II-b$ for two identical detectors with arbitrary orientations located 
diametrically opposite on the surface of the earth, and
$III$ for three identical detectors with identical orientations placed on 
Earth's equator forming an equilateral triangle.
The detector $X_D$ denotes a detector with LIGO-I noise at the location
of the detector $D$. The letters $L$, $H$, $V$, $T$, and $A$ denote the 
detectors, LIGO detector at Louisiana, LIGO detector at Hanford (the one with 
arms of length 4 km each), VIRGO, TAMA and 
AIGO sites, respectively. We assume LIGO-I noise for both the LIGO detectors.
We take the lower limit on the masses to be $0.5 M_{\odot}$,
i.e., $M_1$ and $M_2$ each $\ge 0.5 M_{\odot}$. Thus, we have 
$\xi_{max} \sim 138$ sec. for $f_s = 40$ Hz,  except for the
$LV$ case (where $f_s = 16$ Hz). We consider data trains of $500$ sec. sampled 
at $2$ kHz so that $N \sim 10^6$, $n_S^{\xi} = 4.4 \times 10^3$. 
For the $LV$ network, $\xi_{(1){max}} \sim 1588$ sec. and, therefore, for a
data train of length $5000$ sec., one finds $N \sim 10^7$.}
\label{table2}
\end{table} 

\section{False alarms, Detection probabilities, and Vetoes}
\label{FDV}

\subsection{False alarm and detection probabilities}
\label{FD}

The maximum-likelihood method involves computing the likelihood ratio for 
given data and comparing it with a predetermined threshold. In some cases it is
more useful to replace the likelihood ratio by another quantity derived from 
it. When the likelihood ratio, or the LLR, is a function of several parameters,
it is often possible to maximize it analytically over some of these parameters,
as we have shown here. In our case, 
such a maximization led to a reduced statistic derived from the LLR. We call 
this statistic $\Lambda$. To detect the presence of a signal in the data one 
must, therefore, compute the values that $\Lambda$ takes at all the grid 
points on the space of the 
remaining parameters. Each of these values must then be compared with the 
threshold, $\Lambda_0$, to infer the presence or absence of a signal.
The value of
$\Lambda_0$ can be obtained via the Neyman-Pearson decision criterion
 \cite{Hels1}, given the predetermined value of the false-alarm probability, 
$Q_0$, associated with the
event of detection of the signal. When $\Lambda < \Lambda_0$, we conclude 
that the signal is absent in the data, whereas when $\Lambda > \Lambda_0$,
the detection of the signal is announced.

To compute the false-alarm probability and the detection probability, $Q_d$,
we need to know the probability distribution of $\Lambda$ in the absence of
the signal, i.e., $p_0 (\Lambda)$, and in the presence of the signal, i.e., 
$p_1 (\Lambda)$. We note that $\Lambda$ is a sum of squares of the random 
variables $c_0^+$, $c_{\pi/2}^+$, $c_0^-$, and $c_{\pi/2}^-$.
If our assumed properties of the detector noises are valid, then in the 
absence of a signal, i.e., for hypothesis $H_0$, each of the random variables 
$c_0^+$, $c_{\pi/2}^+$, $c_0^-$, and $c_{\pi/2}^-$
has a mean equal to zero. To see this, let
\be \label{cvec}
{\sf C} = {\sf {c_0}} + i {\sf {c_{\pi/2}}} \ \ ,
\ee
where 
\be
{\sf {c_0}} \equiv \{ c_0^I \} \quad {\rm and} \quad  
{\sf {c_{\pi/2}}} \equiv \{ c_{\pi/2}^I \} \,.
\ee
Further, define
\be
c_0^{\pm} \equiv {\sf {\hat v}}^{\pm} \cdot {\sf {c_0}} \quad {\rm and} \quad  
c_{\pi/2}^{\pm} \equiv {\sf {\hat v}}^{\pm} \cdot {\sf {c_{\pi/2}}} \,.
\ee
Then, from Eq. (\ref{Cstar}) it follows that $c_0^I$, $c_{\pi/2}^I$ and,
therefore, $c_0^{\pm}$ and $c_{\pi/2}^{\pm}$, each has a vanishing mean. 
{}From the assumed independence of noise among the different detectors and the
orthonormality between ${\hat {\sf v}}^+$ and ${\hat {\sf v}}^-$, and also
between $s_o^I$ and $s_{\pi/2}^I$, we obtain the covariances between
$c_0^{\pm}$ and $c_{\pi/2}^{\pm}$ as well as the covariances between 
$c_0^{\pm}$ and $c_{\pi/2}^{\mp}$ to be zero. On the other hand, the variances 
of each of $c_0^{\pm}$ and $c_{\pi/2}^{\pm}$ is unity. Thus, under 
the $H_0$ hypothesis, $\Lambda$ is the sum of squares of the
independent Gaussian random variables [see (\ref{SSnew})] with mean zero 
and unit variance. We conclude from standard literature (see, e.g., Ref.
\cite{Hels}) that the probability density function for
$\Lambda$, under the $H_0$ hypothesis, is a $\chi ^2$ distribution with four
degrees of freedom, and is given by
\be \label{p0}
p_0 (\Lambda) = {\Lambda \over 4} \exp(\Lambda/2) \,.
\ee
The false alarm probability $Q_0$ is then obtained to be
\be \label{q0}
Q_0 = \int_{\Lambda_0}^{\infty} p_0 (\Lambda) d\Lambda = \left(1 + {\Lambda_0 
\over 2}\right) \exp(-\Lambda_0/2) \,.
\ee
The value of $Q_0$, which is inferred from astrophysical estimates of event
rates and detector sensitivities, determines the detection threshold 
$\Lambda_0$ through the above equation.
 
The detection probability is obtained from the probability distribution
$p_1(\Lambda)$. In order to calculate $p_1(\Lambda)$, we need the norm of the 
average network correlation-vector when the template gives a perfect match
with the data. Assuming that the strength of the signal in the data is 
${\sf b}$, i.e., if $x^I(t) = {\sf b} {\hat s^I(t)} + n^I (t)$, then the
average value of the network correlation-vector is $\bar {\sf C} = {\sf b} 
{\sf Q} e^{-i \delta_c}$. Therefore, 
\begin {eqnarray}  \label{aveC}
\parallel {\overline {\sf C}} \parallel^2 &=& {\overline {\sf c}}_0^2 + 
{\overline {\sf c}}_{\pi/2}^2 \nonumber \\
&=& {\overline {c^+}}_0^{2} + {\overline {c^-}}_0^{2} + 
{\overline {c^+}}_{\pi/2}^2 + {\overline {c^-}}_{\pi/2}^2 = {\sf b}^2 \ \ ,
\end {eqnarray}
for which we obtain (see, e.g., Ref. \cite{Hels})
\be \label{p1}
p_1(\Lambda) = {1 \over 2} \left({\sqrt {\Lambda} \over {\sf b}}\right)^{1/2} 
\exp \left[-{{\Lambda +
{\sf b}^2} \over 2} \right] I_1({\sf b}\sqrt{\Lambda}) \ \ ,
\ee
where $I_1$ is the modified Bessel function. The detection probability itself
is
\be
Q_d = \int_{\Lambda_0}^{\infty} p_1(\Lambda) d\Lambda \ \ ,
\ee
which we now obtain in the large SNR limit. 
In terms of the network statistic, $\L \equiv \sqrt{\Lambda}$, this
asymptotic limit amounts to the condition ${\sf b} \L >> 1$, and Eq. 
(\ref{p1}) approximates to a Gaussian distribution:
\be \label{p2}
p_1(\L) = {1 \over {\sqrt {2 \pi}}} \exp \left[-{{\left(\L - 
{\sf b} \right)^2} \over 2}\right] \,.
\ee
Thus, in the large SNR limit the network statistic is a
Gaussian with mean approximately equal to the network strength of
the signal, ${\sf b}$. For the networks considered in Sec. \ref{difnet}, we 
summarize in Table \ref {tab:3} detection thresholds, the resolution achievable
in the direction to the source, and the relative sensitivity of the network 
compared to that of a single detector. We take the false-alarm rate to be one 
per year and the detection probability to be 
95$\%$. Then, assuming a sampling rate of about 2 kHz,
we arrive at a false-alarm probability of 
$Q_0 \sim 1.5 \times 10^{-11}/n_{tot}$.
{}For the sake of this calculation, we assumed that output samples in the
correlation vector are uncorrelated. The correlation between these samples will
reduce $Q_0$ but this does not make appreciable difference to the
thresholds and sensitivities \cite{SVDS}.
The threshold $\Lambda_0$ is then computed using Eq. (\ref{q0}). 

We define the sensitivity of an $M$-detector network relative to that of a 
single detector to be equal to
$\sqrt{\sum_{I=1}^M {g_{(I)}^2/g_{(1)}^2}}
(\L_{0(1)}+ \Delta \L)/(\L_{0(M)} + \Delta \L)$, where 
$\L_{0(M)}$ is the threshold corresponding to a network of $M$
detectors (therefore, $\L_{0(1)}$ is the threshold for a single detector) and 
$\Delta \L$ is the solution of
\be
{1 \over {\sqrt {2 \pi}}} \int_{-\infty}^{\Delta \L} e^{-x^2/2} dx = Q_d \equiv
0.95 \,.
\ee
It yields $\Delta L \approx 1.64$. The sensitivity (which is $> 1$) is roughly 
proportional to the average distance at which one can detect a source of a 
given $\xi$ with a network of $M$ detectors as 
compared to a single detector, with $95\%$ detection probability, where the 
average is taken over all directions and
orientations of the binary. {}For a network of detectors 
with identical orientations the result is exact.

The resolution
in the direction to the binary is obtained from Eq. (\ref{resol}). It is 
obtained by noting that the error in the $n_3-n_1$ plane is given by
\be\label{n1n3error}
b^{-2} \sqrt{\g^{22} \g^{33} -(\g^{23})^2} = \sin\t \cos\phi \>\sigma_\Omega \ \ ,
\ee
where $\sigma_\Omega$ is the resolution given in terms of the angles $\theta$ 
and $\phi$. We take
SNR $\sim 12$, sufficiently above the threshold to guarantee a good detection 
probability. We compute the source-direction resolutions $\sigma_{\theta}$ 
and $\sigma_{\Omega}$ in the case of two- and three-detector networks,
respectively. The  $\sigma_{\Omega}$ is obtained from the covariance matrix
using the above equations.  
In Table \ref {tab:3}, we give the values for source-direction resolution for
a direction normal to the plane in which the detectors lie. The reason we 
choose this direction is because in this direction we expect the resolution 
to be high.
The big difference in the values of $\sigma_\Omega$ between case $III$ and 
$LHV$ is because $III$ is a degenerate case of a network of identical 
detectors, which are merely spatially separated.

\begin{table}[!hbt]
\caption{Total number of templates, false-alarm probabilities, 
detection thresholds, relative sensitivities, and the resolutions in
the direction to the source for various network configurations. We have 
$g_{(1)}^2 / g_{(2)}^2 = 1.58$, where the subscripts 1 and 2 correspond to 
VIRGO and LIGO, respectively. The $g_{(I)}$ are needed to compute network  
sensitivities relative to that of a single detector with LIGO-I noise.}
\begin{center}
\begin{tabular}{cccccc}
Network & $n_{tot}$ & $Q_0$ &  $\L_{0(M)}$ & Relative &
$\sigma_{\theta}$ \\
configuration &{}&{}&{}& sensitivities & or $\sigma_{\Omega}$ \\ 
\hline
$I$&$8.9 \times 10^3$&$1.7 \times 10^{-15}$& {8.7} &1  &{-} \\
$II-a$& $1.6 \times 10^6$ &$ 9.4 \times 10^{-18}$ & 9.3 & 1.3 &{$0.3^{\circ}$}\\
$II-b$& $ 7.7 \times 10^5$ &$ 1.9 \times 10^{-17}$ & 9.2 & 1.3 &{$0.6^{\circ}$}\\
$LH$& $ 1.9 \times 10^5$ &$ 7.9 \times 10^{-17}$ & 9.0 & 1.4 &{$2.5^{\circ}$}\\
$LX_V$& $ 5.3 \times 10^5$ &$ 2.8 \times 10^{-17}$ & 9.2 & 1.4 &{$0.9^{\circ}$}\\
$LX_T$& $ 6.2 \times 10^5$ &$ 2.4 \times 10^{-17}$ & 9.2 & 1.3 &{$0.8^{\circ}$}\\
$LX_A$& $ 8.0 \times 10^5$ &$ 1.9 \times 10^{-17}$ & 9.2 & 1.3 &{$0.6^{\circ}$}\\
$LV$& $ 3.5 \times 10^6$ & $4.3 \times 10^{-18}$ & 9.4& 1.5 &{$0.7^{\circ}$}\\
$III$& $ 3.6 \times 10^8$ &$ 4.2 \times 10^{-20}$ & 9.9 & 1.6 &{$0.15$ sq.deg.}\\
$LHV$& $ 3.5 \times 10^8$ &$ 4.3 \times 10^{-20}$ & 9.9 & 1.7 &{$1.2$ sq.deg.}\\
\end{tabular}
\end{center}
\label{tab:3}
\end{table} 

\subsection{Vetoing non-Gaussian events in detector noises}
\label{subsec:veto}

The assumptions of Gaussianity and stationarity of noise in detectors is an 
idealistic one. The noise in actual detectors will not, in general, satisfy 
these assumptions, but will rather contain a non-Gaussian component arising 
from causes such as sudden strain releases in mechanical structures, 
ringing from electronic servo loops, etc. Their deviations from Gaussianity 
are poorly understood and are difficult to model. But, since such noise
components usually have high amplitude, they can trigger the statistic $\L$ to 
register a ``detection'' within the scope of the methods described so far. 
This is where a ${\chi}^2$-type test described in Ref. \cite{Allen} can be 
used to discriminate against such contingencies
by using the specific spectral profile of a chirp $({|\tilde {S}(f)|}^2 
\propto f^{-7/3})$, which is different from that of a non-Gaussian event, in 
general. We briefly describe the test below.

The frequency bandwidth in each detector, from $f=0$ to $f=f_{\rm Nyquist}$, 
is divided into $p$-subintervals in the following way. Let,
\be 
\langle x^{I}, y^{I}\rangle_{(I)k} = 
2 \Re \int_{f_{k(I)}}^{f_{k+1(I)}} {{ {\tilde{x}}_{I}^{*}(f) {\tilde {y}}^I 
(f)} \over {s_{h(I)} (f)}} df \ \ ,
\ee
where in the integrand the index $I$ is not summed over. Using the above
definition, one partitions the interval $[0,f_{\rm Nyquist}]$ by setting 
$\langle s_0^I, s_0^I \rangle _{(I)k} =\langle s_{\pi/2}^I, s_{\pi/2}^I 
\rangle _{(I)k} = 1/(2 p)$. This way, for each detector, $I$, we get a 
partition $0, f_{1(I)}, f_{2(I)},...,f_{p(I)}=f_{\rm Nyquist}$. Next, one
computes the following correlation over each subinterval $k$:
\be 
\label{C_KI}
C_{Ik}^{*} \equiv \langle S^{I}, x^{I}\rangle_{(I)k} \,.
\ee
The individual detector-correlations can, therefore, be expressed as 
\be
 C^{I} = \sum_{k=1}^{p} C_{k}^I \,.
\ee
Define the deviation in $C_{k}^I$ from the average contribution to $C^I$ from
the $k$-th frequency bin to be
\be
\D C_{k}^I \equiv C_{k}^I - C^{I}/p \ \ ,
\ee
which, by definition, obeys $\sum_{k=1}^{p} \D C_{k}^I = 0$. Then the 
$\chi^{2}$ statistic is given by
\be
\chi^{2}_{(I)} = p \sum_{k=1}^{p} |\D C_{k}^I|^{2} \ \ ,
\ee
which has $2p-2$ degrees of freedom. If the detector $I$ is behaving 
``properly'', that is, if the detector output is mainly the Gaussian noise,
with or without a chirp, then $\chi_{(I)}^{2}$ has a small value. On the other 
hand, a relatively large value of $\chi_{(I)}^{2}$ is taken to indicate 
non-Gaussianity. Choosing $p=20$, as in Ref. \cite{Allen}, we see that 
$\chi_{(I)}^{2}$ has 38 degrees of freedom. Defining $\chi_{*}^{2}$  to be the 
decision threshold, if $\chi_{(I)}^{2} > \chi_{*}^{2}$, then we reject the 
hypothesis that the event is a signal, else we accept that there is a signal 
present. {}For 38 degrees of freedom at $90$ percent confidence level, one
finds $\chi_{*}^{2} \sim 50$.

We apply the test to the network in the following way. Suppose, the statistic
$\L$ exceeds the threshold $\L_0$ for some time lag $\tau = \tau_0$. 
After accounting for the relative time delays, 
we compute the $C_{(k)}^I$ at $\tau_0$ and construct 
$\chi_{(I)}^2$, for all $I$.
Next, we test whether each detector satisfies the assumption
of Gaussianity and stationarity by comparing $\chi_{(I)}^2$ with $\chi_*^2$.
If $\chi_{(I)}^2 < \chi_{*}^2$, for any $I$, then we accept the 
decision that we have actually detected a signal. On the other hand, if 
for some $J = I_1, I_2, I_3,...,I_{M_1}$, $\chi^2_{(J)} > \chi^2_*$, we 
ignore the contribution from these detectors in computing $\L$ 
and construct $\L'$ for the rest of the data from $M-M_{1}$ detectors. 
Now, if $\L'$ crosses the threshold, then we say that the signal is detected,
otherwise it is not. If non-Gaussian events occur relatively rarely, then 
it is unlikely that more then one detector simultaneously will have such 
events. Then $M_1$ on most occasions will be unity. This concludes a simple
generalization of the vetoing used for single detectors to the case of a
networks. 

\section{Concluding Remarks}
\label{conclu}
  
We have presented here a data-analysis strategy based on the maximum-likelihood
method for the detection of GW signals from inspiraling compact binary stars 
with a network of laser interferometric detectors. Our approach is based on a
coherent search of the data from all the detectors in a network and, therefore,
is inherently optimal. The formalism described
is mathematically elegant and simple. In Gaussian noise, the method is
tantamount to matched filtering the signal. However, the noise model 
we consider here is more realistic in that it allows for occasional
non-Gaussian bursts superposed on a predominantly Gaussian noise background.
Sections of data that contain non-Gaussian bursts are vetoed out 
by a $\chi^2$ criterion. For simplicity, we consider the Newtonian inspiral 
waveform, but it is clear that our formalism is as well applicable to 
waveforms depending on a larger number of intrinsic parameters, such as spins 
of the binary members.
In particular, the formalism is extendable to the restricted 2.5 PN 
inspiral waveform. In that event the number of network templates required 
increases essentially by the 
same factor as in the case of a single detector: Assuming LIGO-I noise in 
the detectors and
individual stellar masses of $0.5 M_{\odot}$ or larger, the increase in 
the number of templates is by a factor of about 
4 to 5, when the maximum allowed mismatch is $3 \%$. One would then expect the
computational cost to increase by a similar factor. We expect to look 
into this issue in greater detail in the future.

For the Newtonian case, the online computational speed requirements  
are high - from tens
of Gflops for a network of two detectors to a few tens of Tflops for a network 
of three widely separated detectors around the globe. Clearly, efficient
signal extraction methods are called for. A hierarchical search approach to 
this problem should, therefore, be explored. Alternatively, the search can 
be restricted to 
selected regions in parameter space, the selection of regions being 
based on prior astrophysical information. For example, the search may be 
restricted to the masses of the stars being greater than a solar mass,
the argument being that it is unlikely to find compact objects of
a smaller mass. This would reduce the computational cost by a factor of
4 in the Newtonian case.

The relative sensitivity of a network, on an average, increases by a factor
little less than $\sqrt{M}$, where $M$ is the number of detectors 
having identical noise PSDs. Although the signal energy accessible to the 
network on an average (with the average taken over all the directions and the 
orientations of the binary) increases by a factor of $M$, the change in the 
threshold value of the network statistic, 
which obeys the Rayleigh distribution in the absence of the signal,
is such that the overall factor of increase in sensitivity is a little less
than $\sqrt{M}$. In the case of detectors with different noise curves, the
quantity $g_{(I)}$ plays a central role. In such a case, the relative 
sensitivity
of a network on an average increases as $\sqrt{\sum_{I=1}^M {g_{(I)}^2/
g_{(1)}^2}}$ where the sensitivity is normalized to detector 1. Thus, 
a detector with a larger value of $g_{(I)}$ influences 
the relative sensitivity that much more. 

We also estimate the errors made in determining the direction to the
source by computing the covariance matrix which is just the inverse of the
metric obtained in the parameter space divided by the square of the SNR. 
We find that for a network of detectors spread around the globe, with all 
detectors having LIGO-I noise curve and an SNR
of 12, the resolution is about a fraction of a degree.

Our analysis essentially assumes Gaussian noise (with occasional non-Gaussian 
bursts). The fact that real detectors produce non-Gaussian and non-stationary
noise makes this issue highly relevant. This issue must be addressed more 
thoroughly in the future. Since the signals are generally weak in
nature, it is desirable that the search strategy be optimal. If a simple
enough mathematical model that adequately describes the noise in the
real detectors can be given, then our approach based on the maximum likelihood
method can still be explored. Creighton has already investigated this
approach where the model for noise contains Poisson-distributed bursts
superposed on the usual Gaussian component \cite{CRET}. Such an approach seems 
promising and could be investigated further. Another approach that is
simple, but suboptimal 
is that of matching event lists in each detector and 
putting thresholds on estimated source parameter differences.

These issues for more realistic noises and signals still remain to
be addressed. Here our main goal in this work is to provide a general
framework based on the method of maximum likelihood, 
which uses a coherent  
approach and is therefore optimal. Also
many of the results we obtain here may be scaled up in a straightforward
way to obtain order of magnitude estimates in more general situations.
The extension to the PN waveform is just one such case.

\acknowledgments

The authors would like to thank Bruce Allen, Eric Chassande-Mottin, S. D. 
Mohanty, B. S. Sathyaprakash, and Bernard Schutz for useful discussions. AP 
would like to thank the AEI, Potsdam, for a three month visit, the IAU for a 
travel grant, and the CSIR, India. 
The work of SB was funded in part by PPARC grant PPA/G/S/1997/00276.

\appendix
\section{STF tensors and Gel'fand functions: A representation of 
$SO(3)$}
\label{app:stf}

To understand the relation between the responses of two different detectors 
in a network to the same incoming chirp, it is useful to study the behavior
of the detector and wave tensors under three-dimensional orthogonal 
transformations. This is tantamount to developing an understanding of STF 
tensors (of rank 2) under the action of an element of the rotation group 
SO(3), $g(\a,\b,\g)$, where $(\a,\b,\g)$ are the Euler angles. Since we 
extensively deal with STF tensors of rank 2 in the text, enunciating some 
frequently used properties of such objects is in order.\footnote{For a 
detailed discussion, see Refs. \cite{GMS,KT1}. {}For a more selective reading 
of immediate relevance, we refer to Ref. \cite{DT1}.} 

Any STF tensor of rank $l$ can be expanded in a location-independent basis of
``STF-$l$'' tensors, ${\cal Y}_{lm}^{ij}$, which has a dimension of $2l+1$. 
STF-$l$ tensors, with rank $l=2$ are related to
the spherical harmonics as follows:
\be \label{stf2}
Y_{2m} (\t,\p) = {\cal Y}_{2m}^{ij} n_i n_j \>, \quad {\rm where} \quad
m=\pm 2, \>\pm 1,\> 0\ \ ,
\ee
where ${\bf n} = (\cos\p\sin\t,\> \sin\p\sin\t,\> \cos\t )$. The ``STF-$2$''
tensors defined above are also called spin-weighted spherical harmonics. They
obey the following normalization relation
\be\label{norm-stf2}
{\cal Y}_{2m}^{ij} {\cal Y}^{2m'*}_{ij} = {15\over 8\pi}\d_m^{m'} \,.
\ee
When one makes a passive orthogonal transformation of frames through the Euler
angles $\{\a,\b,\g\}$, the angles $\{\t,\p\}$ get relabeled to, say, 
$\{\t',\p'\}$. Then, the spherical harmonics in the new frame can be
expanded in terms of those in the old frame as,
\be
Y_{2m} (\t',\p') = T_{m}{}^{n} (\a,\b,\g) Y_{2n} (\t,\p) \ \ ,
\ee
where the right-hand side has an implicit summation over $n=0$, $\pm 1$, 
$\pm 2$. Above, the expansion coefficients, $T_m{}^{n}$, are the Gel'fand 
functions of rank $2$.

The group composition law of two elements of the rotation group, say, 
$g_1(\p,\t,\psi)$ and $g_2(\a,\b,\g)$ leads to the following addition 
theorem for the  Gel'fand functions:
\be
\label{add}
T_m{}^n(\p ^{'},\t ^{'},\psi ^{'}) = T_m{}^s(\p,\t,\psi) 
T_s{}^n(\a,\b,\g) \ \ ,
\ee
where once again the summation over $s=0$, $\pm 1$, $\pm 2$ on the 
right-hand side is understood. The transformation of STF tensors under 
rotation is governed by the above theorem for Gel'fand functions.

\section{GW polarization tensors and the detector tensor}
\label{app:poldeten}

The detector tensor for an interferometer is defined as
\be\label{dt}
d_{ij} = {\sin 2\O} \left(n_{1i} n_{1j} -  n_{2i} n_{2j}\right) 
\ \ ,\ee
where ${\bf n}_1$ and ${\bf n}_2$ are the unit vectors along the arms of
the interferometer and $2\O$ is the opening angle, i.e., the angle between 
its two arms \cite{DT1}. Here we shall take the detectors to have orthogonal 
arms. In that event
\be
{\bf n}_1 = \frac{1}{\sqrt 2}(1,-1,0)\>,\quad 
{\bf n}_2 = \frac{1}{\sqrt 2}(1,1,0) \ \ ,
\ee 
in the detector frame. 
When referred to the fide frame, however, ${\bf n}_{1,2}$ depend on
the Euler angles, $\{\a,\b,\g\}$, that give the orientation of the detector
relative to the fide. The dependence of $d_{ij}$ on these angles can be 
expressed in a neat form by realizing that it is a second-rank STF tensor, as
is evident from Eq. (\ref{dt}). Thus, it can be expanded in a basis of STF-2 
tensors. It can be shown that the expansion coefficients in such a basis 
are \cite{DT1} 
\be\label{dtcomp}
d_{ij} {\cal Y}^{ij}_{2n} = -i  \sqrt{15\over 8\pi}
\left(T^{2*}{}_n (\a,\b,\g)- T^{-2*}{}_n (\a,\b,\g)\right) \ \ ,
\ee
in the fide frame. Above, $n=0$, $\pm 1$, $\pm 2$, and the Gel'fand functions 
depend on $\{\a,\b,\g\}$.

Similarly, the corresponding components of $d_{ij}$ can be deduced 
in the wave frame. Apart from depending on the angles $\{\a,\b,\g\}$, these
coefficients will also depend on the orientation of the wave frame relative to
the fide, given by $\{\p,\t,\psi\}$. Using the addition theorem for Gel'fand
functions, these components are 
\be \label{DIp}
d_{ij} {\cal Y}^{ij}_{2n} = -i \sqrt{15\over 8\pi}
T_n{}^s (\p,\t,\psi)\left(T^{2*}{}_s (\a,\b,\g)- T^{-2*}{}_s (\a,\b,\g)\right) 
\equiv \sqrt{15\over 8\pi} D_n\ \ ,
\ee
in the wave frame. The extended beam-pattern function (\ref{ext1}) depends on 
the coefficients, $D_n$, with $n=\pm 2$ and $\psi=0$.  

\section{Noise curves and noise moments}
\label{app:noise}

We define the moments of the $I$-th detector's noise curve as,
\be\label{momdef}
i_{(I)} (q) = s_{h(I)} (f_{0(I)})\int_{1}^{f_{c(I)}/f_{s(I)}}dx
\frac{x^{-q/3}}{s_{h(I)}(xf_{s(I)} )} \ \ ,
\ee
where $f_{0(I)}$ denotes the ``knee'' frequency of that detector; this is the 
frequency at which the sensitivity of the detector is the highest. On the other
hand, $f_{c(I)}$ is its high-frequency cut-off and $f_{s(I)}$ is the seismic
cut-off. The noise moment, $i_{(I)}(7)$, is related to the normalization, 
$g_{(I)}$, as follows:
\be\label{gI7}
g_{(I)}^2 = {8\over 3s_{h(I)} (f_{0(I)})} 
\varrho_{(I)}^{-4/3} i_{(I)} (7) \ \ ,
\ee
where $\varrho_{(I)} \equiv f_{s(I)}/f_s$. 
Since our templates are normalized using the above factor, we find the
following ratio useful in our calculations of the parameter-space metric:
\be\label{jq}
j_{(I)}(q) \equiv i_{(I)} (q)/i_{(I)} (7) \,.
\ee
In this paper, we evaluate these noise moments using the analytical fits to 
noise PSDs of different detectors given in Ref. \cite{OS}. These fits are
presented in Table \ref {table:noise}

\begin{table}[!htb]
\caption{Analytical fits to noise PSDs, $s_h(f)$, of the interferometric 
detectors studied in this paper. Here $s_0$ denotes the minimum value of 
$s_h(f)$, and $f_0$ is the frequency at which this minimum occurs. We 
take $s_h(f)$ to be infinite below the seismic cut-off frequency $f_s$. We 
choose the high frequency cut-off, $f_{c(I)}$, to be $800$ Hz for all $I$.}
\vskip 5pt
\begin{tabular}{lccrr}
Detector & Fit to noise PSD & $s_0$ (Hz$^{-1}$) &
$f_0$ (Hz) & $f_s$ (Hz)\\
\hline
TAMA-300 & $s_0/32\,\left\{(f_0/f)^5+13(f_0/f)+9[1+(f/f_0)^2]\right\}$ &
$2.4\times10^{-44}$ & 400 & 75 \\
GEO600 & $s_0/5\,\left[4(f_0/f)^{3/2}-2+3(f/f_0)^2\right]$ &
$6.6\times10^{-45}$ & 210 & 40 \\
VIRGO & $s_0/4\,\left[290(f_s/f)^5+2(f_0/f)+1+(f/f_0)^2\right]$ &
$1.1\times10^{-45}$ & 475 & 16 \\
LIGO I & $s_0/3\,\left[(f_0/f)^4+2(f/f_0)^2\right]$ &
$8.0\times10^{-46}$ & 175 & 40 
\end{tabular}
\label{table:noise}
\end{table}

There are certain combinations of the noise moments that appear frequently 
in the expression for the metric on the parameter space relevant for a network
(see Sec. \ref{difnet}). In order to simplify these expressions, we define the 
following noise-moment combinations:
\bea \label{k1k2k3}
k_{1(I)} &\equiv& \left[j_{(I)}(1)-j_{(I)}^2(4)\right],\nonumber \\
k_{2(I)} &\equiv& 9\left[j_{(I)}(17)-j_{(I)}^2(12)\right]/25, \nonumber \\
k_{3(I)} &\equiv& 3\left[j_{(I)}(9)-j_{(I)}(4)j_{(I)}(12)\right]/5 \ \ ,
\eea
which are, in general, different for detectors with different noise PSDs. 

The moment functionals $\langle \Phi_\a \rangle_{(I)}$ and
$\langle\Phi_\a \Phi_\b \rangle_{(I)}$ defined in Eq. (\ref{XIf})
for the $I$-th detector, can be given in terms of the moments of its noise 
curve. They are (for $n_2 \neq 0$)
\bea\label{newtmomfunc}
\langle\Phi_0  \rangle_{(I)} &=& 2\pi \varrho_{(I)} ~j_{(I)}(4)\ \ , \no\\
\langle\Phi_1  \rangle_{(I)} &=& 6\pi \varrho_{(I)}^{-5/3}~j_{(I)}(12)/5\ \ , 
 \no\\
\langle\Phi_{2,3}  \rangle_{(I)} &=& 2\pi
\left[r_{(I)3,1} - r_{(I)2} n_{3,1}/n_2\right] 
\varrho_{(I)}
~j_{(I)}(4)\no\\
&=& \left[r_{(I){3,1}} - r_{(I)2} n_{3,1}/n_2\right] \langle\Phi_0  \rangle_{(I)} \ \ , \no\\
\langle \left(\Phi_0\right)^2  \rangle_{(I)} &=& 4\pi^2 \varrho_{(I)}^2
~j_{(I)}(1) \ \ , \no\\
\langle\Phi_0\Phi_1  \rangle_{(I)} &=& 12\pi^2 \varrho_{(I)}^{-2/3}~j_{(I)}(9)
/5 \ \ , \no\\
\langle\Phi_0\Phi_{2,3}  \rangle_{(I)} &=& 4\pi^2 
\left[r_{(I){3,1}} - r_{(I)2} n_{3,1}/n_2\right] 
\varrho_{(I)}^2~j_{(I)}(1)\no\\
&=& \left[r_{(I){3,1}} - r_{(I)2} n_{3,1}/n_2\right] \langle \left(\Phi_0\right)^2  
\rangle_{(I)} \ \ ,\no\\
\langle\left(\Phi_1\right)^2  \rangle_{(I)} &=& 36\pi^2
\varrho_{(I)}^{-10/3}~j_{(I)}(17)/25 \ \ ,\no\\
\langle\Phi_1\Phi_{2,3}  \rangle_{(I)} &=& 12\pi^2 \left[ r_{(I){3,1}} - r_{(I)2} n_{3,1}
/n_2\right] \varrho_{(I)}^{-2/3}~j_{(I)}(9)/5\no\\
&=&  \left[r_{(I){3,1}} - r_{(I)2} n_{3,1}/n_2\right] \langle\Phi_0\Phi_1  \rangle_{(I)} \ \ , \no\\
\langle\left(\Phi_{2,3}\right)^2  \rangle_{(I)} &=& 4\pi^2 \varrho_{(I)}^2
\left[r_{(I){3,1}} - r_{(I)2} n_{3,1}/n_2\right]^2~j_{(I)}(1) \no\\
&=& \left[r_{(I){3,1}} - r_{(I)2} n_{3,1}/n_2\right]^2 \langle \left(\Phi_0\right)^2  \rangle_{(I)} \ \ ,\no\\
\langle\Phi_2\Phi_3  \rangle_{(I)} &=& 4\pi^2 \varrho_{(I)}^2
\left[r_{(I){3}} - r_{(I)2} n_{3}/n_2\right] \left[r_{(I){1}} - r_{(I)2} n_{1}/n_2\right]
j_{(I)}(1) \no\\ 
&=& \left[r_{(I){3}} - r_{(I)2} n_{3}/n_2\right] \left[r_{(I)1} - r_2^I n_{1}/n_2\right] 
\langle \left(\Phi_0\right)^2 \rangle_{(I)} \ \ ,
\eea
which shows that all the moment functionals are expressible in terms of five 
independent noise-moments, $j_{(I)}(q)$. These are the ones corresponding to 
$q= 1,\>4,\>9,\>12,\>17$. The values of these noise moments and the 
combinations (\ref{k1k2k3}) 
for relevant noise PSDs are listed in Table \ref {table5}.
Alternatively, all the moment functionals are determined by five basic 
ones, namely, $\langle\Phi_0  \rangle_{(I)}$, $\langle\Phi_3 \rangle_{(I)}$,
$\langle\Phi_0^2 \rangle_{(I)}$, $\langle\Phi_0 \Phi_3 \rangle_{(I)}$, and 
$\langle\Phi_3^2 \rangle_{(I)}$. 

{}For a network of three or less detectors a plane can always be arranged to
contain the hubs of all the detectors. This makes it possible to choose the
fide frame (or the network frame, in this case) in such a way that 
$r_{(I)2} = 0$ for all $I$. With this choice the moment functionals
reduce to
\bea\label{newtmomfunc3D}
\langle\Phi_0  \rangle_{(I)} &=& 2\pi \varrho_{(I)} ~j_{(I)}(4) \ \ , \no\\
\langle\Phi_1  \rangle_{(I)} &=& 6\pi \varrho_{(I)}^{-5/3}~j_{(I)}(12)/5 
\ \ , \no\\
\langle\Phi_{2,3}  \rangle_{(I)} &=& r_{(I){3,1}}\langle\Phi_0  \rangle_{(I)}
\ \ , \no\\
\langle \left(\Phi_0\right)^2  \rangle_{(I)} &=& 4\pi^2 \varrho_{(I)}^2
~j_{(I)}(1) \ \ , \no\\
\langle\Phi_0\Phi_1  \rangle_{(I)} &=& 12\pi^2 \varrho_{(I)}^{-2/3}~j_{(I)}(9)
/5 \ \ , \no\\
\langle\Phi_0\Phi_{2,3}  \rangle_{(I)} &=& r_{(I){3,1}} 
\langle \left(\Phi_0\right)^2 \rangle_{(I)} \ \ , \no\\
\langle\left(\Phi_1\right)^2  \rangle_{(I)} &=& 36\pi^2
\varrho_{(I)}^{-10/3}~j_{(I)}(17)/25 \ \ , \no\\
\langle\Phi_1\Phi_{2,3}  \rangle_{(I)} &=& r_{(I){3,1}} \langle\Phi_0\Phi_1  
\rangle_{(I)} \ \ , \no\\
\langle\left(\Phi_{2,3}\right)^2  \rangle_{(I)} &=& r_{(I){3,1}}^{2} 
\langle \left(\Phi_0\right)^2  \rangle_{(I)} \ \ , \no\\
\langle\Phi_2\Phi_3  \rangle_{(I)} 
&=& r_{(I)3} r_{(I)1} \langle \left(\Phi_0\right)^2 \rangle_{(I)}\,. 
\eea
The moment functionals simplify significantly in this case.  

\begin{table}[!hbt]
\caption {Noise moments of some of the planned interferometric detectors.
In evaluating these, we take the values of $f_{s(I)}$ and $f_{0(I)}$ as 
given in Table \ref {table:noise}. The upper cut-off frequency, $f_{c(I)}$,
is assumed to be $800$ Hz for all detectors.}
\vskip 5pt
\begin{tabular}{cccccccccc}  
Noise moments&j(1)&j(4)&j(7)&j(9)&j(12)&j(17)&$k_1$&$k_2$&$k_3$\\
\hline
LIGO-I&21.3&4.089&1&0.444&0.157&0.045&4.572&0.007&-0.1197\\
VIRGO&132.4&7.774&1&0.407&0.167&0.068&71.99&0.0145&-0.5347\\
GEO-600&17.99&3.49&1&0.537&0.273&0.136&5.809&0.0222&-0.2493\\
TAMA&20.94&4.111&1&0.443&0.133&0.045&4.039&0.0099&-0.0625\\
WHITE NOISE&12.97&2.574&1&0.677&0.453&0.291&6.351&0.031&-0.293\\
\end{tabular} 
\label{table5}
\end{table}

\section{A network of two identical detectors with the same noise
PSD but different orientations.}
\label{app:2D}

Consider a network of two identical detectors having orientations
$\{\alpha_{(1)}$, $\beta_{(1)}$, and $\gamma_{(1)}\}$ and 
$\{\alpha_{(2)}$, $\beta_{(2)}$, and $\gamma_{(2)}\}$, respectively, which 
we take to be different. Then the beam-pattern functions of each detector 
are dependent on $\{\alpha_{(I)},\beta_{(I)},\gamma_{(I)},\epsilon,\psi,\theta,
\phi\}$, where $I=$ 1, 2. Here, we prove that for a given 
set of values for the 
detector orientations and source-direction, $(\theta,\phi)$, the 
function $|Q^1| |Q^2|$ can always attain the maximum value of $1/2$, with 
$|Q^1|=|Q^2|=1/{\sqrt 2}$. This proof is assumed in obtaining the result in 
Sec. \ref{subsec:2Ddifforient}. 
\par
\noindent {\bf Proof}:-

\noindent The network vector $\sf Q$ lies in the helicity plane ${\cal H}$.
Therefore,
\be \label{qf}
{\sf Q} =  Q^{-2} {\hat {\sf D}}_{-2} + Q^{+2} {\hat {\sf D}}_{+2} \ \ ,
\ee
where 
$Q_{+2} = {\hat {\sf D}}_{+2} \cdot {\sf Q}$
and $Q_{-2} = {\hat {\sf D}}_{-2} \cdot {\sf Q}$.
Alternatively, for a two-detector network we can expand ${\sf Q}$ in the
real basis of $I$'s on the network space.
\be \label{q1}
{\sf Q} = Q^1 {\hat {1}} + Q^2 {\hat {2}} \ \ , 
\ee
with $Q_1 = {\hat 1}\cdot {\sf Q}$ and $Q_2 = {\hat 2} \cdot {\sf Q}$. Using
Eqs. (\ref{qf}) and (\ref{q1}), we find $Q_1$ and $Q_2$ in terms of 
$Q^{-2}$ and $Q^{+2}$ as 
\bea
Q_1 &=& (Q^{-2} {D_{-2}}_1 +Q^{+2} {D_{+2}}_1)/\parallel {\sf D} \parallel
\ \ , \no\\
Q_2 &=& (Q^{-2} {D_{-2}}_2 +Q^{+2} {D_{+2}}_2)/\parallel {\sf D} \parallel
\ \ , 
\eea
where ${D_{\pm 2}}_1 = {\hat 1} \cdot {{\hat{\sf D}}_{\pm 2}}$ and 
${D_{\pm 2}}_2 = {\hat 2} \cdot {\hat{\sf D}_{\pm 2}}$. Thus,
\be \label{q12}
{Q_1 \over Q_2} = {{Q^{-2} {D_{-2}}_1 + Q^{+2} {D_{+2}}_1} 
\over {Q^{-2} {D_{-2}}_2 +Q^{+2} {D_{+2}}_2}}.
\ee
Let us assume that $Q_1$ differs from $Q_2$ by an overall phase factor $e^{i
\omega}$. Then, since $|Q^1|^2 + |Q^2|^2 = 1$ for such a network, we have
$|Q_1| = |Q_2| = 1/{\sqrt 2}$. We shall now prove that for a
given set of values for the detector orientations, source-direction, 
and $\omega$, one
can always find $\epsilon$ and $\psi$, within their physically allowed range,
i.e., $\epsilon \in [0,\pi]$ and $\psi \in [0, 2\pi]$, such that 
our above assumption is met. Equation (\ref{q12}), therefore, leads to
\begin{mathletters}
\label{qffmain}
\bea 
{Q^{+2} \over Q^{-2}} &=& {{{D_{-2}}_1 - {D_{-2}}_2 e^{i \omega}} 
\over {{D_{+2}}_2 e^{i \omega} - {D_{+2}}_1}} \label{qff} \\
&\equiv& \Upsilon \exp (i\upsilon) \label{qff2}\ \ ,
\eea
\end{mathletters}
where $\Upsilon$ and $\upsilon$ are real numbers. Note that the 
RHS of (\ref{qff}) can take any value in the complex plane. More importantly,
this is true also of the LHS of (\ref{qff}) because $ {Q^{+2}/Q^{-2}} = 
T^{-2}_{2}(\psi,\epsilon,0) / T^{2}_{2}(\psi,\epsilon,0)$ can always take 
any value on the Argand plane for astrophysically relevant ranges of $\epsilon$
and $\psi$. Thus, our assumption remains vindicated and, hence, one can always
choose values for $\epsilon$
and $\psi$ that maximize the function $|Q_1||Q_2|$ to attain the value of
1/2. These values corresponding to the maximum are $\psi = 
-\upsilon/4$ and $\epsilon = \cos^{-1} [(\Upsilon^{1/2} - 1)/
(\Upsilon^{1/2} + 1)]$.

\end{document}